\documentclass{aa}
\usepackage[breaklinks,colorlinks,linkcolor=blue,citecolor=blue,urlcolor=blue,hyperfootnotes=true]{hyperref}
\usepackage{graphicx}
\usepackage{txfonts}
\usepackage[figuresright]{rotating}
\usepackage{url}
\usepackage{graphicx}

\usepackage{xcolor}
\usepackage[normalem]{ulem}

\begin{document}

\title{Open clusters housing classical Cepheids in \emph{Gaia} DR3}

\author{C. J. Hao\inst{1,2}, Y. Xu\inst{1,2}, Z. Y. Wu\inst{3,4},
Z. H. Lin \inst{1,2}, S. B. Bian\inst{1,2}, Y. J. Li\inst{1}, D. J. Liu\inst{1,2}}

\institute{Purple Mountain Observatory, Chinese Academy of Sciences, Nanjing 210023, PR China
\email{xuye@pmo.ac.cn}
\and
School of Astronomy and Space Science, University of Science and Technology of China, Hefei 230026, PR China
\and
National Astronomical Observatories, Chinese Academy of Sciences, 20A Datun Road, Chaoyang District, Beijing 100101, PR China
\and
School of Astronomy and Space Science, University of Chinese Academy of Sciences, Beijing 101408, PR China
}

\date{Received 22 July 2022 / Accepted 23 September 2022}
\titlerunning{Open clusters housing classical Cepheids}
\authorrunning{C. J. Hao et al.}

\abstract
{
The latest \emph{Gaia} Data Release 3 provides an opportunity 
to expand the census of Galactic open clusters harboring classical 
Cepheid variables, thereby bolstering the cosmic 
distance scale. 
A comprehensive analysis yielded a total of 50 classical Cepheids 
associated with 45 open clusters, of which 39 open cluster--classical 
Cepheid pairs are considered probable, with the remaining 11 pairs 
considered improbable but worth following up. 
Two previously identified clusters by us possibly 
host classical Cepheids (OC-0125/V1788 Cyg and 
OC-0675/OGLE-BLG-CEP-114). 
In addition, we identify 38 new open cluster candidates within 
the Galactic disk.
}

\keywords{ Galaxy: stellar content -- open clusters and associations: 
general -- stars: variables: Cepheids -- methods: data analysis}

\maketitle
%

%------------------------------------------------------------------$-$
\section{Introduction}

Classical Cepheids are intermediate-mass stars that can exhibit 
regular pulsations in fundamental and overtone modes, and possess 
ages that overlap with those of open clusters (OCs; see 
Table~\ref{table:table1}).
Therefore, classical Cepheids are expected to be in OCs, and 
such pairs are significant objects in astronomy.

Cepheids have a famous period--luminosity relationship (PLR), also 
known as the Leavitt law, securing them as standard candles for 
constituting the cosmic distance scale and as essential indicators,
empowering firmer constraints on Hubble's
law~\citep[e.g.,][]{leavitt1912,feast1999,madore1991}.
However, one of the critical debates in cosmology is the Hubble 
constant ({\it H$_{\rm 0}$}) tension.
Based on the current standard $\Lambda$ cold dark matter ($\Lambda$CDM) 
model, estimates of {\it H$_{\rm 0}$} based on measurements of 
fluctuations in the temperature and polarization of the cosmic 
microwave background (CMB) from Planck and ACT+WMAP 
consistently yield values of 67.4 $\pm$ 0.5 and 67.6 $\pm$ 1.1
km s$^{-1}$ Mpc$^{-1}$, respectively~\citep{planck2020,aiola2020}.
In contrast, \citet{riess2019,riess2021} argued that the Cepheid distance 
scale implies a larger value of {\it H$_{\rm 0}$} = 73.2 $\pm$ 1.3 
km s$^{-1}$ Mpc$^{-1}$, in tension with the CMB result.
Using the tip of the red giant branch in the Large Magellanic 
Cloud, the Carnegie-Chicago Hubble Program put forward an 
{\it H$_{\rm 0}$} = 69.6 $\pm$ 1.9 km s$^{-1}$ Mpc$^{-1}$, close 
to the CMB prediction considering the uncertainty~\citep{freedman2019,
freedman2020}.
Likewise, \citet{majaess2020} pointed out that the accuracy of Cepheid 
distances may be swayed by neglected or inaccurate blending corrections 
for remote targets, which would result in an overestimated {\it H$_{\rm 0}$}.
Therefore, it is not clear so far whether the {\it H$_{\rm 0}$} tension means 
new physics are needed for the $\Lambda$CDM model or that unknown 
systematic errors exist in the above measurements.
Open cluster--Cepheid pairs can help to solve this problem.
As a large number of cluster members makes it possible to derive 
more accurate astronomical parameters than those of individual sources,
OCs harboring Cepheids can optimize the zero point of the PLR for 
Cepheids, and ultimately contribute to reducing the {\it H$_{\rm 0}$} 
uncertainty by providing a Galactic calibration of the Leavitt 
law~\citep[e.g.,][]{turner2002,chen2017,breuval2020}.
Recently, using only 17 reported cluster Cepheids with \emph{Hubble 
Space Telescope} photometry and \emph{Gaia} Early Data Release 3 
(EDR3) parallax, \citet{riess2022} constrained the PLR of Cepheids and 
obtained a 5\%--7\% reduction in the {\it H$_{\rm 0}$} uncertainty.

Open cluster--Cepheid pairs are important in additional ways. 
First, cluster Cepheids can be used to benchmark \emph{Gaia} parallax, 
especially for distant Cepheids.
\emph{Gaia} parallax is affected by a systematic error, the so-called 
``parallax zero point'', which was first clearly identified in Data Release 
2 using quasars~\citep{Brown2018,lindegren2018}.
Also, OC--Cepheid pairs could be an important diagnostic tool for examining
the \emph{Gaia} parallax offset, whereby the cluster distance is tied to 
more stars of perhaps a different mean color than the Cepheid, and 
offsets can be examined as a function of direction, magnitude, color, 
and distance.
For an ongoing investigation using cluster Cepheids to assess the 
systematic errors in the \emph{Gaia} data, we recommend~\citet{riess2022}.
Second, cluster Cepheids are expected to benchmark the stellar evolution 
models and theoretical isochrones~\citep[e.g.,][]{bono2005,turner2006}.
Third, OC--Cepheid pairs can be of service in elucidating the
pulsating nature and dynamical evolution of Cepheids~\citep[e.g.,][]{fry1997,
lemasle2017,dinnbier2022}.
In addition, Cepheids adhere to a period--age relation (PAR), whereby 
longer period Cepheids are more luminous and 
massive~\citep[e.g.,][]{efremov2003,bono2005,turner2012a}.
Therefore, the ages of OC--Cepheid pairs based on isochrone fitting 
to the cluster colour--magnitude diagrams (CMDs) can be used to 
calibrate the PAR of Cepheids~\citep[e.g.,][]{efremov2003,anderson2016,
desomma2020}.
As both OCs and Cepheids are young objects, they can also be 
used as good tracers for studying spiral arms and the detailed disk structure 
of the Milky Way~\citep[e.g.,][]{chen2019,hao2021,poggio2021}.
In particular, \citet{minniti2020} used classical Cepheids to trace the 
gradient of metallicity on both sides of the Galactic disk.
Taking advantage of the \emph{Gaia} data, \citet{hao2020,hao2022} 
found more than 700 new OC candidates.
Although over 2000 new OCs have been identified, the 
Galactic OC census is considered to be far from complete and this 
particularly true along 
heavily obscured sightlines~\citep[e.g.,][]{cantat2020a,castro2022}.
Thanks to the VISTA variables in the Vía Láctea survey 
and other longer wavelength surveys, classical Cepheids are being 
discovered on the other side of the Galaxy~\citep[e.g.,][]{dekany2015,
minniti2021}.
At present, more than 3000 classical Cepheids 
have been discovered in the Milky Way~\citep[e.g.,][]{chen2020,
pietrukowicz2021}.

Despite the significance of OC--Cepheid pairs and the discovery 
of thousands of OCs and classical Cepheids, only a few OCs 
harboring Cepheids have been reported so far.
The presence of Cepheids in Galactic OCs was first discovered
by~\citet{doig1925,doig1926} and probably others, as noted by~\citet{fernie1969}.
By cross-matching the known OCs in \emph{Gaia} with 
a list of Cepheids, \citet{zc2021} reported the largest 
OC--Cepheid sample to date, in which only 33 Cepheids exist in 29 OCs.
Therefore, the identification of more OC--Cepheid pairs is an indispensable 
undertaking with great importance.

The latest \emph{Gaia} data release, DR3, includes astrometric and 
photometric measurements of 1.8 billion objects, and the determination 
of the mean radial velocity ($RV$) of 33 million objects~\citep{vallenari2022}.
Meanwhile, the data quality of \emph{Gaia} has been further improved.
Previous studies identifying OC-Cepheid pairs mainly 
cross-matched Cepheids with known OCs. However, as 
mentioned above, the OC census of the Milky Way is incomplete.
Hence, based on the aforementioned comprehensive datasets, this 
work aims to find OC--Cepheid pairs in the regions of Cepheids 
using the cluster-search method employed in our previous 
studies~\citep{hao2020,hao2022}.

We organize this paper as follows. 
The data used in this work are described in Sect.~\ref{sect:data}. 
Section~\ref{sect:search} presents the method applied to the search
for OCs holding classical Cepheids. The results are displayed in 
Sect.~\ref{sect:results} and we summarize this work in 
Sect.~\ref{sect:summary}.
%

%%%%%%%%%%%%%%%%%%%%%%%%%%%%%%%%%%%%%%%%%%%%%%%%

\section{Data}
\label{sect:data}

The classical Cepheids adopted in this work come from the 
catalog compiled by~\citet{pietrukowicz2021}, who carefully inspected 
candidate Cepheids from many surveys
based on their long-term experience in the field of variable stars. 
The surveys related to the sources in this catalog are as follows: 
General catalog of Variable Stars,
All Sky Automated Survey, 
Northern Sky Variability Survey,
Optical Gravitational Lensing Experiment, 
Asteroid Terrestrial$-$impact Last Alert System, 
All$-$Sky Automated Survey for Supernovae,
Wide$-$field Infrared Survey Explorer,
\emph{Gaia} astrometric mission,
VISTA Variables in the Via Lactea, Zwicky Transient Facility, 
and so on~\citep[for more details, see][and references within]{pietrukowicz2021}.
This catalog, with a purity exceeding 97\%, contains a total of
3 352 classical Cepheid variables, which are available at the 
OGLE Internet Data 
Archive\footnote{\url{https://www.astrouw.edu.pl/ogle/ogle4/OCVS/allGalCep.listID}}.
In addition, \citet{pietrukowicz2021} not only presented detailed 
information on these classical Cepheids but also cross-matched 
them with the \emph{Gaia} EDR3 catalog.
In this work, considering that the present-day uncertainty on parallax in 
the \emph{Gaia} area can reach 0.02--0.03 mas~\citep{lindegren2021}, 
we decided to mainly concentrate on the objects within 5 kpc of 
the Sun, resulting in a subsample that contains 1 085 classical Cepheids.
For each of these Cepheids, we have obtained detailed information such
as astrometric and photometric parameters in the \emph{Gaia} 
DR3 dataset.

We used the dataset from the \emph{Gaia} DR3 catalog, which can be 
found at the \emph{Gaia} 
archive\footnote{\url{https://gea.esac.esa.int/archive/}}.
The astrometric and photometric parameters of stars in \emph{Gaia} 
DR3 are the same as those in \emph{Gaia} EDR3~\citep{gaia2020}, 
which contain
stellar celestial positions, parallaxes, proper motions ($l$, $b$, 
$\varpi$, $\mu_{\alpha^{*}}$, and $\mu_{\delta}$), and three 
photometric bands ($G$, $G_{\rm BP}$, and $G_{\rm RP}$).
Meanwhile, \emph{Gaia} DR3 includes 33 million objects with new 
determinations on their mean radial velocities.
As in many previous works searching 
for OCs~\citep[e.g.,][]{cantat2018,hao2022,castro2022}, we only 
chose sources brighter than $G$ = 18. 
For sources with five-parameter solutions at this magnitude, the
median parallax uncertainty on them is 0.120 mas, and the median 
uncertainties on proper motions in $\mu_{\alpha^{*}}$ and 
$\mu_{\delta}$ are 0.123 and 0.111 mas yr$^{-1}$, 
respectively~\citep{lindegren2021}. We also rejected sources that 
present negative parallaxes or large proper motions with 
$|\mu_{\alpha^{*}}|$, $|\mu_{\delta}|$ > 30~mas~yr$^{-1}$.
As the aim of this work is to find OCs containing classical Cepheids,
we only focus on the 1 085 spatial regions where selected Cepheids 
are present; our experimental setup is described in the following section.
%

%%%%%%%%%%%%%%%%%%%%%%%%%%%%%%%%%%%%%%%%%%%%%%%%

\section{Methods}
\label{sect:search}

The sample-based clustering search method was successfully 
adopted in our previous studies and we found more than 700
new OC candidates in the \emph{Gaia} area~\citep{hao2020,hao2022}.
This method consists primarily of four steps: determining spatial 
regions, searching for stellar clusters in the five-dimensional (5D) 
parametric space ($l$, $b$, $\varpi$, $\mu_{\alpha^{*}}$, and 
$\mu_{\delta}$), selecting targets based on proper-motion dispersion, 
and visually inspecting the multiple distributions.
In this section, we describe the method and process of 
searching for OCs containing classical Cepheids.

{\it Sample determination.} Under the selection criteria of stars 
described in Sect.~\ref{sect:data}, we first extracted stars around the 
spatial locations of classical Cepheids.
The key aspect of this step is to decide the spatial size and distance 
boundaries of the different regions where Cepheids exist.
In this work, by referencing our previous experience in searching for 
OCs~\citep{hao2020,hao2022}, 
the celestial sizes and distance boundaries of different spatial regions 
are set in the range of [$1^{\circ}$, $10^{\circ}$] and [0.5, 5] kpc,
respectively, mainly according to the distances of classical Cepheids 
and the spatial density of regions.
Thus, we obtained 1 085 original stellar samples. 
In addition, if in the following search there are star clusters
located at the borders of regions, we change the sizes of the regions and 
re-detect these star clusters.

{\it Clustering algorithm.} The powerful unsupervised clustering algorithm, 
known as the density-based spatial clustering of applications with 
noise~\citep[DBSCAN,][]{ester1996}, was applied in the multidimensional 
space (i.e., $l$, $b$, $\varpi$, $\mu_{\alpha^{*}}$, and $\mu_{\delta}$) 
to find spatial over-density structures in the regions. For DBSCAN, the 
parameters $\epsilon$ and $minPts$ are needed, which specify the
radius of a neighborhood with respect to some point and the desired
minimum cluster size, respectively. 
We let the parameter $minPts$ vary in the range of six to ten stars.
The corresponding parameters $\epsilon$ of each sample 
were then automatically calculated using the z-score standardization, 
a Gaussian kernel-density-estimation method~\citep{lampe2011}, and the 
k-nearest ($k$ = $minPts $-$ 1$) neighbors algorithm~\citep{Altman1992}; 
see~\citet{hao2020} for elaborate details on the determination 
of $\epsilon$.
After preparing the parameter pairs for all samples, we conducted the 
search for statistical stellar clusters in all regions of existing classical 
Cepheids.

We noticed that there are some classical Cepheids located close to 
each other in spatial locations, which means regional overlaps exist 
among some initial samples, resulting in several stellar clusters being 
found repeatedly. 
We resolved this case by identifying those stellar clusters that have 
comparable mean parameters within 3$\sigma_{i}$ ($\sigma$ is the 
standard deviation, $i$ = {$l$, $b$, $\varpi$, $\mu_{\alpha^{*}}$, and 
$\mu_{\delta}$) in the 5D parametric space. 
Subsequently, we visually inspected these stellar clusters and filtered 
out the repetitions.

{\it Selection based on proper-motion dispersion.} The apparent 
proper-motion dispersion of a stellar cluster can serve as an 
effective means to distinguish reasonable OCs from 
implausible ones, and this technique has been used in some work 
to select potentially real OCs from spatial over-density  
structures~\citep[e.g.,][]{hunt2021,hao2022}.
The gravitationally bound OC systems generally have small internal
velocity dispersions.
Here, as in our previous work~\citep{hao2022}, we selected 
the possibly genuine OCs using the following proper-motion criterion:
%%%%%%%%%%%%%
\begin{equation} 
\sqrt{\sigma_{\mu_{\alpha^{*}}}^{2}+\sigma_{\mu_{\delta}}^{2}} \leq 
\begin{cases}
0.5 \ \rm mas \ yr^{-1} & \text{if \ $\varpi <\ 1 \ \rm mas$}\\
2\sqrt{2} \frac{\varpi}{4.7404} \ \rm mas \ yr^{-1} & \text{if \ $\varpi \geq \ 1 \ \rm mas$}
\end{cases}
.\end{equation}
%%%%%%%%%%%%%
Here, $\sigma_{\mu_{\alpha^{*}}}$, $\sigma_{\mu_{\delta}}$, and 
$\varpi$ are the standard deviations of two proper motions and the 
mean parallax of statistical stellar clusters.
For the property of proper-motion dispersions of OCs identified in 
the \emph{Gaia} area, we recommend~\citet{cantat2020b}.

{\it Visual inspection and confirmation.} For the potentially real 
OCs, we visually inspected their multiple characters, such as their 
members, the density relative to the background stellar distribution, 
the distribution of radial velocities when available, and especially 
the CMDs.
As the member stars of an OC are almost simultaneously formed 
in the same molecular cloud, their distribution on the CMD can exhibit 
the feature of some empirical isochrone.
These measures were mainly taken for the potentially genuine OCs 
that have not been reported to select the more reliable OC 
candidates.

We computed the mean $RV$, and the weighted standard deviation 
of $RV$ for each OC candidate using the method described 
in~\citet{soubiran2018}.
We also estimated the ages, line-of-sight extinction 
(absorption, $A_{\rm G}$), and distance modulus (DM) of the proposed 
OC candidates using the photometric parameters ($G$, 
$G_{\rm BP}$, and $G_{\rm RP}$) provided in \emph{Gaia} to 
construct the theoretical isochrones in their CMDs.  
It should be noted that the differential reddening seen in some OCs can 
complicate matters considerably,
in that a tight obvious main sequence and evolved pattern may not 
be seen. In this case, differential reddening procedures need to be 
employed, potentially using a color--color 
diagram~\citep[e.g.,][see Figure 2 and Figure 3 for NGC 1545]{turner1994}.
The isochrones we adopted are derived from the PARSEC library 
\citep{bressan2012}, which has been updated for the \emph{Gaia} 
passbands with the photometric calibration of~\cite{evans2018}
and contains logarithmic ages (i.e., log(age/yr)) ranging from 5.92 to 
10.13 and metal fractions ($z$) ranging from 0.015 to 0.029, as well 
as a series of isochrones that are generated with steps of 
$\Delta$log(age/yr) = 0.02 and $\Delta$$z$ = 0.001.
We used the least square fitting method to obtain an isochrone that is 
as reliable as possible for each OC candidate. Meanwhile,
the extinction and reddening that we adopted are corrected using 
an extinction law of $R_{\rm v}$ = 3.1 \citep{ODonnell1994} for the 
Milky Way galaxy, and the applied approximate relation is 
${\it E}({\it G}_{\rm BP} - {\it G}_{\rm RP})$ = 0.50 {\it A}$_{\rm v}$
as recommended by \citet{Andrae2018}. 
We refer the reader to Sect.~3.4 in \citet{hao2022} for 
more details on the determinations.
%

%%%%%%%%%%%%%%%%%%%%%%%%%%%%%%%%%%%%%%%%%%%%

\section{Results}
\label{sect:results}
%%%%%%%%%%%%%%%%%%%%%%%%%%%%%%%%%%%%%%%%%%%%

After implementing the processes described in Sect.~\ref{sect:search}, we 
cross-matched the stellar clusters found in this work with 
reported OCs to identify the OCs that are already known and those that 
are new findings. 
In total, 635 known OCs were re-detected, and we proposed 38 
stellar clusters as newly found OC candidates. 
The classical Cepheids used in this work were cross-matched 
with the above OCs, and we found 45 OCs harboring a total
of 50 Cepheids.
For the newly found OCs, we determined their parameters 
and investigated their characteristics.
%

%%%%%%%%%%%%%%%%%%%%%%%%%%%%%%%%%%%%%%%%%%%%%%%%%% Fig. 1
\begin{figure*}
\centering
\includegraphics[width=0.327\linewidth]{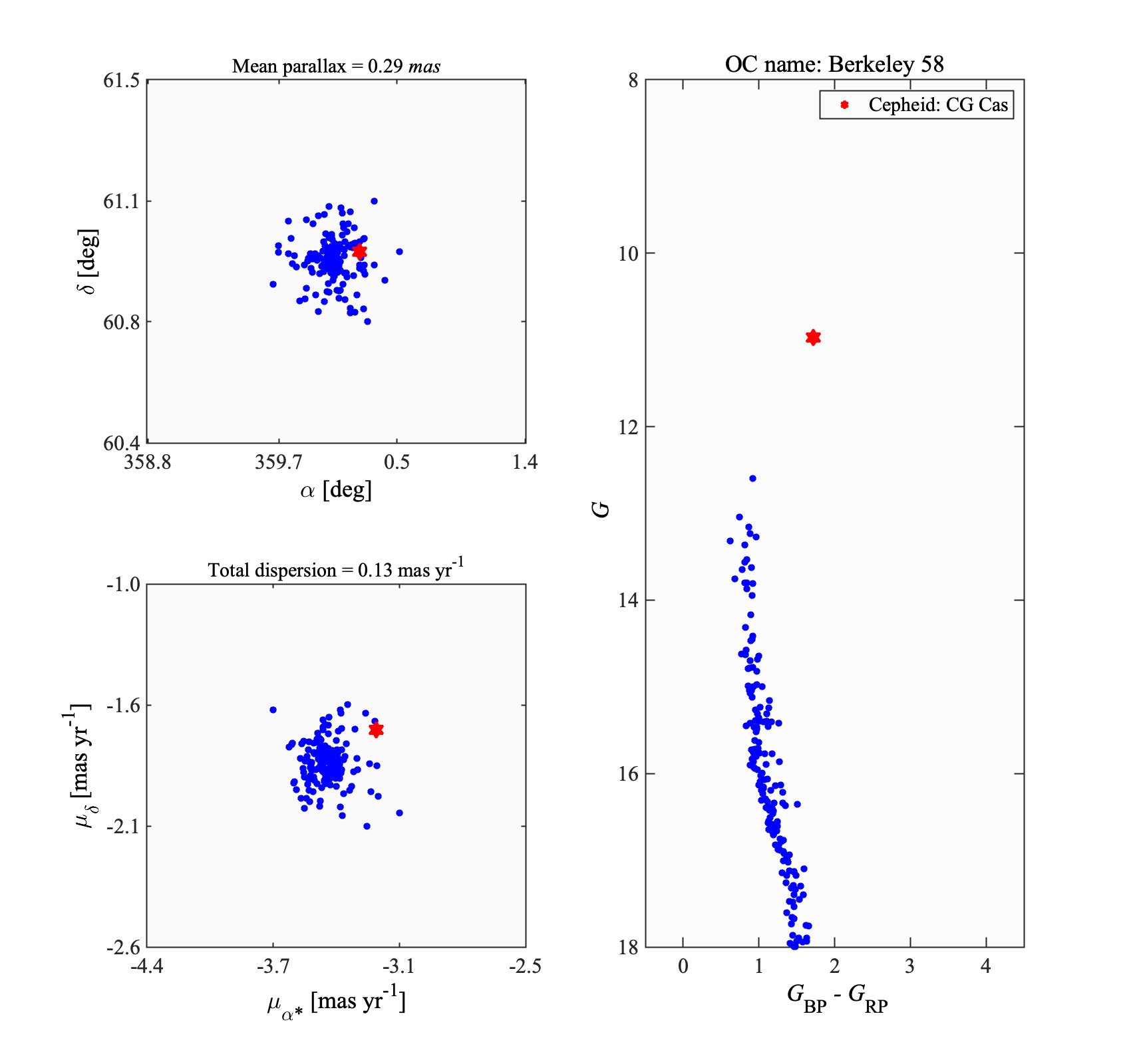} \hspace{0.0cm}
\includegraphics[width=0.327\linewidth]{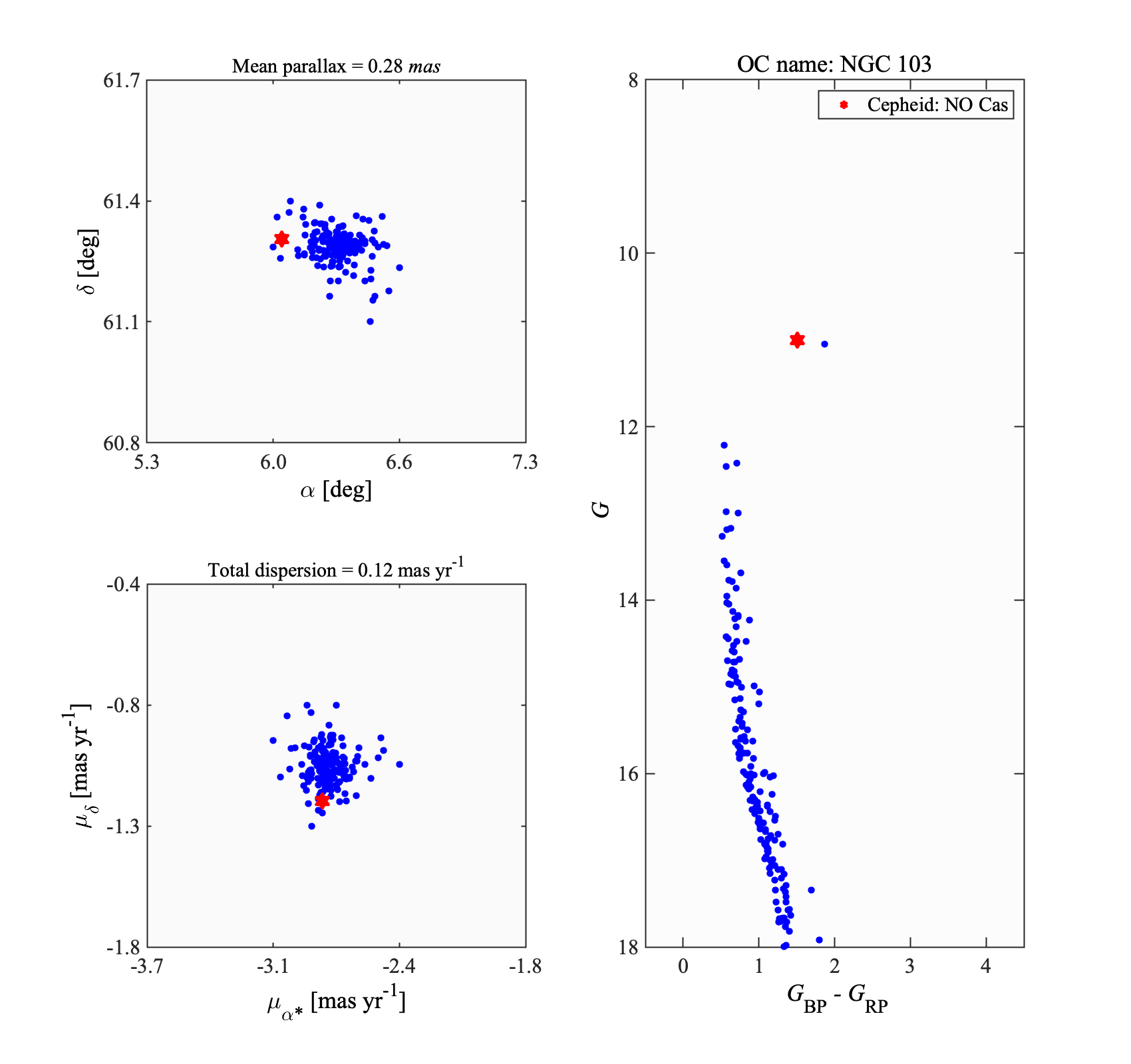}  \hspace{0.0cm}
\includegraphics[width=0.327\linewidth]{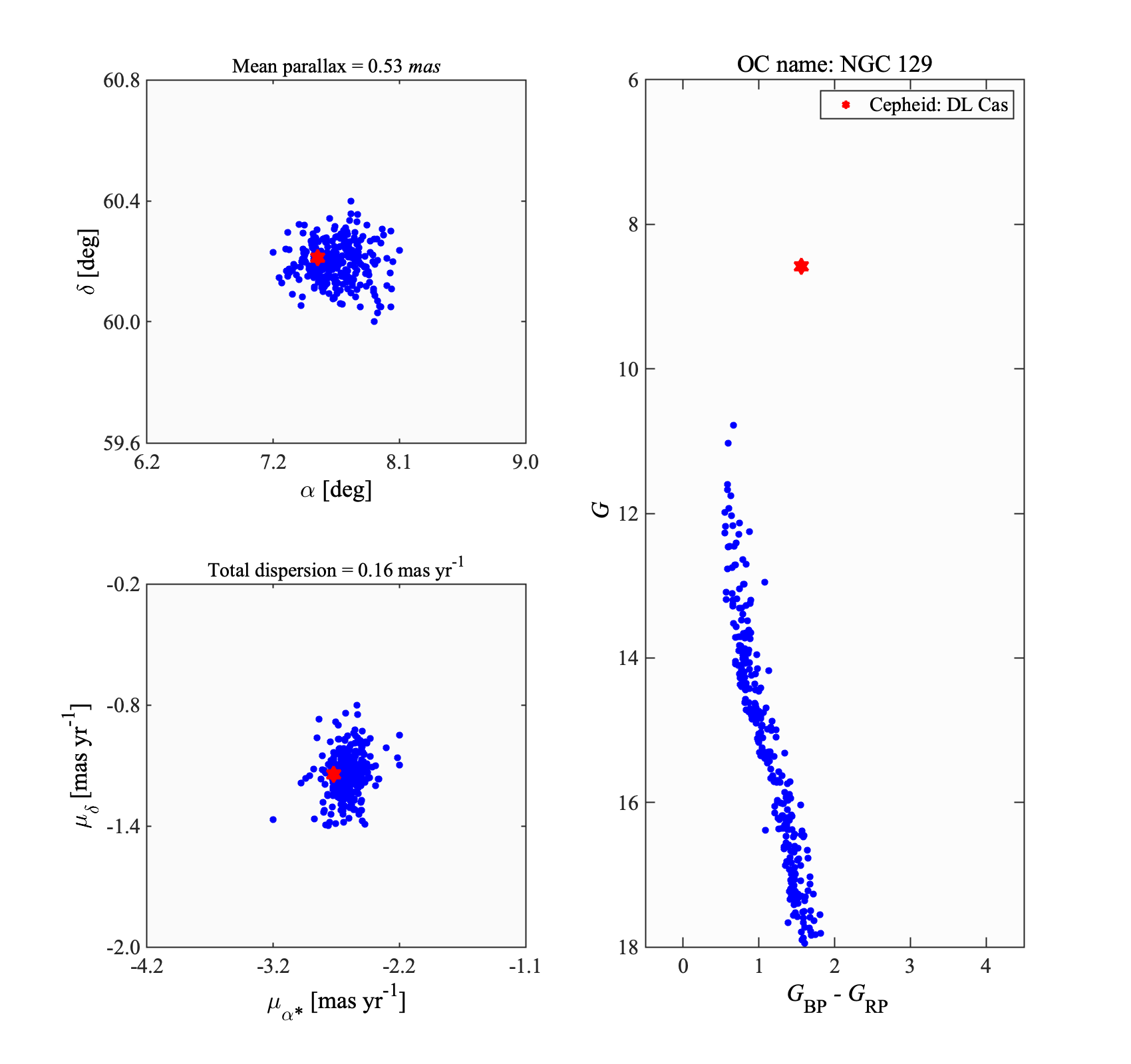} \hspace{0.0cm}
\caption{Examples of OCs (blue dots) harboring classical Cepheids (red hexagram). The columns of each panel represent the distributions of the member stars of OCs and classical Cepheids for position in $RA$ ($\alpha$) and $Dec$ ($\delta$), proper motions in $\mu_{\alpha^{*}}$ and $\mu_{\delta}$, and the CMD in $G$ vs. $G_{\rm BP}$ $-$ $G_{\rm RP}$, as well as the mean parallax and total proper-motion dispersion of OCs. Here, the listed OCs are Berkeley 58, NGC 103, and NGC 129.}
\label{fig:OCs_Ceps}
\end{figure*}
%%%%%%%%%%%%%%%%%%%%%%%%%%%%%%%%%%%%%%%%%%%%%%%

%%%%%%%%%%%%%%%%%%%%%%%%

\subsection{Cross-matching with known OCs}
\label{sect:cross-match}

The stellar clusters found in this work were first cross-matched
with previously reported OCs in the \emph{Gaia} area:
\citet{cantat2018}, \citet{castro2018}, \citet{cantat2019}, \citet{castro2019}, \citet{castro2020}, 
\citet{sim2019}, \citet{liu2019}, \citet{ferreira2019}, \citet{ferreira2020}, \citet{ferreira2021}, 
\citet{hao2020}, \citet{he2021}, \citet{hunt2021}, \citet{hao2022}, and \citet{castro2022}.
In general, two stellar clusters can be considered a match if their mean 
parameters are within 3$\sigma_{i}$ ($i$ = {$l$, $b$, $\varpi$, 
$\mu_{\alpha^{*}}$ and $\mu_{\delta}$}) in the 5D 
parametric space.
Thus, in this step, we obtained 594 previously known OCs in 
the \emph{Gaia} data.

Subsequently, stellar clusters obtained by us were cross-matched with OCs  
that were known before \emph{Gaia}, which mainly derive from the
catalogs compiled by \citet[][V3.5, 2015 edition]{dias2002} 
and \citet{kharchenko2013}, which contain approximately 2000 and 3000 objects, 
respectively, while they have repetitive sources.
The criterion used in this step is the same as in our previous 
work~\citep{hao2022}.
We first selected star clusters that fall within a circle of 
$0.5^{\circ}$ radii of objects listed in these two catalogs.
If the mean proper motions ($\mu_{\alpha^{*}}$ and $\mu_{\delta}$) 
of the listed objects are within three times the standard deviations
of the stellar clusters found in this work, and their distances are 
compatible, they can be considered matched.
Here, the distance of a star cluster found in this work is the 
inverse of the mean parallax, and the corresponding error comes
from three times the standard deviation of the parallax.
\cite{cantat2018} took into account many objects
in these two catalogs, but we re-detected 24 known OCs
in \citet{dias2002} and 21 known OCs in \citet{kharchenko2013},
respectively, and 4 OCs among them are repetitive.
Similar to the OC catalogs, we also cross-matched our findings 
to the globular cluster catalog reported by~\citet{baumgardt2019},
but we did not explore any coincidences.

For the remaining stellar clusters found in this work that were not 
yet cross-matched with known OCs, we carefully inspected their 
multidimensional nature as described in Sect.~\ref{sect:search} and 
proposed 38 reliable ones to be OC candidates, which are numbered 
from OC--0705 in order to be the continuation of our previous 
work~\citep{hao2022}.
Combining these 38 newly found OC candidates with 594 OCs 
in \emph{Gaia} and 41 OCs in the catalogs before \emph{Gaia}, 
we searched for a total of 673 OCs in this work.
Subsequently, we investigated these OCs to identify OCs
hosting classical Cepheids.

\subsection{Open clusters housing classical Cepheids}

We regarded OCs and classical Cepheids as combinations in two 
scenarios: Cepheids are member stars of OCs; or the 5D astrometric 
parameters ($l$, $b$, $\varpi$, $\mu_{\alpha^{*}}$, and $\mu_{\delta}$) 
of Cepheids and OCs are concordant within three times the corresponding 
standard deviations.
As described in Sect.~\ref{sect:data}, we obtained 
\emph{Gaia} DR3 \texttt{source\_id} for each classical Cepheid.
For the first case, we cross-matched the list of classical Cepheids 
with the member stars of the 673 OCs found in this work.
The cross-match was performed using \emph{Gaia} DR3 
\texttt{source\_id} and returned 30 OCs holding 32 Cepheids. 
The 5D astrometric parameters of the remaining classical 
Cepheids were then compared with all found OCs, resulting in 18 
Cepheids associated with 18 OCs, of which 3 OCs are the same 
ones as above.
Thus, we identified 45 OCs housing 50 classical Cepheids,
and the names of these OC-Cepheid pairs are listed in 
Table~\ref{table:table1} in the Appendix.
%

%%%%%%%%%%%%%%%%%%%%%%%%%%%%%%%%%%%%%%%%%%%%%%%
\begin{figure*}
\centering
\includegraphics[width=0.90\linewidth]{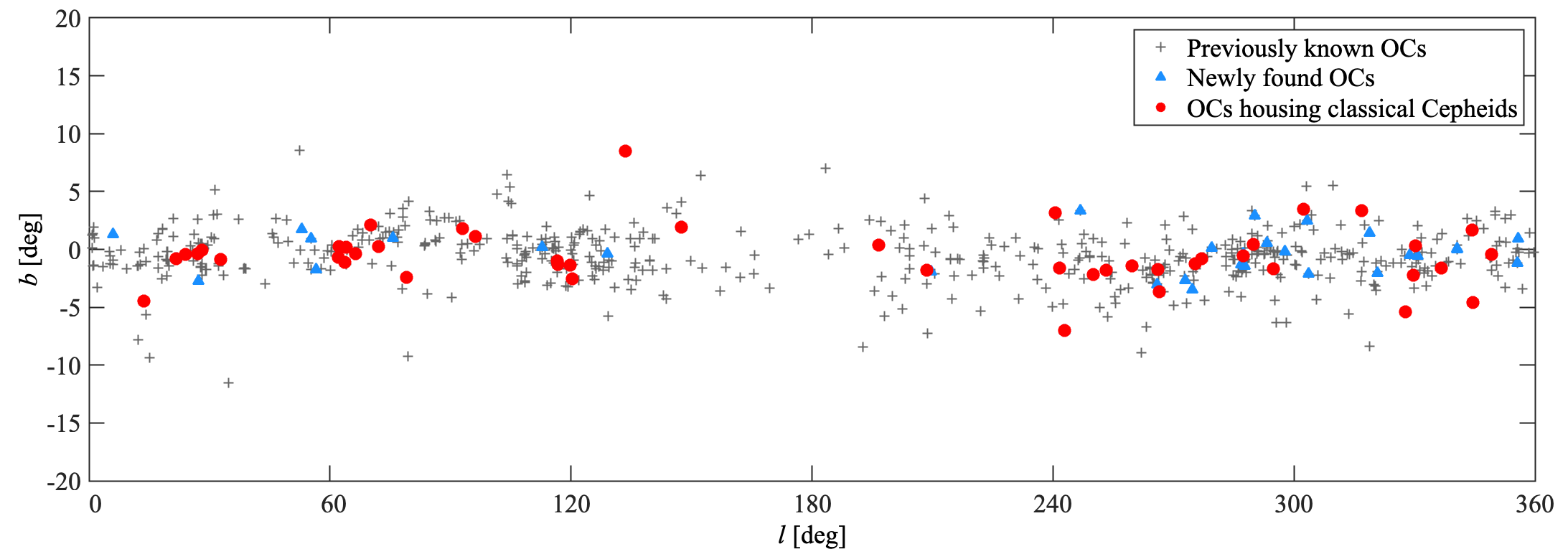} \hspace{0.0cm}
\caption{Distributions in Galactic coordinates of the OCs housing classical Cepheids (red dots), newly found OCs (blue triangles) and previously known OCs (black plus) obtained in this work. }
\label{fig:lb}
\end{figure*}
%%%%%%%%%%%%%%%%%%%%%%%%%%%%%%%%%%%%%%%%%%%%%%%

%%%%%%%%%%%%%%%%%%%%%%%%%%%%%%%%%%%%%%%%%%%%%%%
\begin{figure}
\centering
\includegraphics[width=0.85\linewidth]{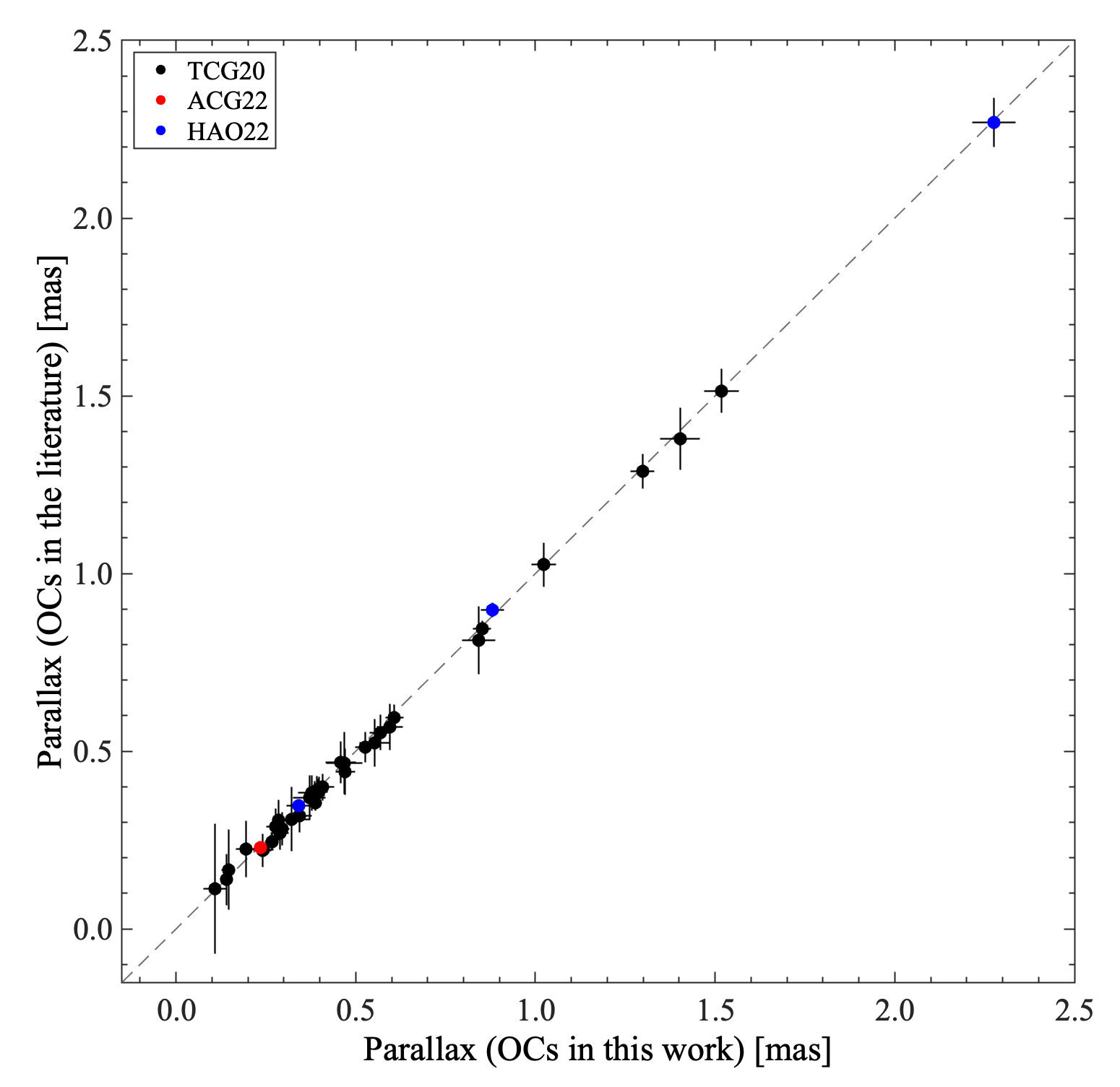} \hspace{0.0cm}
\caption{Parallactic comparison between OCs containing classical Cepheids found in this work
and those previously reported in \citet[][TCG20]{cantat2020a}, \citet[][ACG22]{castro2022}, and 
\citet[][HAO22]{hao2022}}
\label{fig:parallax_known}
\end{figure}
%%%%%%%%%%%%%%%%%%%%%%%%%%%%%%%%%%%%%%%%%%%%%%%

%%%%%%%%%%%%%%%%%%%%%%%%%%%%%%%%%%%%%%%%%%%%%%%%%%
\begin{figure*}
\centering
\includegraphics[width=0.327\linewidth]{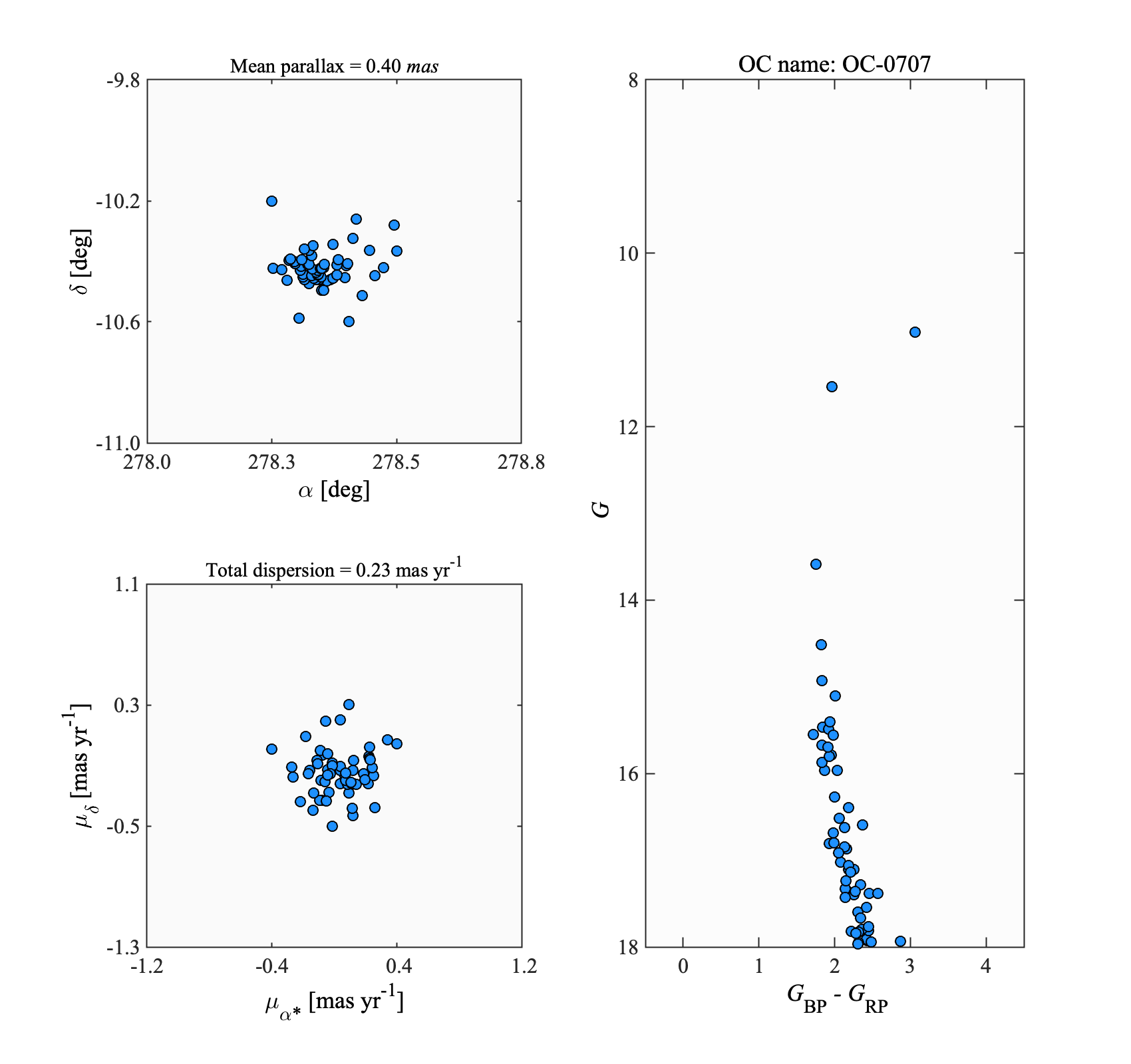} \hspace{0.0cm}
\includegraphics[width=0.327\linewidth]{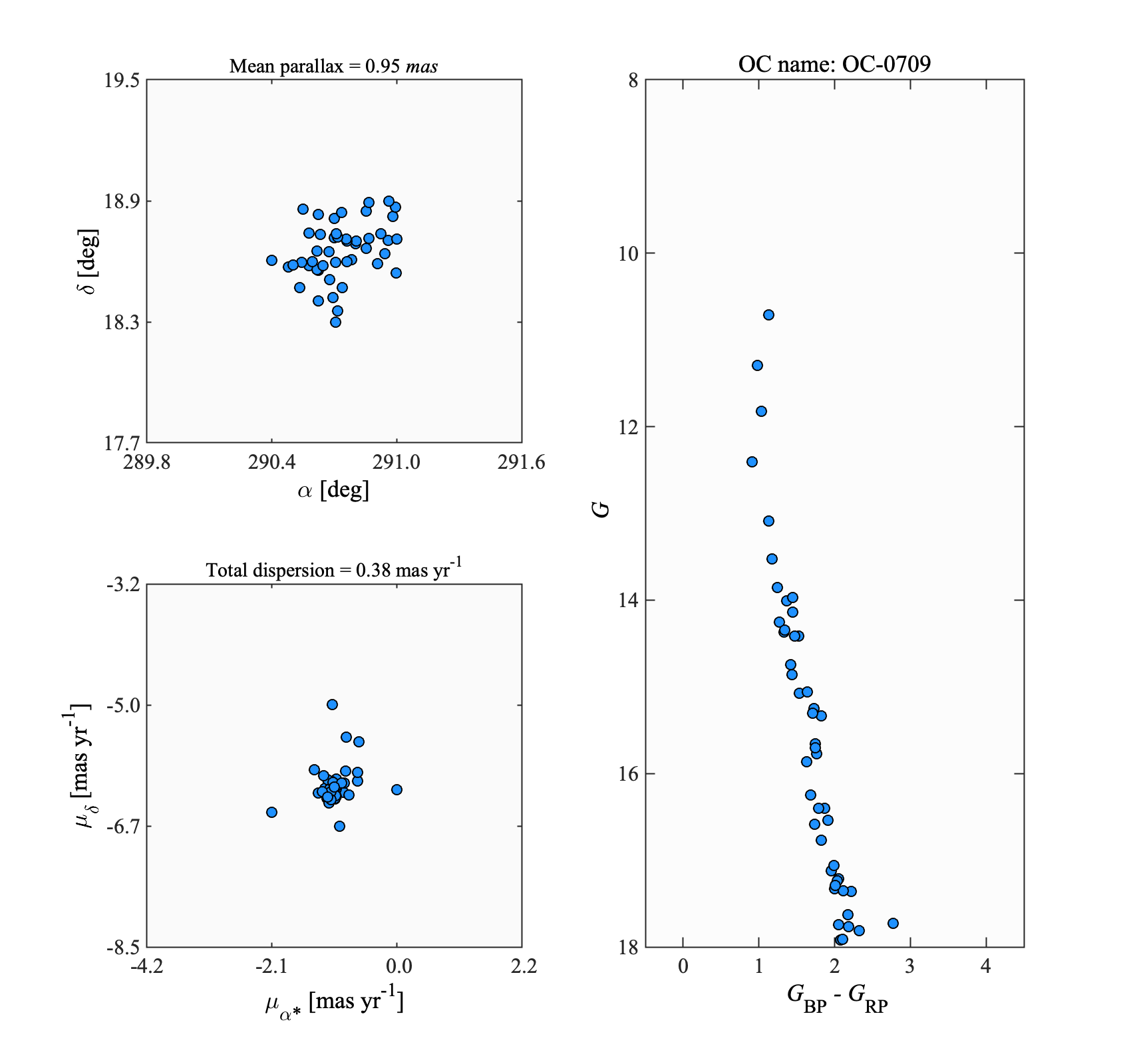}  \hspace{0.0cm}
\includegraphics[width=0.327\linewidth]{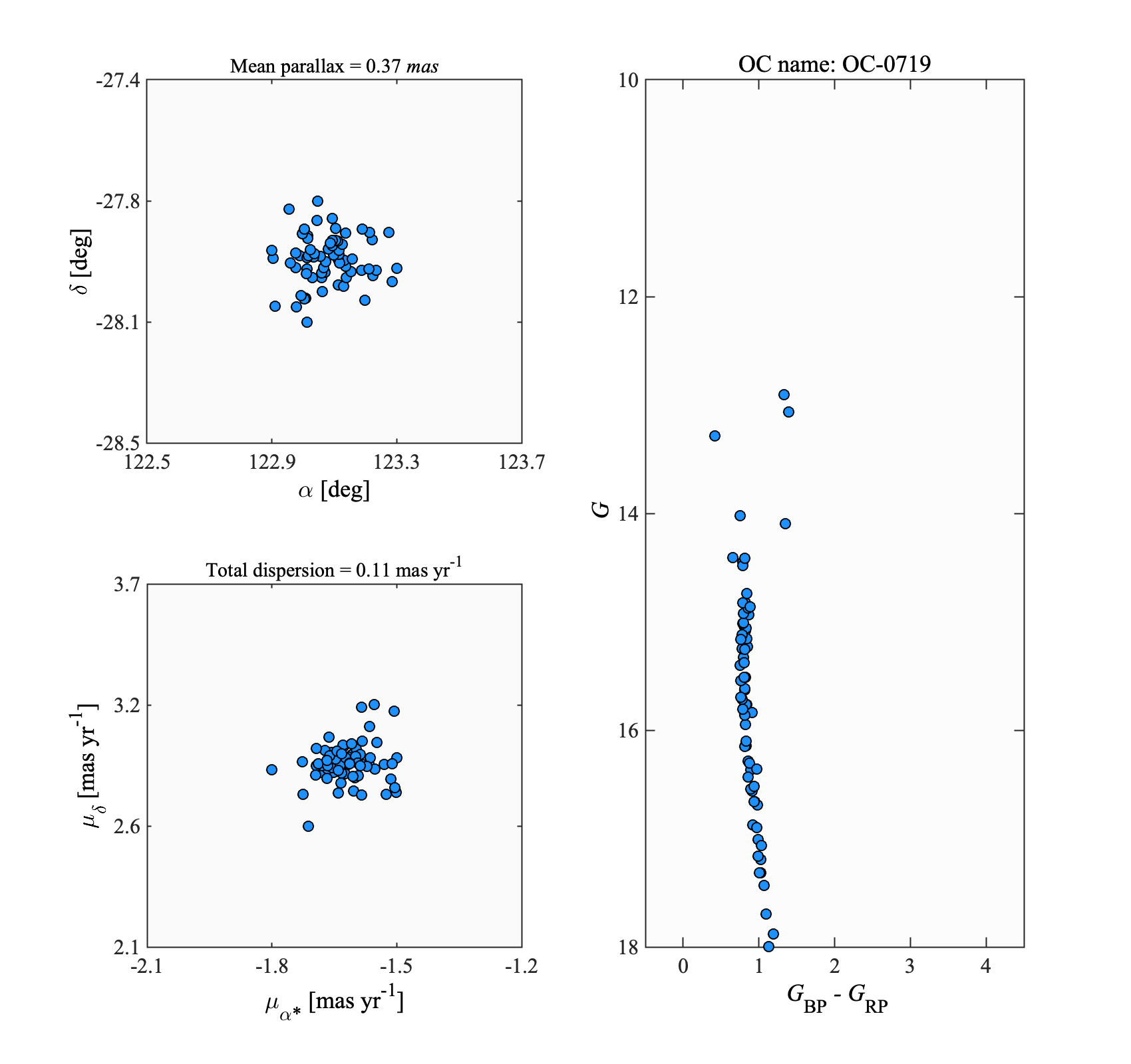} \hspace{0.0cm}
\includegraphics[width=0.327\linewidth]{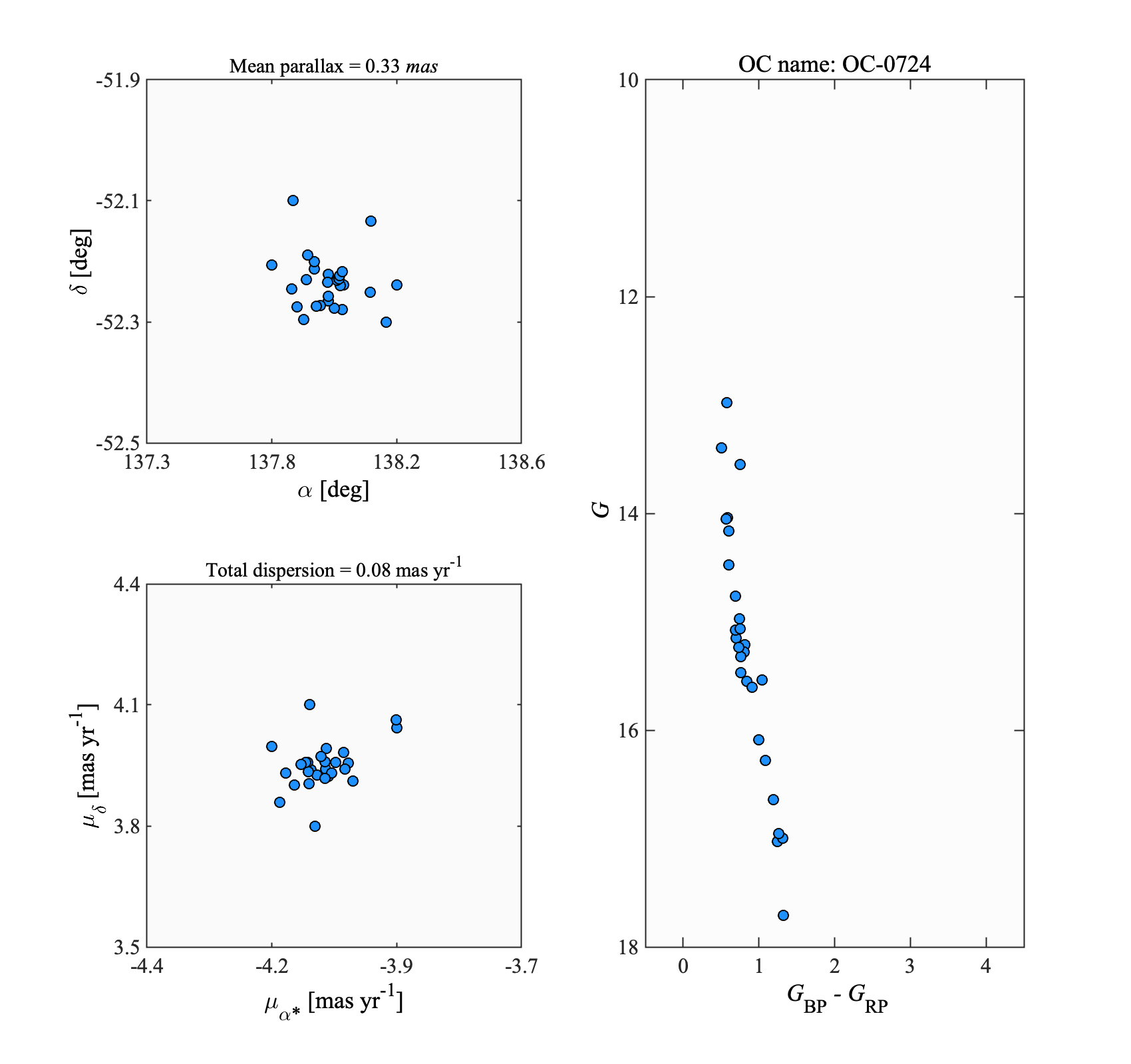} \hspace{0.0cm}
\includegraphics[width=0.327\linewidth]{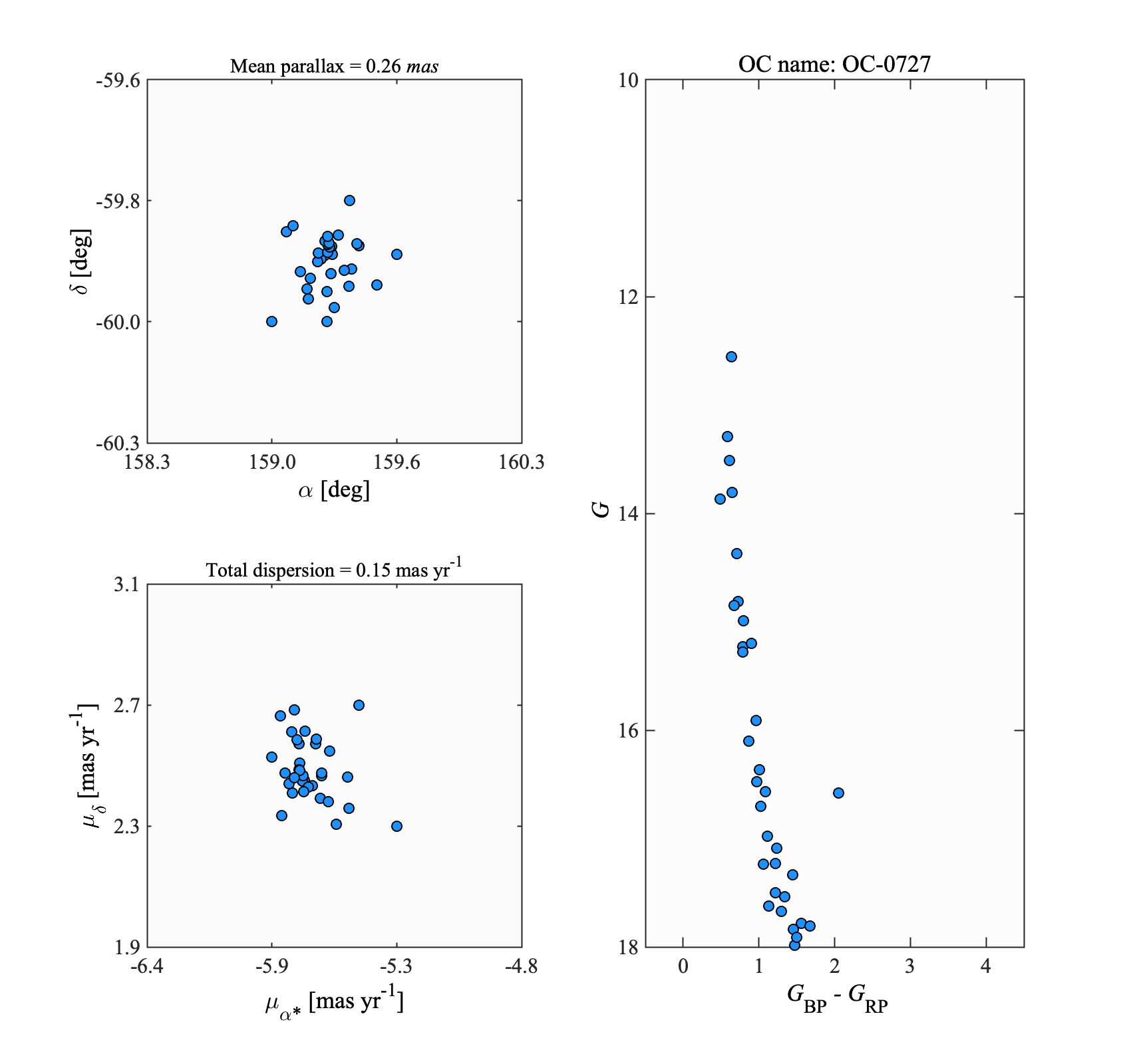}  \hspace{0.0cm}
\includegraphics[width=0.327\linewidth]{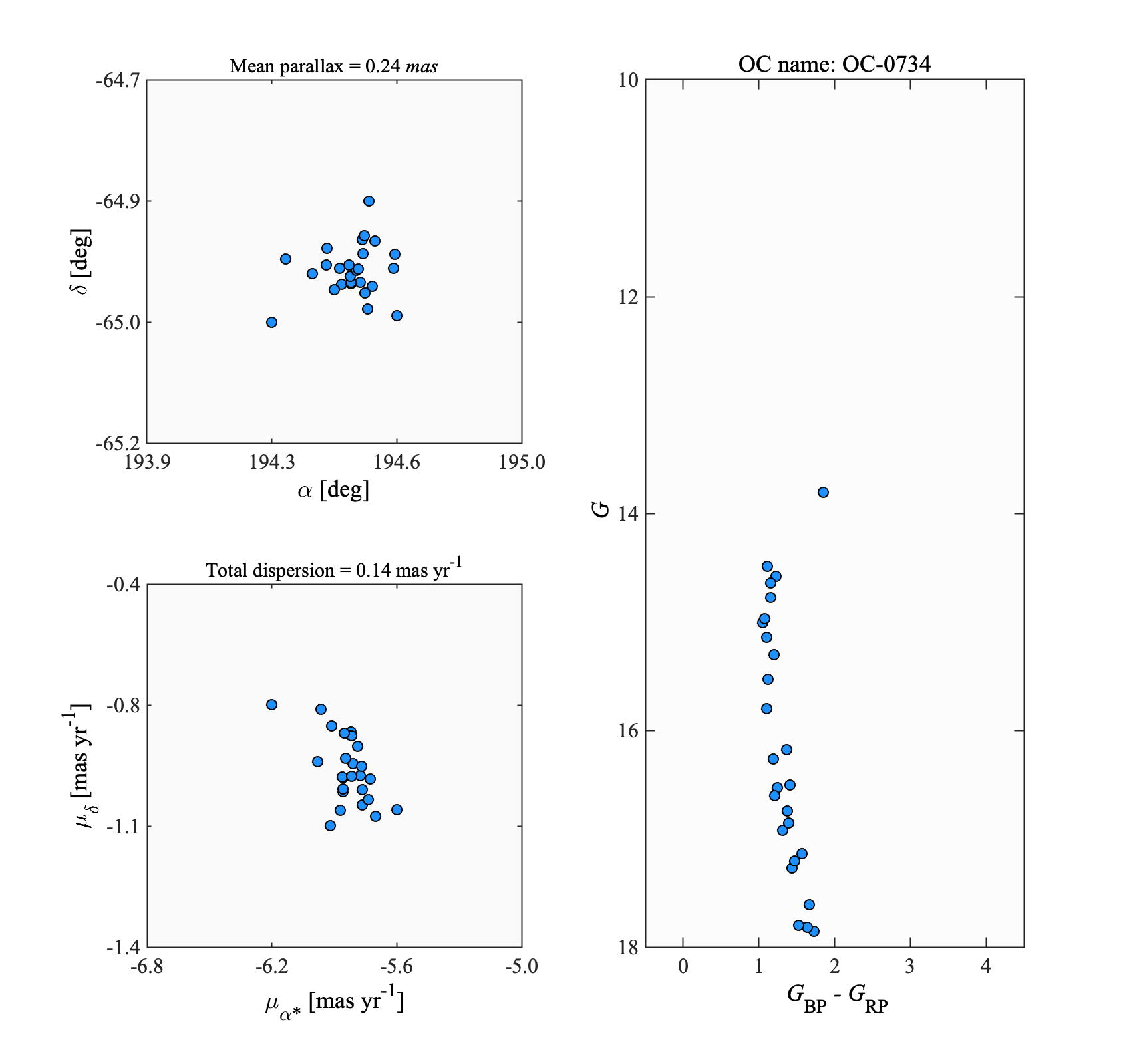} \hspace{0.0cm}
\includegraphics[width=0.327\linewidth]{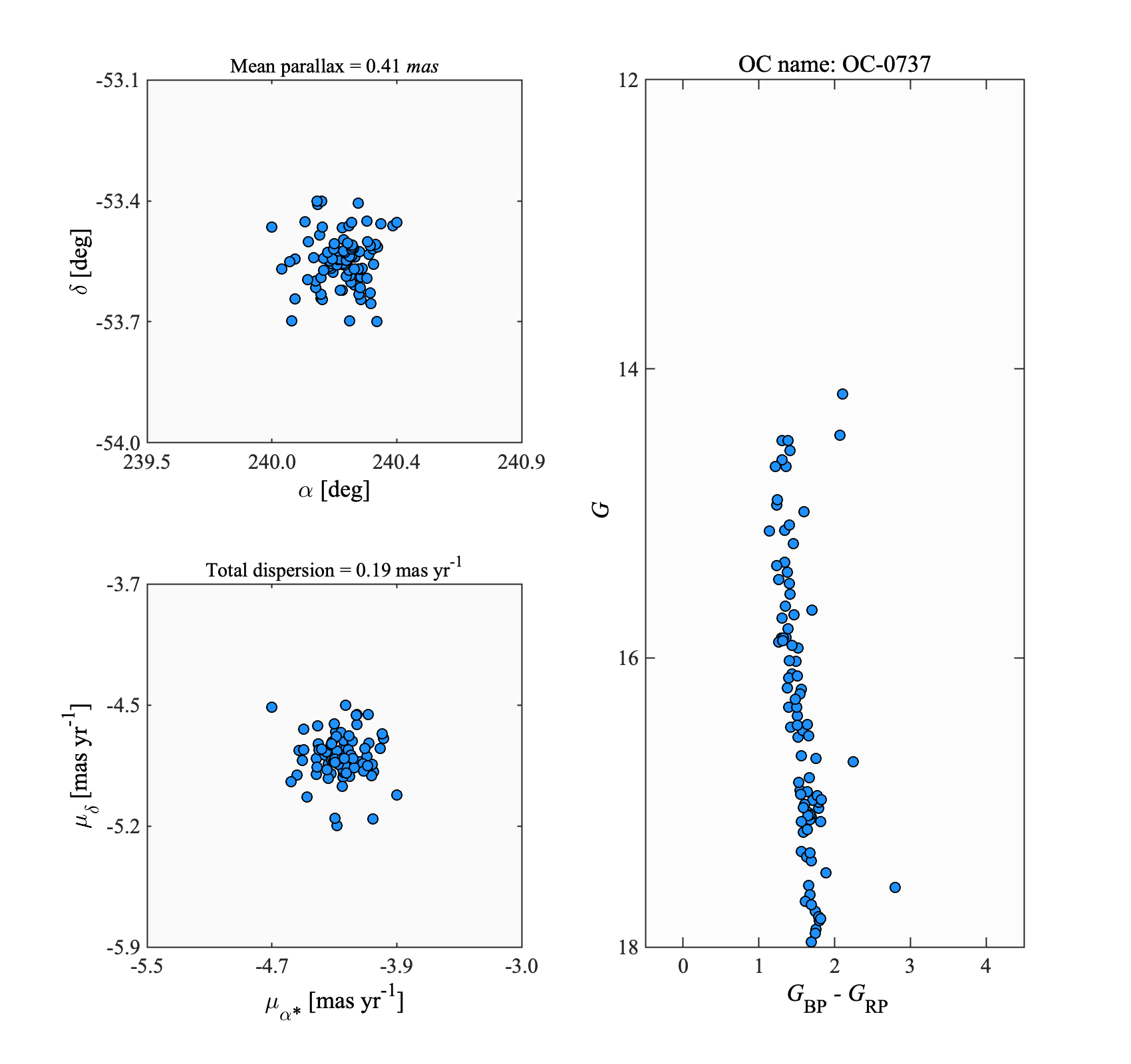}  \hspace{0.0cm}
\includegraphics[width=0.327\linewidth]{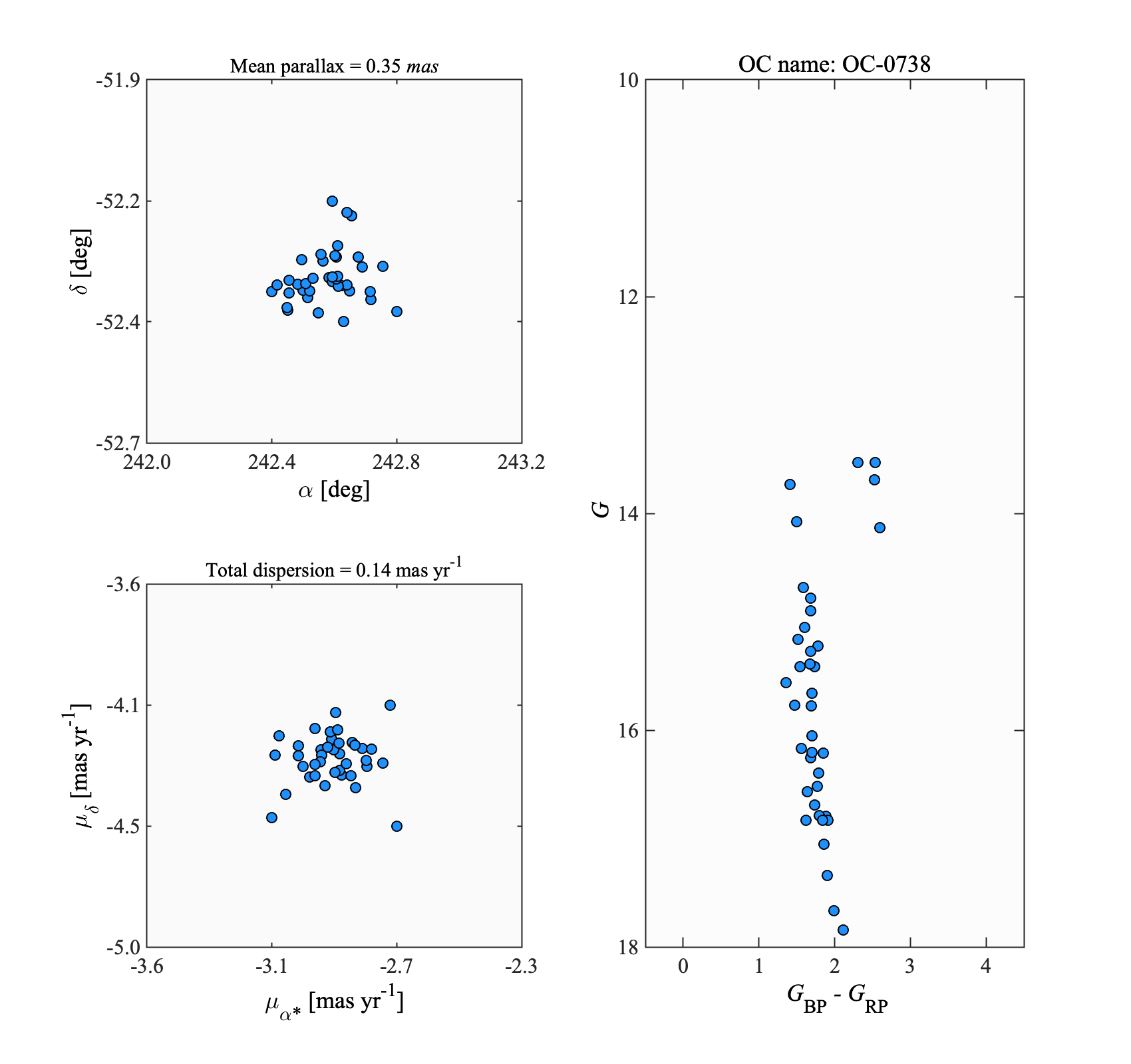} \hspace{0.0cm}
\includegraphics[width=0.327\linewidth]{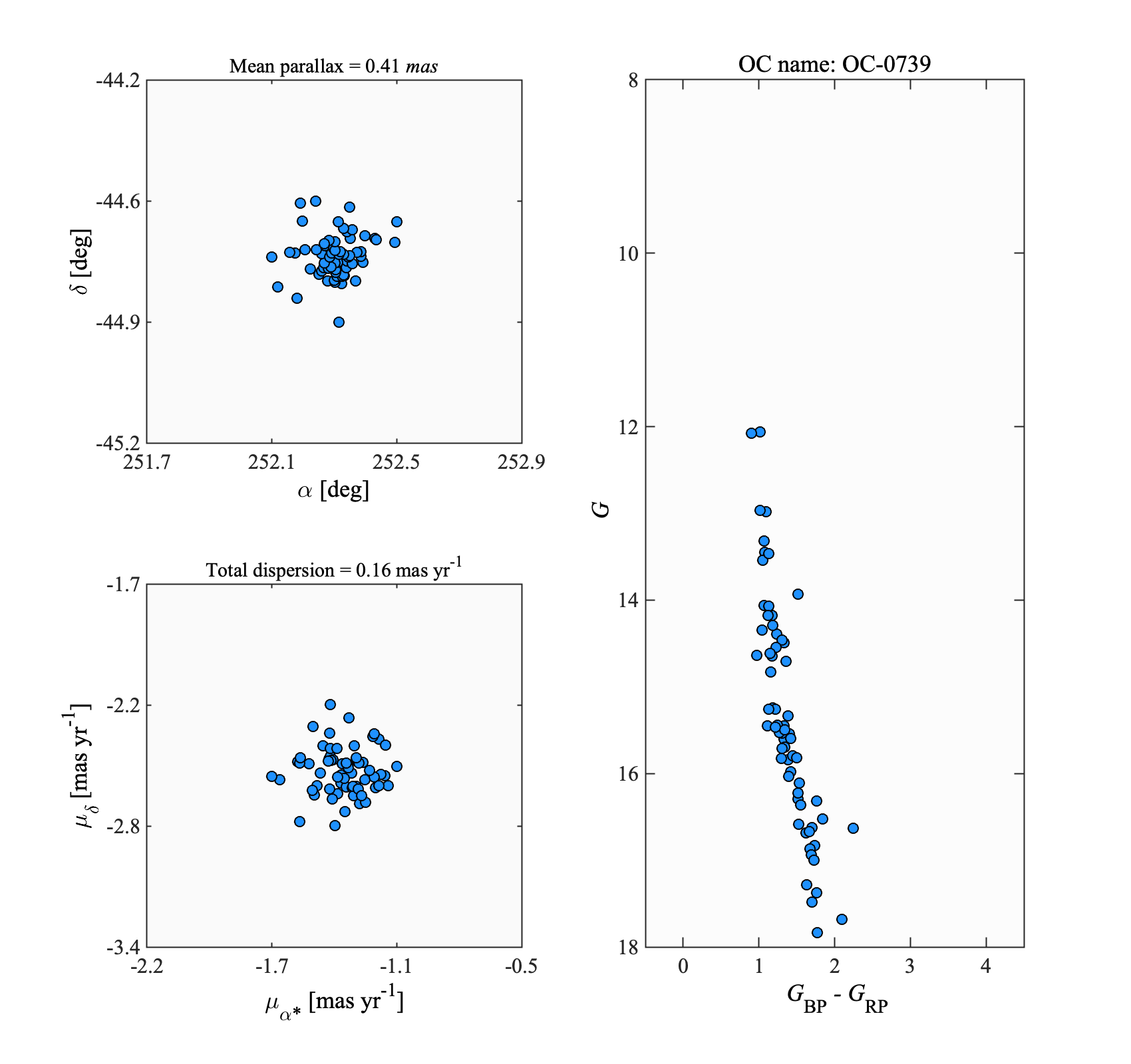} \hspace{0.0cm}
\caption{Examples of newly found OCs in this work. The columns of each panel are the same as Figure~\ref{fig:OCs_Ceps}. Here, the listed OCs are OC--0707, OC--0709, OC--0719, OC--0724, OC--0727, OC--0734, OC--0737, OC--0738, and OC--0739.}
\label{fig:OCs_new}
\end{figure*}
%%%%%%%%%%%%%%%%%%%%%%%%%%%%%%%%%%%%%%%%%%%%%%%

We cross-matched the OC--Cepheid pairs found in this work with 
those in~\citet{zc2021}, who reported the discovery of 29 OCs containing 
33 Cepheids.
Finally, we were able to re-detect 27 OCs holding 30 Cepheids.
We noticed that the differences in proper motion in $\mu_{\delta}$} 
between the remaining two OCs (Ruprecht 97 and UBC 375) and their two 
associated classical Cepheids proposed by~\citet{zc2021} are 
significantly large (10.9$\sigma$ and 7.3$\sigma$, respectively), 
resulting in the two pairs not being identified in this work.
There is another Cepheid (V379 Cas) not being matched by us because 
its spatial location is about one degree away from the target OC 
(NGC 129) in the sky.
The dynamical evolution proposed by~\citet{turner1985} suggests
that instances of Cepheids may be possible members of cluster 
coronae. 
\citet{coulson1985} and \citet{majaess2013} both reported that 
the Cepheid QZ Nor is a coronal member of the OC NGC 6067. 
Therefore, the Cepheid V379 Cas may be an ejected member of
the NGC 129 cluster. 
Although V379 Cas is not followed up here, it is still worth investigating.
We also recovered some interesting OCs that contain more than one
classical Cepheid: for example, the widely known Galactic OC NGC 7790 hosting 
three Cepheids, and the OC Kronberger 84 holding two Cepheids reported 
by~\citet{zc2021}.
The cross-match using \emph{Gaia} DR3 \texttt{source\_id} shows
that the Cepheid V733 Cyg is a non-member of Kronberger 84, but that the 
5D astrometric parameters of V733 Cyg are consistent with the cluster, 
within three times the corresponding standard deviations.
For the 45 OCs containing classical Cepheids, 35 are previously known
OCs compiled by~\citet{cantat2020a}, 
two OCs (OC--0125 and OC--0675) were reported by~\citet{hao2022}, 
one OC (Alessi 95) was reported by~\citet{turner2012b} 
and~\citet{majaess2012},
one OC (UBC 1424) was reported by~\citet{castro2022}, and the last 
six are newly discovered in this work.
Based on the member stars of the 45 OCs in \emph{Gaia} DR3,
we determined their mean astrometric parameters ($\alpha$, $\delta$, 
$\varpi$, $\mu_{\alpha^{*}}$, and $\mu_{\delta}$) and the corresponding 
standard deviations, as presented in Table~\ref{table:table1}.
The apparent angular radius ($\theta$) of each OC was estimated as 
the square root of the quadratic sum of $\sigma_{l}$ and 
$\sigma_{b}$~\citep[e.g.,][]{castro2020,hao2022,castro2022}.
The cluster apparent diameters are then determined as two times the 
$\theta$.
The ages and uncertainties of the known OCs are from the previous 
determinations in \citet{sana2010}, \citet{turner2012b}, \citet{cantat2020a}, 
\citet{hao2022}, and \citet{castro2022},
and those of newly found OCs were estimated by the method described in 
Sect.~\ref{sect:search}.
Table~\ref{table:table1} also presents the astrometric parameters 
($\alpha$, $\delta$, $\varpi$, $\mu_{\alpha^{*}}$, and $\mu_{\delta}$) of 
Classical Cepheids, which are from the dataset of \emph{Gaia} DR3, and 
the listed periods of Cepheids are from \citet{pietrukowicz2021}.
The Cepheid ages listed in Table~\ref{table:table1} are 
taken from the mean calculation of the
Cepheid PAR reported in following three works:
\citet{efremov2003}, \citet{bono2005}, and \citet{turner2012a}, and the 
deviation is cited as the uncertainty.
We note that \citet{efremov2003} and \citet{bono2005} only present 
the PAR for fundamental and first-overtone classical Cepheids,
and so the second-overtone pulsating mode of the Cepheid 
OGLE-GD-CEP-1673 means that its age is unavailable from the two works, 
and its age uncertainty is not listed in Table~\ref{table:table1}.
However, it should be noted that the obtained ages of Cepheids are 
uncertain parameters, and are affected by reddening, distance, photometric 
calibrations, and so on, as pointed out by~\citet{bono2005}.
We also estimated the angular separation between the classical Cepheid
and its host OC, where the cluster centre is derived from the mean 
astrometric positions of member stars.
We noticed that some OCs may exhibit pre-main sequence (PMS) characteristics, 
and therefore would be unrelated to Cepheids, particularly the lower period 
ones.
For example, the readily discernible PMS of Trumpler 14 was studied 
by~\citet{sana2010}, and the presence of O-type stars in NGC 6193 was 
reported by~\citet{skinner2005}, suggesting that the two OCs are not 
related to Cepheids.
Indeed, OCs with ages younger than ten million years are unlikely to be 
related to Cepheids, and those with ages near ten million 
years also tend to be related to very long-period Cepheids~\citep[e.g.,][]{anderson2016}.
We investigated all the identified OC--Cepheid pairs by considering  
the OC ages and the periods of Cepheids, and found that 11 Cepheids 
may be unrelated to OCs.
However, it is interesting that the astrometric matches of these Cepheids 
and their host OCs still emerge, and so these Cepheids are worthy of follow-up 
investigation in order to come to a firm conclusion on their nature.
In Table~\ref{table:table1}, class A Cepheids are related to OCs and have 
been identified as cluster members, 
Cepheids in class B are related to OCs but have not been  
identified as cluster members,
and class C Cepheids may be unrelated to OCs but they possess similar 
astrometric parameters.
Table~\ref{table:table1} can be found online at the CDS, together with the 
table reporting the member stars of OCs harboring classical 
Cepheids.

Figure~\ref{fig:OCs_Ceps} shows three examples of OCs holding 
classical Cepheids, including the members of OCs and Cepheids in the 
astrometric spaces as well as CMDs.
The remaining cases are presented in Figure~\ref{fig:OCs_Ceps_s1} in the Appendix.
As intermediate-mass stars, classical Cepheids have evolved away from 
the main sequence and are in the stage of 
shell hydrogen burning or core helium burning~\citep[e.g.,][]{turner2006}.
Hence, in the CMDs of bona fide OCs harboring classical Cepheids, 
Cepheids are generally located on the classical instability strip and 
are brighter than the main sequence stars.
We visually inspected the CMDs of the OC--Cepheid pairs identified 
in this work and found 38 Cepheids with the above features.  
For the last 12 classical Cepheids that may not be on the 
classical instability strip of their host OCs, these classical Cepheids 
require further investigation, and we associated asterisks with the names 
of these classical Cepheids in Table~\ref{table:table1}.

Figure~\ref{fig:lb} presents the distribution of the 45 OCs hosting classical 
Cepheids in Galactic $l$ and $b$ coordinates.
The vast majority ($\sim$ 93\%) of them are located within the Galactic 
disk at latitudes of $|b| \le $ 5$^{\circ}$.
We also made a comparison between the parallaxes of OCs containing 
classical Cepheids found in this work and those previously reported 
in \citet{cantat2020a}, \citet{castro2022}, and \citet{hao2022}, as shown in 
Fig.~\ref{fig:parallax_known}. 
The mean parallaxes of the known OCs re-detected in this work are 
in good agreement with previously published values.
However, compared with the previous results, the smaller errors of 
parallaxes obtained here demonstrate the improved data quality of 
\emph{Gaia} DR3 over \emph{Gaia} data release 2.
We also noticed that 43 of these OCs have $RV$ 
measurements, of which 40 are based on more than one member 
star, and 35 are based on more than two member stars.
The OCs marked with asterisks in Table~\ref{table:table1} have an 
$RV$ measurement.
Therefore, in addition to providing useful indications for optimizing the 
zero-point of the Cepheid period--luminosity relationship and period--age 
relation, it is hopeful that 
these OCs can be used to unveil the nature and especially the 
dynamical evolution of classical Cepheids.

\subsection{Newly found OCs}

As described in Sect.~\ref{sect:cross-match}, aside from re-detecting 
635 previously known OCs, we are fortunate to propose 
38 reliable ones as new OC candidates in this work.
Figure~\ref{fig:OCs_new} displays the distributions of some examples of 
newly found OCs. Here we present the features of their member stars in 
astrometric spaces and CMDs, indicating that these proposed new OCs 
show high concentrations of member stars in the five astrometric 
parameters, and the members in the CMDs are concentrated on
potentially theoretical isochrones.  
Table~\ref{table:table2} of the Appendix contains a list of the parameters 
of the new OC candidates,
including their mean astrometric parameters ($\alpha$, $\delta$, 
$l$, $b$, $\varpi$, $\mu_{\alpha^{*}}$, and $\mu_{\delta}$) and standard 
deviations, $RV$s, and the weighted standard deviations of $RV$s 
when available, as well as the estimated ages, line-of-sight 
extinction ($A$$_{\rm G}$), and DM.
In addition, the apparent angular sizes of these new OC candidates were 
computed as the square root of the quadratic sum 
of $\sigma_{l}$ and $\sigma_{b}$ following \citet{castro2020}, \citet{hao2022}, 
and \citet{castro2022}. 
The table can be found online at the CDS, including the table of member 
stars.

The new OC candidates are generally at larger distances, of which 
$\sim$ 95\% are further than 1~kpc, and only two OCs are located within 1 kpc.
For the newly found OCs, 31 of them have member stars with $RV$s in 
\emph{Gaia} DR3, more than half of which contain more than one star 
with an $RV$ measurement, 
and the names of these OCs have been placed with asterisks in 
Table~\ref{table:table2}.
The distributions of these OCs in the Galactic plane are also shown in 
Fig.~\ref{fig:lb}.
The newly found OCs are mostly found at low latitudes within the 
Galactic disk, with the vast majority of new OC candidates ($\sim$ 97\%) 
located at $|b| \le $ 4$^{\circ}$, and only one at $b \approx$ 7$^{\circ}$.
%
%

%%%%%%%%%%%%%%%%%%%%%%%%%%%%%%%%%%%%%%%%%%%%%%%%%%

\section{Summary}
\label{sect:summary}

Combining the high-precision data provided by \emph{Gaia} DR3 and 
known classical Cepheids with our previously practiced clustering 
search method, we conducted a search for OCs 
containing classical Cepheids.
We identified 50 classical Cepheids associated with 45 OCs in 
5D astrometric space. 
After investigating the ages of the OCs  and the periods of the Cepheids, 
we found 39 OC-Cepheid pairs to be probable, and a remaining 11 
Cepheids to be improbable and possibly unrelated to OCs but worthy of 
follow-up investigation.
For all of these OC--Cepheid pairs, we provide their astrometric 
and astrophysical parameters.
In addition to re-detecting 635 known OCs in the literature, 
we also propose 38 newly found OC candidates within the Galactic 
disk based on a detailed inspection of the multidimensional distributions of their 
member stars. 
For each of these new OC candidates, we present its mean astrometric 
values and radial velocity when available, together with estimations 
of age, line-of-sight extinction, and distance modulus.
The new OCs are generally located at greater distances from the Sun,
which can help to improve the completeness of the Galactic OC 
census.

Our current search mainly focuses on the regions within 5 kpc 
of the Sun. 
Hence, it is expected that more OC--classical Cepheid pairs 
will be detected in our Galaxy along with the improvements in parallax 
and proper motion precision of the future data releases of \emph{Gaia}.
%

%%%%%%%%%%%%%%%%%%%%%%%%%%%%%%%%%%%%%%%%%%%%%%%%%%
\section*{Acknowledgements}
We sincerely appreciate the anonymous referee for the instructive comments 
which help us to improve the paper.
This work was funded by NSFC Grants 11933011, 11873019, 12203104, and 
the Key Laboratory for Radio Astronomy. 
YJL thanks support from the Natural Science Foundation of Jiangsu Province 
(grant number BK20210999) and the Entrepreneurship and Innovation Program 
of Jiangsu Province.
Our work has made use of data from the European Space Agency (ESA) 
mission Gaia (\url{https://www.cosmos.esa.int/gaia}), processed by the Gaia 
Data Processing and Analysis Consortium 
(DPAC, \url{https://www. cosmos.esa.int/web/gaia/dpac/consortium}). 
Funding for the DPAC has been provided by national institutions, in particular 
the institutions participating in the Gaia Multilateral Agreement.

% WARNING
%------------------------------------------------------------------$-$
% Please note that we have included the references to the file aa.dem

% order to compile it, but we ask you to:
%
% $-$ use BibTeX with the regular commands:
  %\bibliographystyle{aa} % style aa.bst
 %\bibliography{refGaia.bib} % your references Yourfile.bib
%
% $-$ join the .bib files when you upload your source files
%------------------------------------------------------------------$-$

%%%%%%%%%%%%%%%%%%%%%%%%%%%%%%%%%%%%%%%%%%%%

\begin{appendix}
\section{Additional tables and figures}

%%%%%%%%%%%%%%%%%%%%%%%%%%%%%%%%%%%%%%%%%%%% Table.1
%%%%%%%%%%%%%%%%%%%%%%%%%%%%%%%%%%%%%%%%%%%%
\begin{sidewaystable*}[htp]
\centering
\caption{Parameters of the OC-Cepheid pairs identified in this work.}
\scriptsize
\setlength{\tabcolsep}{0.5mm}
\renewcommand\arraystretch{1.0}
\begin{tabular}{l  r@{}l  r@{}l  c l r@{}l  r@{}l  r  r  r@{}l  r@{}l  r@{}l  r@{}l  r@{}l  r@{}l   r@{}l  r@{}l r r}
\hline \hline 
\multicolumn{1}{c}{Cep name} &
\multicolumn{2}{c}{Cep: $\alpha$} &
\multicolumn{2}{c}{Cep: $\delta$} &
\multicolumn{1}{c}{Cep: $\log$(P)} &
\multicolumn{1}{c}{OC name} &
\multicolumn{2}{c}{OC: $\alpha$} &
\multicolumn{2}{c}{OC: $\delta$} &
\multicolumn{1}{c}{Separation} &
\multicolumn{1}{c}{OC: AD} &
\multicolumn{2}{c}{Cep: $\mu_{\alpha^{*}}$} &
\multicolumn{2}{c}{Cep: $\mu_{\delta}$} &
\multicolumn{2}{c}{OC: $\mu_{\alpha^{*}}$} &
\multicolumn{2}{c}{OC: $\mu_{\delta}$} &
\multicolumn{2}{c}{Cep: $\varpi$} &
\multicolumn{2}{c}{OC: $\varpi$} &
\multicolumn{2}{c}{Cep: $\log$(age)} &
\multicolumn{2}{c}{OC: $\log$(age)} &
\multicolumn{1}{c}{OC: $N$} &
\multicolumn{1}{c}{{\rm Num}}
\\
\multicolumn{1}{c}{} &
\multicolumn{2}{c}{[deg]} &
\multicolumn{2}{c}{[deg]} &
\multicolumn{1}{c}{day} &
\multicolumn{1}{c}{} &
\multicolumn{2}{c}{[deg]} &
\multicolumn{2}{c}{[deg]} &
\multicolumn{1}{c}{[arcmin]} &
\multicolumn{1}{c}{[arcmin]} &
\multicolumn{2}{c}{[mas yr$^{-1}$]} &
\multicolumn{2}{c}{[mas yr$^{-1}$]} &
\multicolumn{2}{c}{[mas yr$^{-1}$]} &
\multicolumn{2}{c}{[mas yr$^{-1}$]} &
\multicolumn{2}{c}{[mas]} &
\multicolumn{2}{c}{[mas]} &
\multicolumn{2}{c}{[yr]} &
\multicolumn{2}{c}{[yr]} &
\multicolumn{1}{c}{} &
\multicolumn{1}{c}{}
 \\
 \hline
\multicolumn{27}{c}{\hspace{3.0cm}  class A } \\ \hline
CG Cas &   0.25 & (0.01) &  60.96 & (0.01) & 0.64 &   Berkeley 58(*) &   0.07 & (0.13) &  60.94 & (0.06) &   5.3 &   10.8 &  -3.24 & (0.01) &  -1.67 & (0.01) &  -3.48 & (0.09) &  -1.81 & (0.09) &   0.27 & (0.01) &   0.29 & (0.02) & 8.01 & (0.02) & 7.78 & (0.16) &  173  & 1\\
NO Cas &   6.02 & (0.01) &  61.34 & (0.01) & 0.41 &     NGC 103(*) &   6.31 & (0.12) &  61.32 & (0.04) &   8.4 &    8.4 &  -2.83 & (0.01) &  -1.21 & (0.01) &  -2.80 & (0.09) &  -1.08 & (0.07) &   0.27 & (0.01) &   0.28 & (0.02) & 8.04 & (0.06) & 8.01 & (0.16) &  173  & 2     \\ 
DL Cas &   7.49 & (0.02) &  60.21 & (0.02) & 0.90 &     NGC 129(*) &   7.62 & (0.17) &  60.20 & (0.07) &    4.0 &   13.2 &  -2.71 & (0.03) &  -1.19 & (0.03) &  -2.59 & (0.12) &  -1.19 & (0.11) &   0.55 & (0.03) &   0.53 & (0.03) & 7.81 & (0.03) & 8.11 & (0.16) &  296  & 3    \\ 
SU Cas &  42.99 & (0.04) &  68.89 & (0.04) & 0.29 &   Alessi 95(*) &  42.98 & (0.77) &  68.89 & (0.21) &    0.5 &   42.0 &   3.10 & (0.04) &  -8.15 & (0.05) &   1.46 & (0.34) &  -7.94 & (0.41) &   2.17 & (0.06) &   2.28 & (0.06) & 8.14 & (0.04) & 8.20 & (0.10) &  104 & 4 \\ 
RS Ori &  95.55 & (0.03) &  14.68 & (0.02) & 0.88 &      FSR 0951(*) &  95.58 & (0.09) &  14.62 & (0.10) &    3.5 &   16.8 &   0.20 & (0.04) &   0.01 & (0.03) &   0.22 & (0.10) &   0.03 & (0.11) &   0.56 & (0.03) &   0.57 & (0.03) & 7.83 & (0.02) & 8.72 & (0.17) &  146  & 5 \\ 
CV Mon &  99.27 & (0.01) &   3.06 & (0.01) & 0.73 &     vdBergh 1(*) &  99.28 & (0.04) &   3.08 & (0.03) &    1.1 &    6.0 &   0.35 & (0.02) &  -0.67 & (0.01) &   0.40 & (0.13) &  -0.71 & (0.12) &   0.57 & (0.01) &   0.55 & (0.04) & 7.94 & (0.02) & 7.61 & (0.15) &   59  & 6 \\ 
WX Pup & 115.50 & (0.01) & -25.88 & (0.01) & 0.95 &     OC-0717(*) & 115.29 & (0.24) & -26.03 & (0.28) &   14.2 &   42.0 &  -2.16 & (0.01) &   2.56 & (0.01) &  -2.15 & (0.25) &   2.25 & (0.43) &   0.37 & (0.02) &   0.37 & (0.01) & 7.77 & (0.03) & 8.28 & (0.17) &   71   & 7 \\ 
V335 Pup & 119.24 & (0.01) & -22.83 & (0.01) & 0.69 &     UBC 229(*) & 119.28 & (0.05) & -22.82 & (0.05) &    2.0 &    8.4 &  -2.97 & (0.01) &   2.89 & (0.02) &  -2.98 & (0.06) &   2.90 & (0.11) &   0.42 & (0.02) &   0.38 & (0.03) & 7.81 & (0.11) & 8.05 & (0.16) &   57   & 8 \\ 
J075840-3330.2 & 119.67 & (0.01) & -33.50 & (0.01) & 0.64 &   UBC 1424(*) & 119.70 & (0.09) & -33.58 & (0.07) &    5.2 &   12.0 &  -2.34 & (0.01) &   3.42 & (0.01) &  -2.34 & (0.07) &   3.27 & (0.16) &   0.24 & (0.01) &   0.24 & (0.01) & 8.00 & (0.02) & 7.66 & (0.15) &   37  & 9 \\ 
CS Vel & 145.29 & (0.01) & -53.82 & (0.01) & 0.77 &  Ruprecht 79(*) & 145.26 & (0.06) & -53.84 & (0.04) &    1.9 &    6.0 &  -4.57 & (0.01) &   3.13 & (0.01) &  -4.59 & (0.11) &   3.02 & (0.09) &   0.26 & (0.01) &   0.24 & (0.02) & 7.91 & (0.02) & 7.79 & (0.16) &  185   & 10 \\ 
V Cen & 218.14 & (0.01) & -56.89 & (0.01) & 0.74 &    NGC 5662(*) & 218.77 & (0.37) & -56.67 & (0.21) &   24.7 &   34.8 &  -6.70 & (0.02) &  -7.07 & (0.02) &  -6.48 & (0.27) &  -7.19 & (0.25) &   1.39 & (0.02) &   1.30 & (0.03) & 7.93 & (0.02) & 8.30 & (0.17) &  244  & 11 \\ 
V340 Nor & 243.32 & (0.02) & -54.23 & (0.01) & 1.05 &  NGC 6067(*) & 243.31 & (0.08) & -54.23 & (0.05) &    0.6 &    8.4 &  -2.07 & (0.03) &  -2.63 & (0.02) &  -1.97 & (0.14) &  -2.58 & (0.14) &   0.47 & (0.03) &   0.47 & (0.03) & 7.70 & (0.03) & 8.10 & (0.16) &  633  & 12 \\ 
S Nor & 244.72 & (0.02) & -57.90 & (0.01) & 0.99 &   NGC 6087(*) & 244.77 & (0.18) & -57.87 & (0.11) &    2.2 &   16.8 &  -1.61 & (0.02) &  -2.14 & (0.02) &  -1.63 & (0.18) &  -2.42 & (0.20) &   1.08 & (0.02) &   1.02 & (0.03) & 7.74 & (0.03) & 8.00 & (0.16) &  157  & 13 \\ 
OGLE-BLG-CEP-114(*) & 260.81 & (0.04) & -44.33 & (0.03) & 0.27 &   OC-0675(*) & 260.78 & (0.06) & -44.39 & (0.04) &    3.6 &    7.2 &   2.15 & (0.07) &  -3.34 & (0.05) &   2.12 & (0.17) &  -3.26 & (0.15) &   0.91 & (0.05) &   0.88 & (0.03) & 8.16 & (0.03) & 7.70 & (0.18) &   37  & 14 \\ 
U Sgr & 277.97 & (0.02) & -19.13 & (0.02) & 0.83 &   IC 4725(*) & 277.97 & (0.17) & -19.12 & (0.15) &    0.4 &   26.4 &  -1.80 & (0.02) &  -6.13 & (0.02) &  -1.69 & (0.23) &  -6.17 & (0.26) &   1.57 & (0.02) &   1.52 & (0.05) & 7.87 & (0.02) & 8.05 & (0.16) &  399  & 15  \\ 
V367 Sct & 278.40 & (0.02) & -10.43 & (0.02) & 0.80 &   NGC 6649(*) & 278.36 & (0.05) & -10.40 & (0.05) &    3.0 &    8.4 &   0.08 & (0.02) &  -0.27 & (0.02) &   0.03 & (0.15) &  -0.12 & (0.15) &   0.42 & (0.02) &   0.47 & (0.05) & 7.89 & (0.02) & 7.85 & (0.16) &  507  & 16 \\ 
EV Sct & 279.17 & (0.01) &  -8.18 & (0.01) & 0.49 &   NGC 6664(*) & 279.13 & (0.06) &  -8.18 & (0.05) &    2.7 &    9.6 &  -0.21 & (0.02) &  -2.55 & (0.01) &  -0.07 & (0.14) &  -2.59 & (0.13) &   0.49 & (0.02) &   0.46 & (0.04) & 7.97 & (0.07) & 8.35 & (0.17) &  214  & 17 \\ 
GQ Vul & 296.99 & (0.01) &  26.00 & (0.02) & 1.10 &  FSR 0158(*) & 297.00 & (0.05) &  26.04 & (0.05) &    2.2 &    8.4 &  -2.81 & (0.02) &  -5.47 & (0.02) &  -2.77 & (0.08) &  -5.50 & (0.10) &   0.14 & (0.02) &   0.14 & (0.01) & 7.66 & (0.03) & 7.81 & (0.25) &   33   & 18 \\ 
X Vul & 299.37 & (0.01) &  26.56 & (0.02) & 0.80 &   UBC 129(*) & 299.09 & (0.16) &  26.48 & (0.16) &   15.7 &   26.4 &  -1.35 & (0.02) &  -4.25 & (0.02) &  -1.00 & (0.15) &  -4.35 & (0.20) &   0.84 & (0.02) &   0.85 & (0.02) & 7.89 & (0.02) & 7.72 & (0.15) &  145  & 19 \\ 
GI Cyg & 299.89 & (0.01) &  33.75 & (0.01) & 0.76 &   UBC 135(*) & 299.82 & (0.06) &  33.73 & (0.04) &    3.6 &    7.2 &  -3.45 & (0.01) &  -6.58 & (0.02) &  -3.52 & (0.08) &  -6.45 & (0.10) &   0.25 & (0.01) &   0.24 & (0.03) & 7.92 & (0.02) & 7.38 & (0.15) &   88  & 20 \\ 
J201151.18+342447.2 & 302.96 & (0.01) &  34.41 & (0.01) & 0.99 & Berkeley 51(*) & 303.01 & (0.05) &  34.39 & (0.05) &    2.6 &    7.2 &  -3.18 & (0.01) &  -4.91 & (0.02) &  -3.06 & (0.10) &  -4.84 & (0.12) &   0.19 & (0.01) &   0.15 & (0.02) & 7.74 & (0.03) & 7.98 & (0.16) &   64  & 21 \\ 
V1788 Cyg & 310.65 & (0.02) &  38.46 & (0.02) & 1.15 &   OC-0125(*) & 310.77 & (0.09) &  38.45 & (0.05) &    5.3 &   10.8 &  -2.71 & (0.02) &  -4.45 & (0.02) &  -2.74 & (0.17) &  -4.28 & (0.17) &   0.35 & (0.02) &   0.34 & (0.03) & 7.63 & (0.04) & 7.50 & (0.17) &   74  & 22 \\ 
J211659.94+514556.7 & 319.25 & (0.02) &  51.77 & (0.01) & 0.77 &   Berkeley 55(*) & 319.24 & (0.06) &  51.77 & (0.03) &    0.5 &    6.0 &  -3.91 & (0.02) &  -4.72 & (0.02) &  -4.07 & (0.15) &  -4.65 & (0.14) &   0.32 & (0.02) &   0.32 & (0.05) & 7.91 & (0.02) & 8.30 & (0.17) &  104  & 23 \\ 
J213533.70+533049.3 & 323.89 & (0.01) &  53.51 & (0.01) & 0.51 &  Kronberger 84(*) & 323.91 & (0.18) &  53.53 & (0.11) &    1.2 &   18.0 &  -2.88 & (0.01) &  -3.11 & (0.01) &  -2.92 & (0.17) &  -3.03 & (0.15) &   0.19 & (0.01) &   0.19 & (0.03) & 7.96 & (0.08) & 8.46 & (0.17) &   93  & 24 \\ 
CE Cas B & 359.54 & (0.01) &  61.21 & (0.01) & 0.65 &  NGC 7790(*) & 359.61 & (0.08) &  61.21 & (0.03) &    2.1 &    6.0 &  -3.30 & (0.01) &  -1.81 & (0.02) &  -3.23 & (0.09) &  -1.73 & (0.08) &   0.31 & (0.02) &   0.29 & (0.03) & 8.00 & (0.02) & 8.11 & (0.16) &  159 & 25 \\ 
CE Cas A & 359.54 & (0.01) &  61.21 & (0.01) & 0.71 &   NGC 7790(*) & 359.61 & (0.08) &  61.21 & (0.03) &    2.1 &    6.0 &  -3.30 & (0.01) &  -1.87 & (0.02) &  -3.23 & (0.09) &  -1.73 & (0.08) &   0.31 & (0.02) &   0.29 & (0.03) & 7.96 & (0.02) & 8.11 & (0.16) &  159 \\ 
CF Cas & 359.57 & (0.01) &  61.22 & (0.01) & 0.69 &  NGC 7790(*) & 359.61 & (0.08) &  61.21 & (0.03) &    1.3 &    6.0 &  -3.24 & (0.01) &  -1.77 & (0.01) &  -3.23 & (0.09) &  -1.73 & (0.08) &   0.29 & (0.01) &   0.29 & (0.03) & 7.97 & (0.02) & 8.11 & (0.16) &  159 \\ 
 \hline
\multicolumn{27}{c}{\hspace{3.0cm}  class B} \\ \hline                 
V423 CMa(*) & 110.81 & (0.01) & -29.72 & (0.01) & 0.52 &   OC-0718(*) & 110.65 & (0.21) & -29.86 & (0.14) &   11.5 &   27.6 &  -1.05 & (0.01) &   1.83 & (0.02) &  -0.73 & (0.11) &   1.87 & (0.21) &   0.13 & (0.02) &   0.12 & (0.01) & 8.09 & (0.01) & 8.65 & (0.17) &   45 & 26 \\ 
BH Vel(*) & 127.45 & (0.01) & -41.22 & (0.01) & 0.86 &    OC-0721(*) & 127.19 & (0.10) & -41.19 & (0.14) &   12.1 &   19.2 &  -3.24 & (0.01) &   4.10 & (0.02) &  -2.92 & (0.12) &   3.49 & (0.28) &   0.27 & (0.01) &   0.25 & (0.01) & 7.84 & (0.03) & 7.58 & (0.15) &   22 & 27   \\ 
J084951-4627.2 & 132.46 & (0.01) & -46.45 & (0.01) & 0.58 &   OC-0723(*) & 132.34 & (0.11) & -46.56 & (0.11) &   8.4 &   16.8 &  -3.66 & (0.02) &   3.45 & (0.02) &  -3.25 & (0.15) &   3.61 & (0.17) &   0.21 & (0.01) &   0.19 & (0.02) & 7.90 & (0.10) & 7.46 & (0.15) &  116 &  28   \\ 
DP Vel & 142.57 & (0.01) & -53.06 & (0.01) & 0.74 &   UBC 491(*) & 142.62 & (0.33) & -53.01 & (0.12) &    3.6 &   27.6 &  -4.31 & (0.02) &   3.35 & (0.01) &  -4.35 & (0.16) &   3.34 & (0.12) &   0.33 & (0.01) &   0.27 & (0.02) & 7.93 & (0.02) & 8.26 & (0.17) &  199 &  29   \\ 
X Cru & 191.59 & (0.01) & -59.12 & (0.01) & 0.79 &   UBC 290(*) & 191.88 & (0.30) & -59.37 & (0.11) &   17.3 &   22.8 &  -5.93 & (0.02) &  -0.17 & (0.02) &  -5.94 & (0.22) &  -0.22 & (0.17) &   0.64 & (0.02) &   0.61 & (0.02) & 7.89 & (0.03) & 8.28 & (0.17) &  229 &  30 \\ 
VdBH222-505 & 259.70 & (0.03) & -38.29 & (0.02) & 1.37 &   BH 222(*) & 259.66 & (0.08) & -38.29 & (0.04) &    1.8 &    9.6 &  -1.94 & (0.04) &  -3.42 & (0.03) &  -1.83 & (0.17) &  -3.18 & (0.19) &   0.17 & (0.04) &   0.11 & (0.03) & 7.46 & (0.05) & 7.73 & (0.15) &   72 &  31 \\ 
CM Sct & 280.61 & (0.01) &  -5.34 & (0.01) & 0.59 &   UBC 106(*) & 280.47 & (0.05) &  -5.42 & (0.05) &    9.7 &    9.6 &  -1.06 & (0.02) &  -1.41 & (0.01) &  -1.04 & (0.12) &  -1.36 & (0.13) &   0.41 & (0.02) &   0.41 & (0.03) & 8.04 & (0.02) & 8.20 & (0.16) &  294 & 32 \\ 
CN Sct & 280.63 & (0.03) &  -4.33 & (0.02) & 1.00 &   Trumpler 35(*) & 280.74 & (0.06) &  -4.22 & (0.06) &    9.3 &    9.6 &  -1.04 & (0.03) &  -2.26 & (0.03) &  -0.97 & (0.13) &  -2.22 & (0.16) &   0.36 & (0.03) &   0.34 & (0.03) & 7.74 & (0.03) & 7.59 & (0.15) &  183 & 33 \\ 
J194806.54+260526.1 & 297.03 & (0.01) &  26.09 & (0.01) & 0.82 &  FSR 0158(*) & 297.00 & (0.05) &  26.04 & (0.05) &    3.6 &    8.4 &  -2.87 & (0.01) &  -5.55 & (0.02) &  -2.77 & (0.08) &  -5.50 & (0.10) &   0.17 & (0.02) &   0.14 & (0.01) & 7.69 & (0.14) & 7.81 & (0.25) &   35 &   \\ 
SV Vul & 297.88 & (0.01) &  27.46 & (0.02) & 1.65 &  UBC 130(*) & 298.06 & (0.06) &  27.44 & (0.06) &   9.7 &    9.6 &  -2.16 & (0.02) &  -5.96 & (0.02) &  -2.12 & (0.09) &  -5.87 & (0.13) &   0.37 & (0.02) &   0.40 & (0.02) & 7.25 & (0.06) & 7.44 & (0.15) &  101 &  34 \\ 
J300.0102+29.1869(*) & 300.01 & (0.02) &  29.19 & (0.03) & 1.27 &  FSR 0172 & 300.00 & (0.03) &  29.22 & (0.03) &    1.9 &    3.6 &  -2.50 & (0.03) &  -6.17 & (0.03) &  -2.52 & (0.08) &  -6.01 & (0.10) &   0.34 & (0.04) &   0.29 & (0.01) & 7.54 & (0.04) & 8.20 & (0.16) &   41 & 35 \\ 
 V733 Cyg & 323.51 & (0.01) &  53.31 & (0.01) & 0.66 &  Kronberger 84(*) & 323.91 & (0.18) &  53.53 & (0.11) &   19.4 &   18.0 &  -2.71 & (0.01) &  -3.34 & (0.01) &  -2.92 & (0.17) &  -3.03 & (0.15) &   0.22 & (0.01) &   0.19 & (0.03) & 7.99 & (0.02) & 8.46 & (0.17) &   93 &    \\ 
  \hline
\multicolumn{27}{c}{\hspace{3.0cm}  class C} \\ \hline    
MQ Cam &  60.38 & (0.01) &  55.05 & (0.01) & 0.82 &   OC-0715(*) &  59.94 & (0.40) &  55.54 & (0.27) &   33.3 &   42.0 &   0.26 & (0.02) &   0.08 & (0.02) &  -0.01 & (0.11) &  -0.05 & (0.11) &   0.25 & (0.02) &   0.22 & (0.02) & 7.87 & (0.03) & 6.80 & (0.14) &  307 & 36 \\ 
 OGLE-GD-CEP-0177(*) & 122.16 & (0.05) & -36.07 & (0.05) & 0.26 & OC-0720(*) & 122.19 & (0.11) & -36.16 & (0.17) &    5.9 &   22.8 &  -7.37 & (0.07) &  12.49 & (0.07) &  -7.51 & (0.20) &  11.41 & (0.51) &   2.79 & (0.06) &   2.78 & (0.06) & 8.17 & (0.03) & 6.32 & (0.13) &   15   & 37 \\ 
 OGLE-GD-CEP-0270(*) & 130.89 & (0.03) & -47.75 & (0.03) & 0.21 &   IC 2395(*) & 130.47 & (0.26) & -48.07 & (0.15) &   25.7 &   27.6 &  -4.77 & (0.05) &   3.36 & (0.04) &  -4.43 & (0.23) &   3.33 & (0.24) &   1.38 & (0.04) &   1.40 & (0.05) & 8.21 & (0.02) & 7.31 & (0.15) &  284 & 38  \\ 
OGLE-GD-CEP-1673(*) & 161.00 & (0.05) & -59.57 & (0.05) & 0.66 &  Trumpler 14(*) & 161.01 & (0.07) & -59.55 & (0.03) &    1.6 &    6.0 &  -6.56 & (0.07) &   2.07 & (0.07) &  -6.50 & (0.20) &   2.08 & (0.17) &   0.34 & (0.06) &   0.39 & (0.03) & 7.78 & (--) & 5.60 & (0.10) &  342  & 39 \\ 
OGLE-GD-CEP-1676(*) & 166.11 & (0.13) & -59.73 & (0.12) & 0.01 &    UBC 266 & 166.29 & (0.10) & -59.73 & (0.05) &   5.7 &    8.4 &  -6.59 & (0.17) &   1.52 & (0.16) &  -6.71 & (0.09) &   1.77 & (0.08) &   0.22 & (0.15) &   0.39 & (0.01) & 8.65 & (0.08) & 7.11 & (0.14) &   78 & 40 \\ 
OGLE-GD-CEP-1688(*) & 174.87 & (0.01) & -63.49 & (0.01) & 0.31 &    BH 121(*) & 174.76 & (0.24) & -63.41 & (0.07) &    5.6 &   15.6 &  -5.91 & (0.01) &   0.48 & (0.01) &  -5.99 & (0.15) &   0.66 & (0.14) &   0.41 & (0.01) &   0.40 & (0.01) & 8.12 & (0.04) & 6.42 & (0.13) &  153  & 41 \\ 
 TW Nor & 241.23 & (0.02) & -51.95 & (0.01) & 1.03 &   Lynga 6(*) & 241.19 & (0.09) & -51.96 & (0.05) &    1.6 &    9.6 &  -1.89 & (0.02) &  -2.81 & (0.02) &  -1.91 & (0.12) &  -2.74 & (0.11) &   0.32 & (0.02) &   0.38 & (0.04) & 7.71 & (0.03) & 6.49 & (0.13) &  180  & 42 \\ 
OGLE-GD-CEP-1175(*) & 250.20 & (0.05) & -48.70 & (0.03) & 0.30 &   NGC 6193(*) & 250.33 & (0.09) & -48.78 & (0.05) &    7.1 &    9.6 &   0.94 & (0.07) &  -3.92 & (0.04) &   1.28 & (0.24) &  -3.99 & (0.19) &   0.78 & (0.06) &   0.84 & (0.05) & 8.13 & (0.04) & 6.71 & (0.13) &  184 &  43 \\ 
OGLE-GD-CEP-1194(*) & 253.43 & (0.11) & -41.30 & (0.07) & 0.28 &   UBC 323(*) & 253.83 & (0.23) & -40.77 & (0.21) &   36.6 &   32.4 &  -0.55 & (0.15) &  -1.76 & (0.12) &  -0.30 & (0.34) &  -1.35 & (0.37) &   0.52 & (0.12) &   0.53 & (0.01) & 8.15 & (0.03) & 6.95 & (0.14) &  116 &  44 \\ 
OGLE-GD-CEP-1196(*) & 253.62 & (0.10) & -41.27 & (0.06) & 0.01 &   UBC 323(*) & 253.83 & (0.23) & -40.77 & (0.21) &   31.8 &   32.4 &  -0.59 & (0.14) &  -1.85 & (0.11) &  -0.30 & (0.34) &  -1.35 & (0.37) &   0.80 & (0.10) &   0.53 & (0.01) & 8.51 & (0.06) & 6.95 & (0.14) &  116 &   \\ 
J297.7863+25.3136 & 297.79 & (0.01) &  25.31 & (0.01) & 0.47 &  Czernik 41(*) & 297.75 & (0.06) &  25.29 & (0.06) &    2.7 &    9.6 &  -3.03 & (0.01) &  -6.29 & (0.02) &  -2.94 & (0.11) &  -6.15 & (0.13) &   0.32 & (0.02) &   0.37 & (0.04) & 7.99 & (0.07) & 7.11 & (0.14) &  236  & 45 \\ 
 \hline
\end{tabular}
 \tablefoot{Class A: Cepheids are related to OCs and have been identified as cluster members; 
 Class B: Cepheids are related to OCs but have not been identified as cluster members.; 
 Class C: Cepheids may be unrelated to OCs but they possess similar astrometric parameters.
The astrometric parameters for each classical Cepheid (Cep) are from \emph{Gaia} DR3. 
Separation is the angular distance between a Cepheid and its host OC.
For each OC, its astrometric parameters, corresponding standard deviations, and apparent angular diameter (AD) are derived from its member stars ($N$). 
The ages and uncertainties of known OCs are derived from the literature, and those of newly found OCs were estimated in this work.
Asterisks associated with Cepheid names indicate that these Cepheid may not fall within the instability strip.
OC names with asterisks show that they have some \emph{Gaia} RV measurements.
The table can be found at the CDS.}
 \label{table:table1}
\end{sidewaystable*}
%%%%%%%%%%%%%%%%%%%%%%%%%%%%%%%%%%%%%%%%%%%%

%%%%%%%%%%%%%%%%%%%%%%%%%%%%%%%%%%%%%%%%%%%% Table 2
%%%%%%%%%%%%%%%%%%%%%%%%%%%%%%%%%%%%%%%%%%%%
\begin{table*}
\centering
\caption{Parameters of the proposed newly found OCs ordered by increasing $l$.}
\scriptsize
\setlength{\tabcolsep}{1.20mm}
\renewcommand\arraystretch{1.80}
\begin{tabular}{l  r@{}l  r@{}l  r@{}l  r@{}l  c  l@{}r  r@{}l  r@{}l  r@{}l  c  c  r@{}l  r@{}l}
\hline \hline \\
\multicolumn{1}{c}{ID} &
\multicolumn{2}{c}{$\alpha$} &
\multicolumn{2}{c}{$\delta$} &
\multicolumn{2}{c}{$l$} &
\multicolumn{2}{c}{$b$} &
\multicolumn{1}{c}{$\theta$} &
\multicolumn{2}{c}{$\varpi$} &
\multicolumn{2}{c}{$\mu_{\alpha^{*}}$} &
\multicolumn{2}{c}{$\mu_{\delta}$} &
\multicolumn{2}{c}{$\log$(age)} &
\multicolumn{1}{c}{$A$$_{\rm G}$} &
\multicolumn{1}{c}{DM} &
\multicolumn{2}{c}{$V_{r}$} &
\multicolumn{2}{c}{$N$}
\\
\multicolumn{1}{c}{} &
\multicolumn{2}{c}{[deg]} &
\multicolumn{2}{c}{[deg]} &
\multicolumn{2}{c}{[deg]} &
\multicolumn{2}{c}{[deg]} &
\multicolumn{1}{c}{[deg]} &
\multicolumn{2}{c}{[mas]} &
\multicolumn{2}{c}{[mas yr$^{-1}$]} &
\multicolumn{2}{c}{[mas yr$^{-1}$]} &
\multicolumn{2}{c}{[yr]} &
\multicolumn{1}{c}{[mag]} &
\multicolumn{1}{c}{[mag]} &
\multicolumn{2}{c}{[km s$^{-1}$]} &
\multicolumn{2}{c}{$(N_{r})$}
 \\ \\
 \hline \\
          OC$-$0705(*) & 268.56 & (0.11) & $-$23.14 & (0.09) &   5.97 & (0.09) &   1.31 & (0.09) & 0.13 &   0.74 & (0.01) &   0.10 & (0.23) &  $-$2.64 & (0.25) &   7.50 & (0.17) &   2.48 &  10.86 & $-$5.40 & (--) &   30 & (1) \\
          OC$-$0706(*) & 278.38 & (0.04) & $-$10.40 & (0.04) &  21.65 & (0.04) &  $-$0.80 & (0.04) & 0.06 &   0.54 & (0.01) &  $-$0.01 & (0.17) &  $-$0.12 & (0.17) &   8.46 & (0.19) &   4.04 &  10.95 &  $-$18.24 & (--) &   26 & (1) \\
          OC$-$0707(*) & 278.37 & (0.06) & $-$10.39 & (0.06) &  21.65 & (0.06) &  $-$0.79 & (0.06) & 0.09 &   0.40 & (0.01) &   0.03 & (0.17) &  $-$0.14 & (0.16) &   7.72 & (0.18) &   4.04 &  11.19 &  $-$9.37 & (--) &   58 & (1) \\
          OC$-$0708 & 282.68 & (0.05) &  $-$6.35 & (0.05) &  27.19 & (0.05) &  $-$2.73 & (0.05) & 0.07 &   0.37 & (0.01) &  $-$1.56 & (0.13) &  $-$4.23 & (0.13) &   6.92 & (0.16) &   2.26 &  12.53 &     -- & (--) &   39 & (0) \\
          OC$-$0709(*) & 290.72 & (0.15) &  18.67 & (0.14) &  53.06 & (0.16) &   1.75 & (0.12) & 0.20 &   0.95 & (0.02) &  $-$1.01 & (0.27) &  $-$6.14 & (0.27) &   7.88 & (0.18) &   2.62 &   9.90 &  $-$10.80   & (15.39) &   47 & (5) \\
          OC$-$0710(*) & 292.64 & (0.04) &  20.26 & (0.05) &  55.32 & (0.06) &   0.92 & (0.03) & 0.06 &   0.28 & (0.01) &  $-$2.81 & (0.08) &  $-$6.44 & (0.11) &   8.32 & (0.19) &   3.66 &  12.20 &  24.56 & (22.24) &   32 & (4) \\
          OC$-$0711(*) & 295.73 & (0.18) &  20.03 & (0.18) &  56.54 & (0.18) &  $-$1.72 & (0.18) & 0.25 &   1.09 & (0.05) &  $-$0.50 & (0.40) &  $-$5.45 & (0.43) &   7.74 & (0.18) &   1.78 &   9.41 &  $-$18.07 & (34.56) &   55 & (11) \\
          OC$-$0712(*) & 304.72 & (0.06) &  37.75 & (0.06) &  75.72 & (0.07) &   0.98 & (0.04) & 0.08 &   0.30 & (0.01) &  $-$2.95 & (0.13) &  $-$4.67 & (0.09) &   8.00 & (0.18) &   3.76 &  12.65 & $-$35.07 & (--) &   34 & (1) \\
          OC$-$0713(*) & 351.27 & (0.33) &  61.32 & (0.10) & 112.76 & (0.17) &   0.16 & (0.08) & 0.19 &   0.30 & (0.01) &  $-$3.94 & (0.15) &  $-$2.17 & (0.11) &   7.00 & (0.16) &   2.68 &  12.60 & 63.16  & (8.10) &   26 & (2) \\
          OC$-$0714 &  26.04 & (0.13) &  61.88 & (0.05) & 129.10 & (0.06) &  $-$0.36 & (0.05) & 0.08 &   0.38 & (0.01) &  $-$1.12 & (0.12) &  $-$0.31 & (0.11) &   7.22 & (0.17) &   2.70 &  11.89 &     -- & (--) &   51 & (0) \\
          OC--0715(*) &  59.94 & (0.40) &  55.54 & (0.27) & 147.43 & (0.30) &   1.90 & (0.18) & 0.35 &   0.22 & (0.02) &  $-$0.01 & (0.11) &  $-$0.05 & (0.11) &   6.80 & (0.14) &  3.08 &  13.05 & $-$38.52 & (10.95) &  307 & (3) \\
          OC$-$0716(*) &  99.64 & (0.12) &   2.08 & (0.10) & 209.61 & (0.12) &  $-$1.91 & (0.11) & 0.16 &   0.51 & (0.01) &  $-$1.09 & (0.15) &  $-$2.02 & (0.13) &   9.32 & (0.21) &   1.24 &  11.26 &  35.10 & (6.27) &   63 & (4) \\
          OC--0717(*) & 115.29 & (0.24) & $-$26.03 & (0.28) & 241.55 & (0.29) &  $-$1.60 & (0.21) & 0.35 &   0.37 & (0.01) &  $-$2.15 & (0.25) &   2.25 & (0.43) &   8.28 & (0.17) &  1.14 &  11.95 &  55.82 & (31.47) &   71 & (4) \\
          OC--0718(*) & 110.65 & (0.21) & $-$29.86 & (0.14) & 242.96 & (0.13) &  $-$7.03 & (0.20) & 0.23 &   0.12 & (0.01) &  $-$0.73 & (0.11) &   1.87 & (0.21) &   8.65 & (0.17)  &  0.38 &  14.70 &  91.92 & (121.58) &   45 & (3) \\
          OC$-$0719(*) & 123.11 & (0.09) & $-$27.93 & (0.07) & 246.80 & (0.07) &   3.35 & (0.08) & 0.11 &   0.37 & (0.01) &  $-$1.60 & (0.06) &   2.90 & (0.09) &   9.20 & (0.21) &   0.54 &  11.94 &  27.71 & (8.29) &   71 & (8) \\
          OC--0720(*) & 122.19 & (0.11) & $-$36.16 & (0.17) & 253.27 & (0.18) &  $-$1.79 & (0.06) & 0.19 &   2.78 & (0.06) &  $-$7.51 & (0.20) &  11.41 & (0.51) &   6.32 & (0.13) &  2.40 &   7.58 & 30.73 & (31.17) &   15 & (5) \\
          OC--0721(*) & 127.19 & (0.10) & $-$41.19 & (0.14) & 259.62 & (0.13) &  $-$1.40 & (0.09) & 0.16 &   0.25 & (0.01) &  $-$2.92 & (0.12) &   3.49 & (0.28) &   7.58 & (0.15) &  3.16 &  12.82 &  73.17 & (--) &   22 & (1) \\
          OC$-$0722(*) & 130.66 & (0.06) & $-$47.20 & (0.05) & 265.93 & (0.04) &  $-$3.02 & (0.04) & 0.06 &   0.41 & (0.01) &  $-$2.72 & (0.12) &   5.22 & (0.13) &   8.46 & (0.19) &   1.52 &  11.96 &  19.07 & (--) &   41 & (1) \\
          OC--0723(*) & 132.34 & (0.11) & $-$46.56 & (0.11) & 266.15 & (0.11) &  $-$1.72 & (0.08) & 0.14 &   0.19 & (0.02) &  $-$3.25 & (0.15) &   3.61 & (0.17) &   7.46 & (0.15) & 2.32 &  13.40 & 96.90 & (4.86) &  116 & (2) \\
          OC$-$0724 & 137.95 & (0.10) & $-$52.23 & (0.04) & 272.82 & (0.05) &  $-$2.70 & (0.05) & 0.07 &   0.33 & (0.01) &  $-$4.06 & (0.06) &   3.95 & (0.05) &   8.38 & (0.19) &   1.48 &  12.20 &     -- & (--) &   27 & (0) \\
          OC$-$0725(*) & 139.16 & (0.09) & $-$54.16 & (0.04) & 274.71 & (0.05) &  $-$3.50 & (0.05) & 0.07 &   0.20 & (0.01) &  $-$3.58 & (0.05) &   2.72 & (0.09) &   8.86 & (0.20) &   1.22 &  13.30 &   17.18  & (9.11) &   37 & (10) \\
          OC$-$0726(*) & 149.41 & (0.11) & $-$54.60 & (0.06) & 279.46 & (0.07) &   0.15 & (0.06) & 0.09 &   0.23 & (0.01) &  $-$5.01 & (0.07) &   4.48 & (0.07) &   8.98 & (0.21) &   1.46 &  12.97 &  29.67 & (6.21) &   47 & (7) \\
          OC$-$0727 & 159.27 & (0.13) & $-$59.92 & (0.05) & 286.83 & (0.06) &  $-$1.32 & (0.06) & 0.08 &   0.26 & (0.01) &  $-$5.70 & (0.11) &   2.48 & (0.11) &   7.00 & (0.16) &   2.08 &  12.94 &     -- & (--) &   33 & (0) \\
          OC$-$0728(*) & 161.18 & (0.05) & $-$59.35 & (0.03) & 287.40 & (0.03) &  $-$0.35 & (0.03) & 0.04 &   0.38 & (0.01) &  $-$6.13 & (0.14) &   2.10 & (0.12) &   6.74 & (0.16) &   1.74 &  12.12 &   $-$26.28 & (--) &   33 & (1) \\
          OC$-$0729(*) & 160.61 & (0.15) & $-$60.43 & (0.09) & 287.66 & (0.08) &  $-$1.44 & (0.08) & 0.11 &   0.24 & (0.01) &  $-$5.58 & (0.11) &   2.61 & (0.08) &   8.76 & (0.20) &   1.20 &  13.48 &   22.35 & (--) &   32 & (1) \\
          OC$-$0730(*) & 168.63 & (0.10) & $-$57.56 & (0.04) & 290.20 & (0.06) &   2.90 & (0.04) & 0.07 &   0.46 & (0.01) &  $-$1.65 & (0.09) &  $-$1.22 & (0.09) &   9.44 & (0.22) &   0.62 &  11.90 &   $-$24.12  & (13.14) &   30 & (2) \\
          OC$-$0731(*) & 172.62 & (0.07) & $-$60.76 & (0.03) & 293.20 & (0.03) &   0.58 & (0.03) & 0.04 &   0.22 & (0.01) &  $-$6.86 & (0.08) &   1.37 & (0.08) &   8.96 & (0.21) &   1.52 &  12.71 &  3.17 & (3.27) &   31 & (6) \\
          OC$-$0732(*) & 181.53 & (0.09) & $-$62.62 & (0.03) & 297.75 & (0.04) &  $-$0.21 & (0.03) & 0.05 &   0.28 & (0.01) &  $-$8.28 & (0.10) &   0.95 & (0.06) &   8.66 & (0.20) &   1.56 &  12.53 &  $-$2.87  & (53.36) &   28 & (2) \\
          OC$-$0733 & 193.44 & (0.14) & $-$60.36 & (0.07) & 303.22 & (0.07) &   2.51 & (0.07) & 0.10 &   0.54 & (0.01) &  $-$4.65 & (0.20) &  $-$1.07 & (0.16) &   7.42 & (0.17) &   1.08 &  11.15 &     -- & (--) &   37 & (0) \\
          OC$-$0734 & 194.51 & (0.08) & $-$64.96 & (0.03) & 303.63 & (0.04) &  $-$2.10 & (0.03) & 0.05 &   0.24 & (0.01) &  $-$5.84 & (0.11) &  $-$0.94 & (0.09) &   8.48 & (0.20) &   2.52 &  12.48 & -- & (--) &   26 & (0) \\
          OC$-$0735(*) & 223.62 & (0.20) & $-$57.63 & (0.09) & 318.83 & (0.11) &   1.40 & (0.08) & 0.14 &   0.89 & (0.02) &  $-$3.34 & (0.14) &  $-$2.62 & (0.15) &   7.86 & (0.18) &   2.08 &  10.46 &  $-$31.08 & (5.21) &   26 & (3) \\
          OC$-$0736(*) & 230.27 & (0.11) & $-$59.60 & (0.06) & 320.85 & (0.07) &  $-$2.04 & (0.05) & 0.08 &   0.30 & (0.01) &  $-$3.94 & (0.13) &  $-$3.55 & (0.12) &   6.70 & (0.15) &   3.94 &  13.02 &   $-$8.85 & (137.18) &   30 & (2) \\
          OC$-$0737(*) & 240.21 & (0.09) & $-$53.54 & (0.06) & 328.85 & (0.06) &  $-$0.48 & (0.06) & 0.08 &   0.41 & (0.01) &  $-$4.24 & (0.14) &  $-$4.78 & (0.13) &   8.82 & (0.20) &   2.40 &  11.73 &  $-$30.69 & (35.72) &   92 & (3) \\
          OC$-$0738(*) & 242.58 & (0.09) & $-$52.33 & (0.06) & 330.72 & (0.06) &  $-$0.53 & (0.05) & 0.08 &   0.35 & (0.01) &  $-$2.93 & (0.11) &  $-$4.27 & (0.10) &   8.42 & (0.19) &   3.42 &  11.66 & $-$18.24 & (7.77) &   38 & (6) \\
          OC$-$0739(*) & 252.33 & (0.07) & $-$44.73 & (0.06) & 340.66 & (0.06) &   0.02 & (0.05) & 0.08 &   0.41 & (0.01) &  $-$1.32 & (0.12) &  $-$2.54 & (0.11) &   7.72 & (0.18) &   2.70 &  11.33 &  $-$38.73 & (--) &   67 & (1) \\
          OC$-$0740(*) & 252.32 & (0.07) & $-$44.71 & (0.05) & 340.68 & (0.05) &   0.04 & (0.06) & 0.07 &   0.39 & (0.01) &  $-$1.33 & (0.09) &  $-$2.54 & (0.11) &   8.00 & (0.18) &   2.64 &  11.47 &  $-$72.58  & (48.38) &   32 & (2) \\
          OC$-$0741(*) & 264.91 & (0.07) & $-$33.21 & (0.06) & 355.70 & (0.06) &  $-$1.16 & (0.06) & 0.08 &   0.38 & (0.01) &   0.12 & (0.15) &  $-$0.76 & (0.12) &   6.98 & (0.16) &   4.04 &  11.49 & 111.11 & (160.56) &   61 & (6) \\
          OC$-$0742 & 262.93 & (0.09) & $-$31.97 & (0.09) & 355.84 & (0.08) &   0.91 & (0.09) & 0.12 &   0.35 & (0.01) &  $-$0.82 & (0.23) &  $-$1.81 & (0.22) &   8.00 & (0.18) &   4.04 &  12.26 &     -- & (--) &   27 & (0) \\
 \\ \hline
\end{tabular}
 \tablefoot{The listed parameters of each OC are mean astrometric parameters, corresponding standard 
deviations, and the apparent angular radius ($\theta$), all of which are from the member stars ($N$) and stars with RV measurements ($N_{r}$). Ages ($\log$(age)), line-of-sight extinction ($A$$_{\rm G}$), and DM are also provided.
OC names marked with an asterisk have a \emph{Gaia} RV measurement.
The table is available at the CDS.}
 \label{table:table2}
\end{table*}
%%%%%%%%%%%%%%%%%%%%%%%%%%%%%%%%%%%%%%%%%%%%

%\section{Additional figures}

%%%%%%%%%%%%%%%%%%%%%%%%%%%%%%%%%%%%%%%%%%%%%%%%%% Fig. 1
\begin{figure*}
\centering
\includegraphics[width=0.327\linewidth]{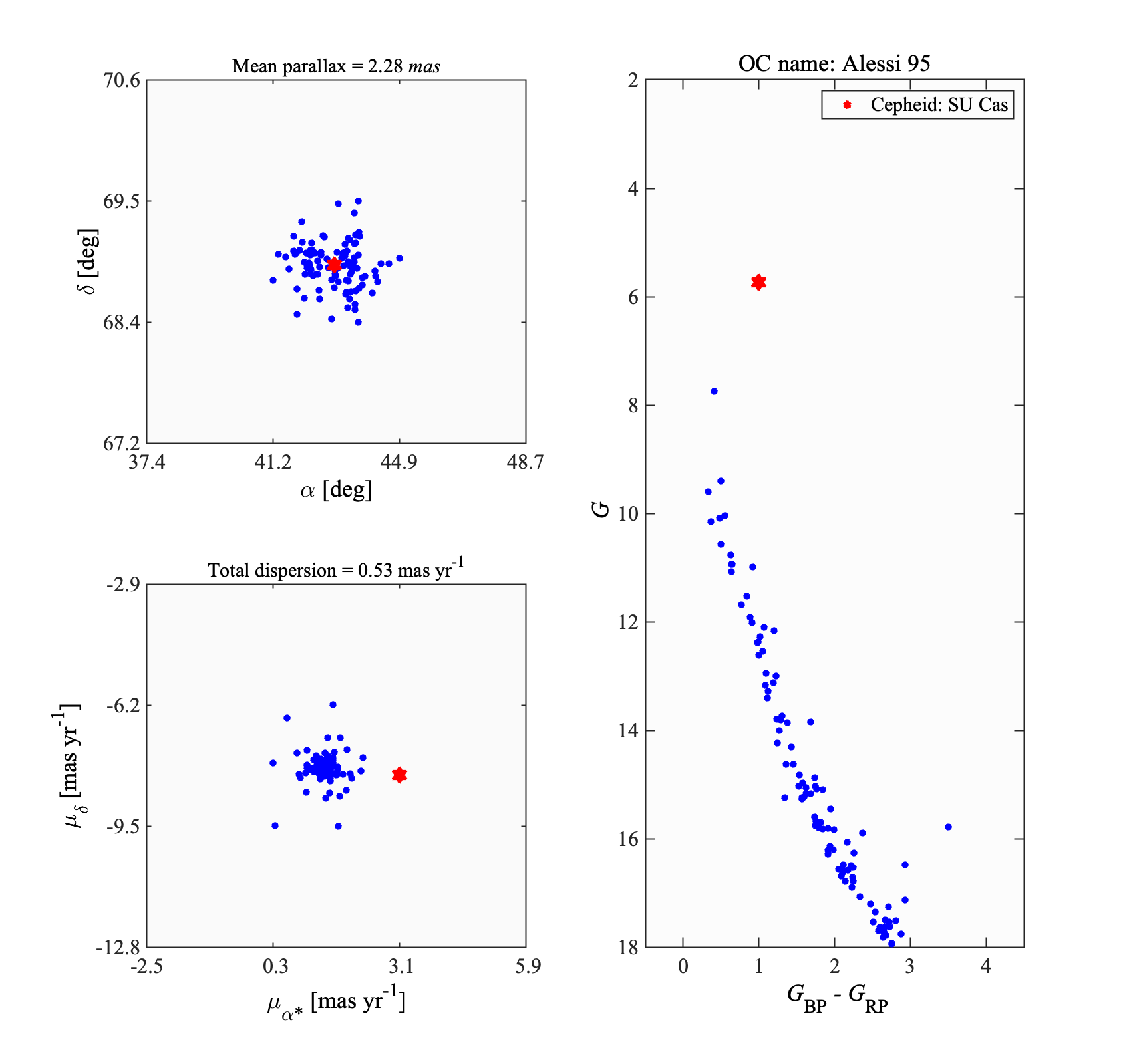} \hspace{0.0cm}
\includegraphics[width=0.327\linewidth]{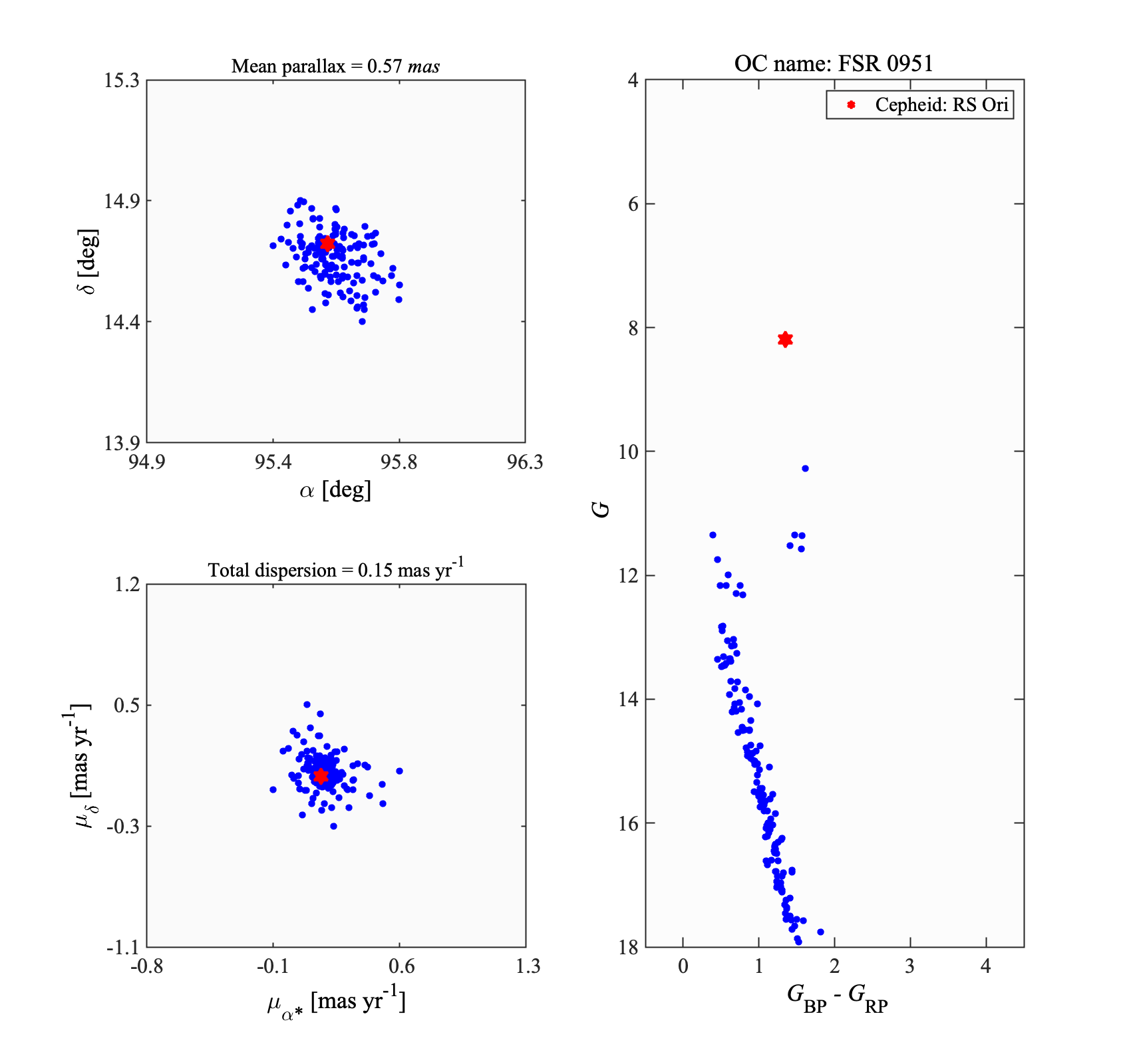}  \hspace{0.0cm}
\includegraphics[width=0.327\linewidth]{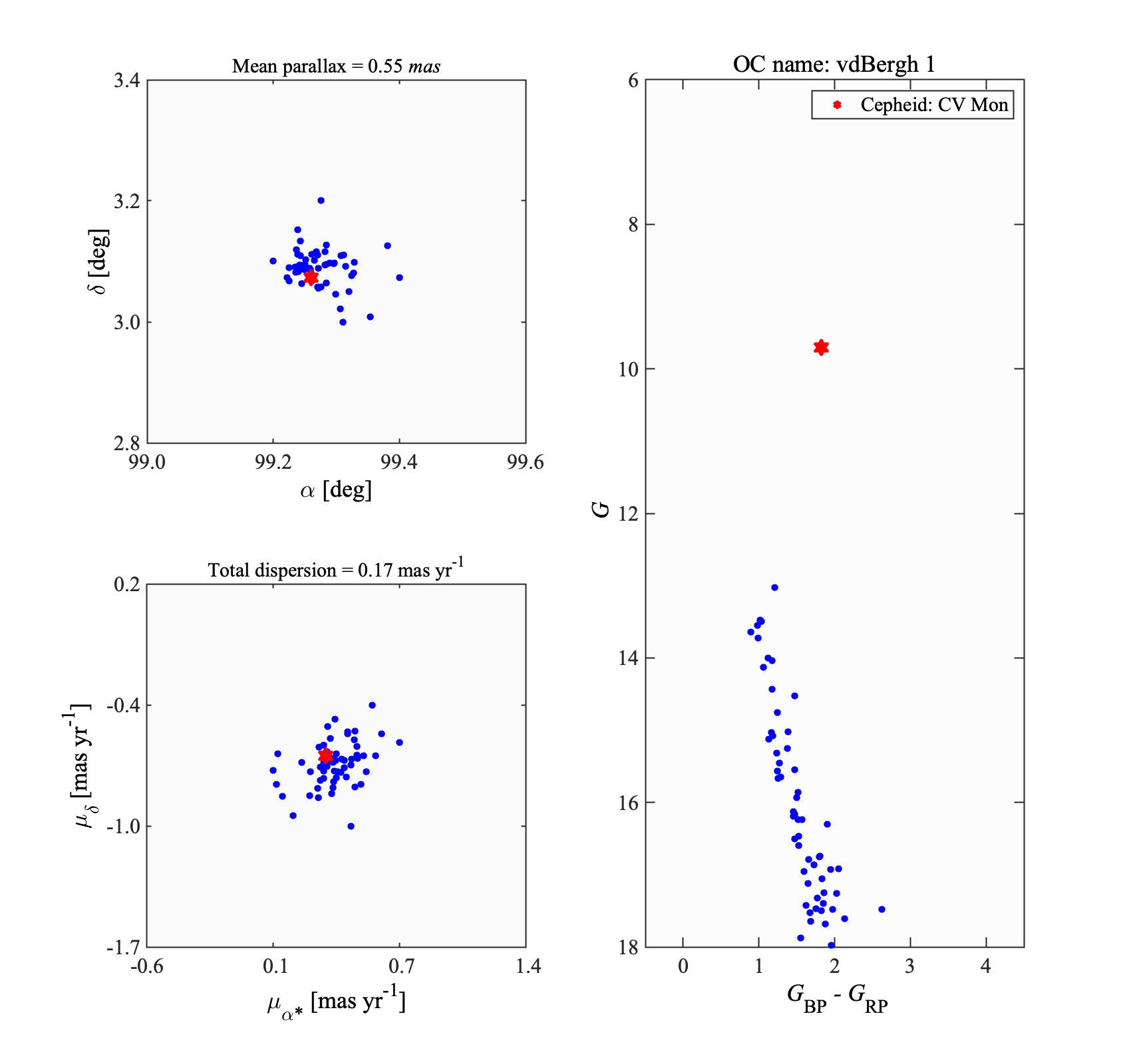} \hspace{0.0cm}
\includegraphics[width=0.327\linewidth]{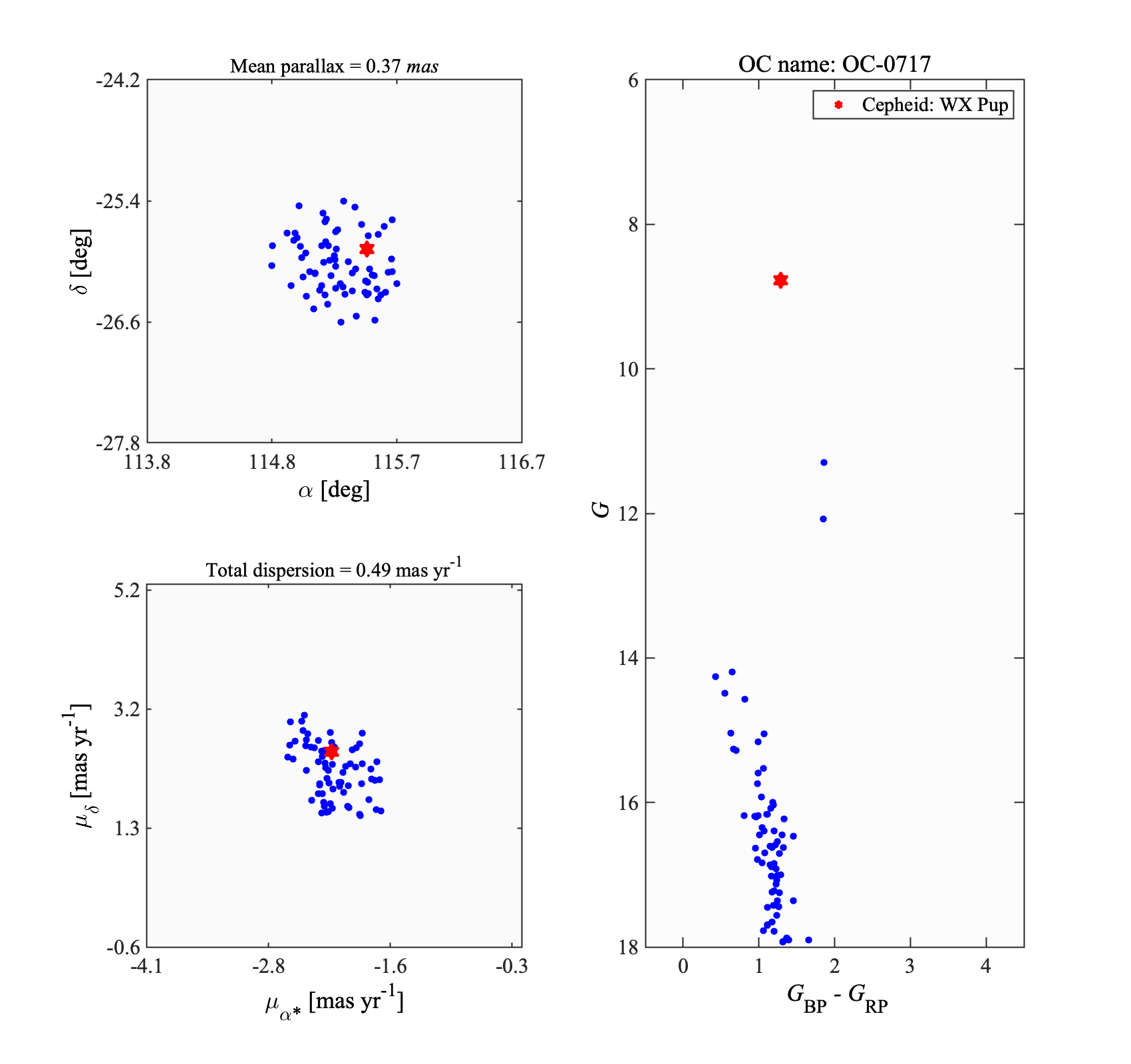}  \hspace{0.0cm}
\includegraphics[width=0.327\linewidth]{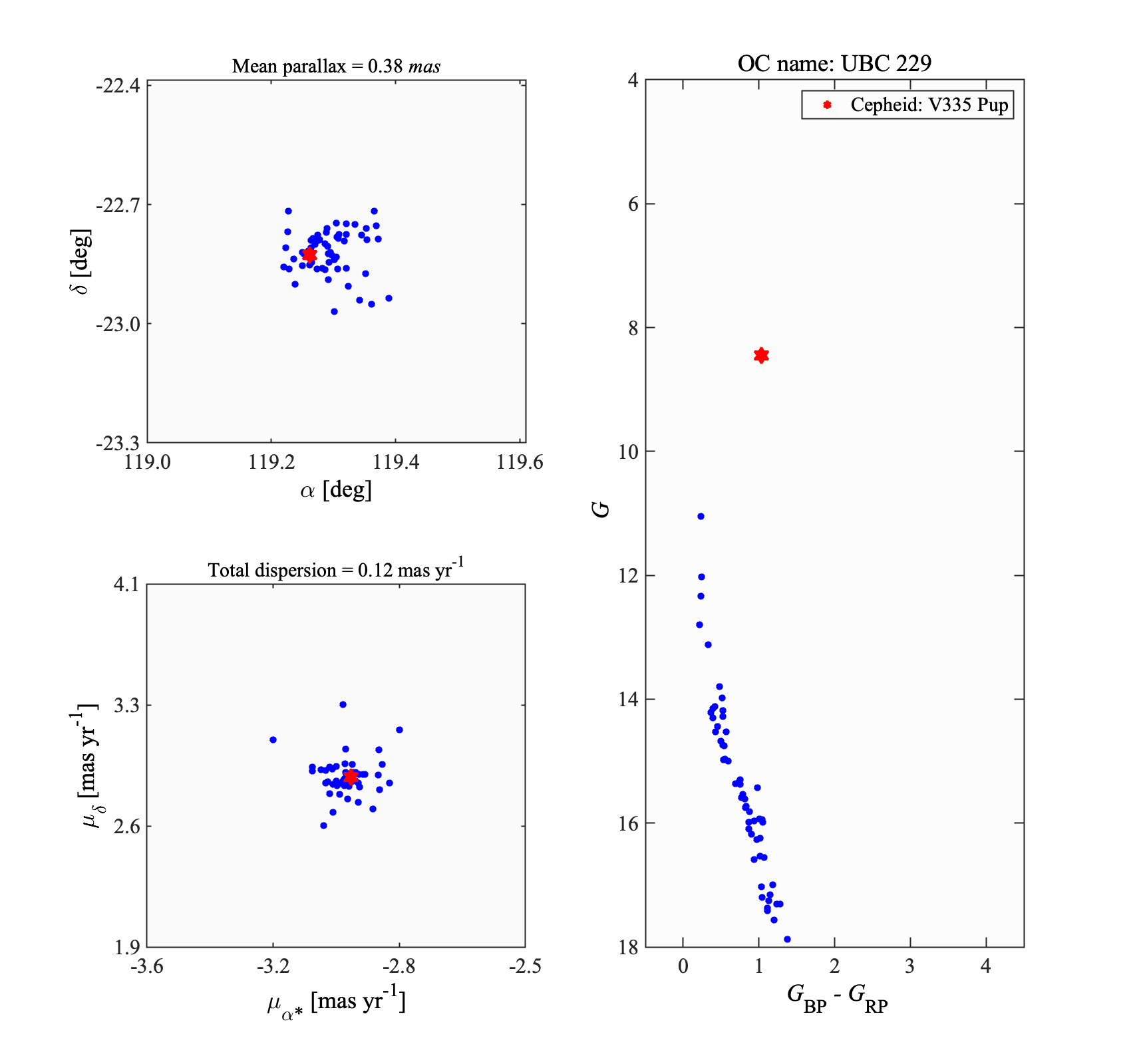} \hspace{0.0cm}
\includegraphics[width=0.327\linewidth]{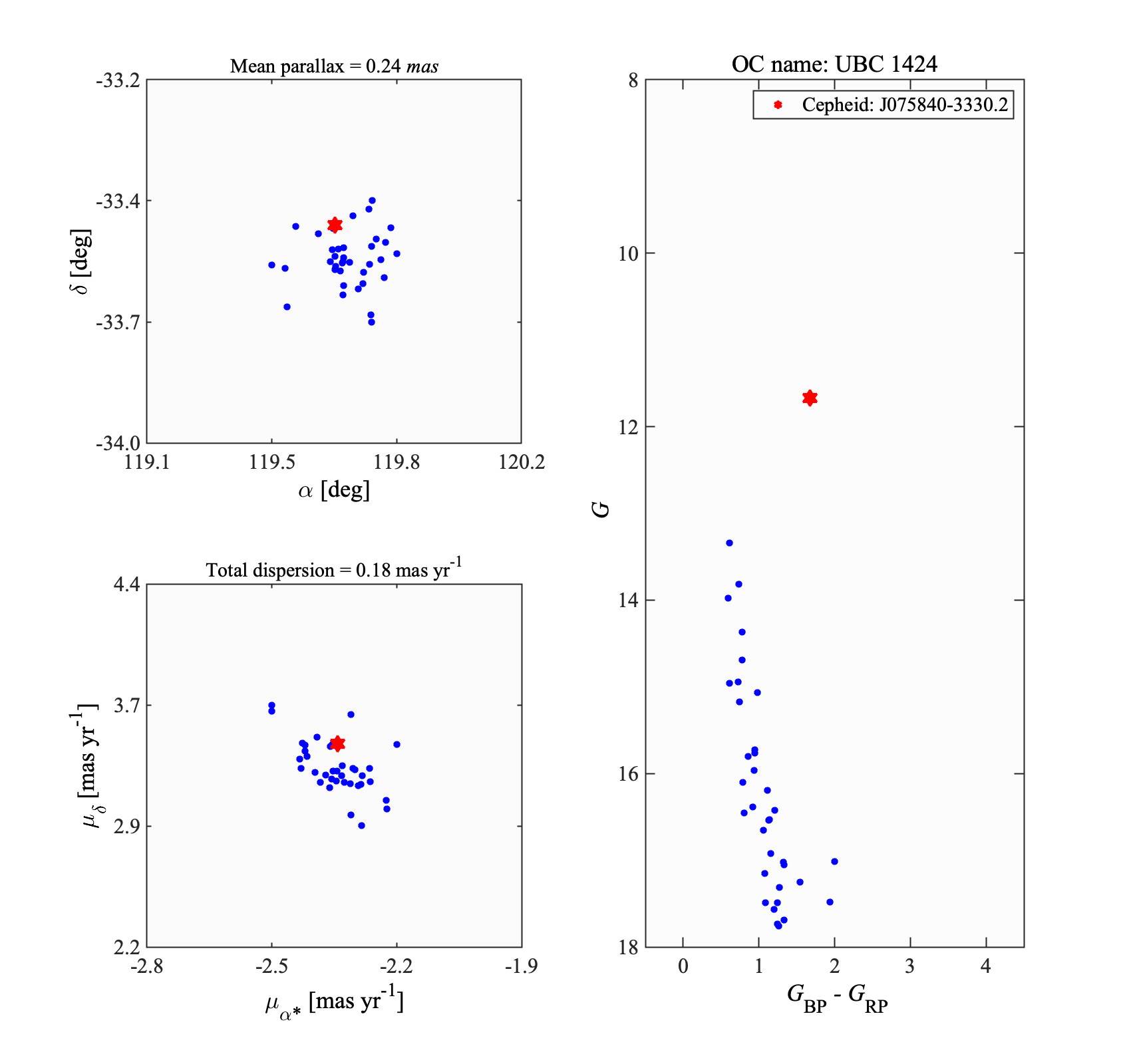} \hspace{0.0cm}
\includegraphics[width=0.327\linewidth]{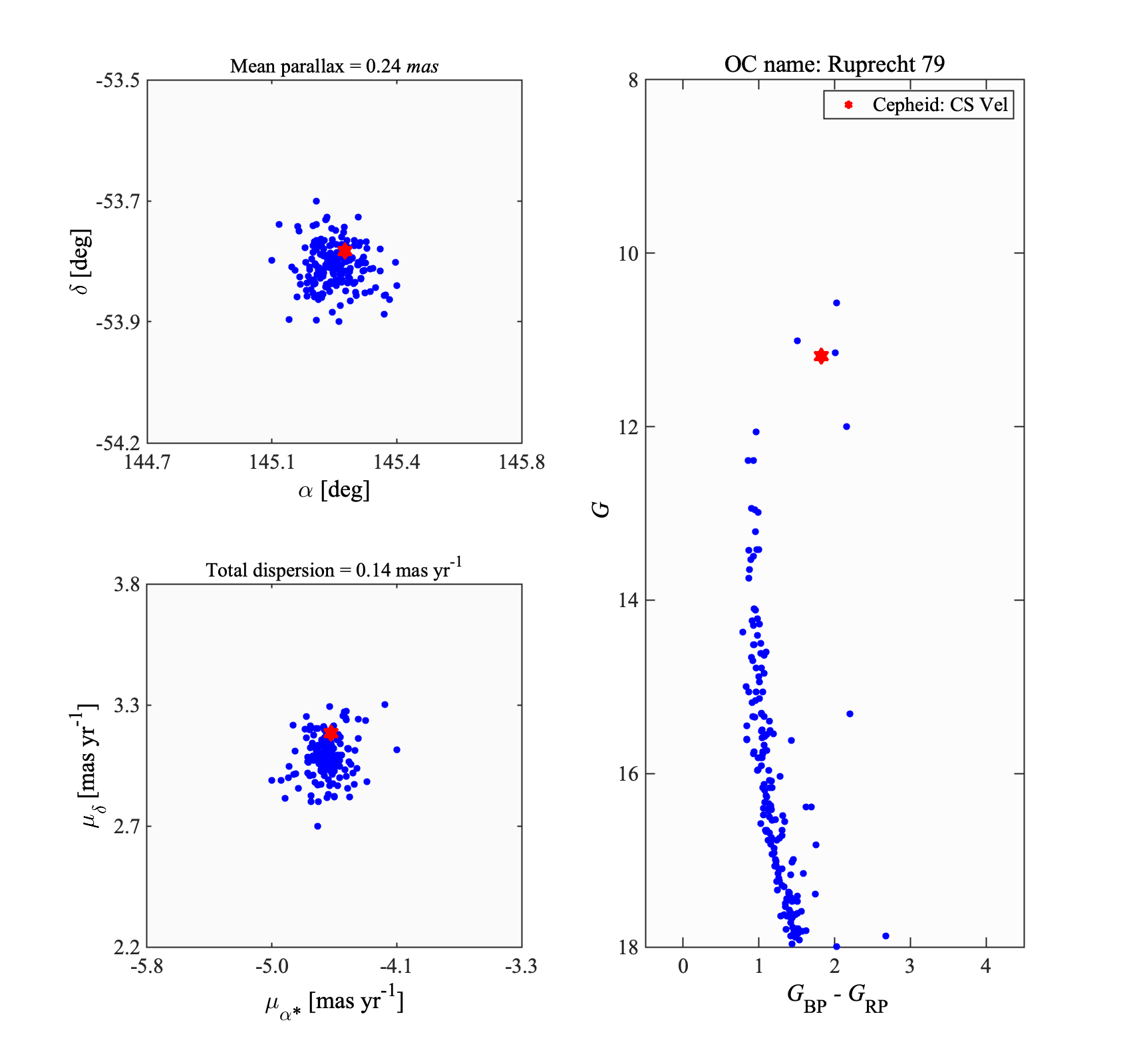} \hspace{0.0cm}
\includegraphics[width=0.327\linewidth]{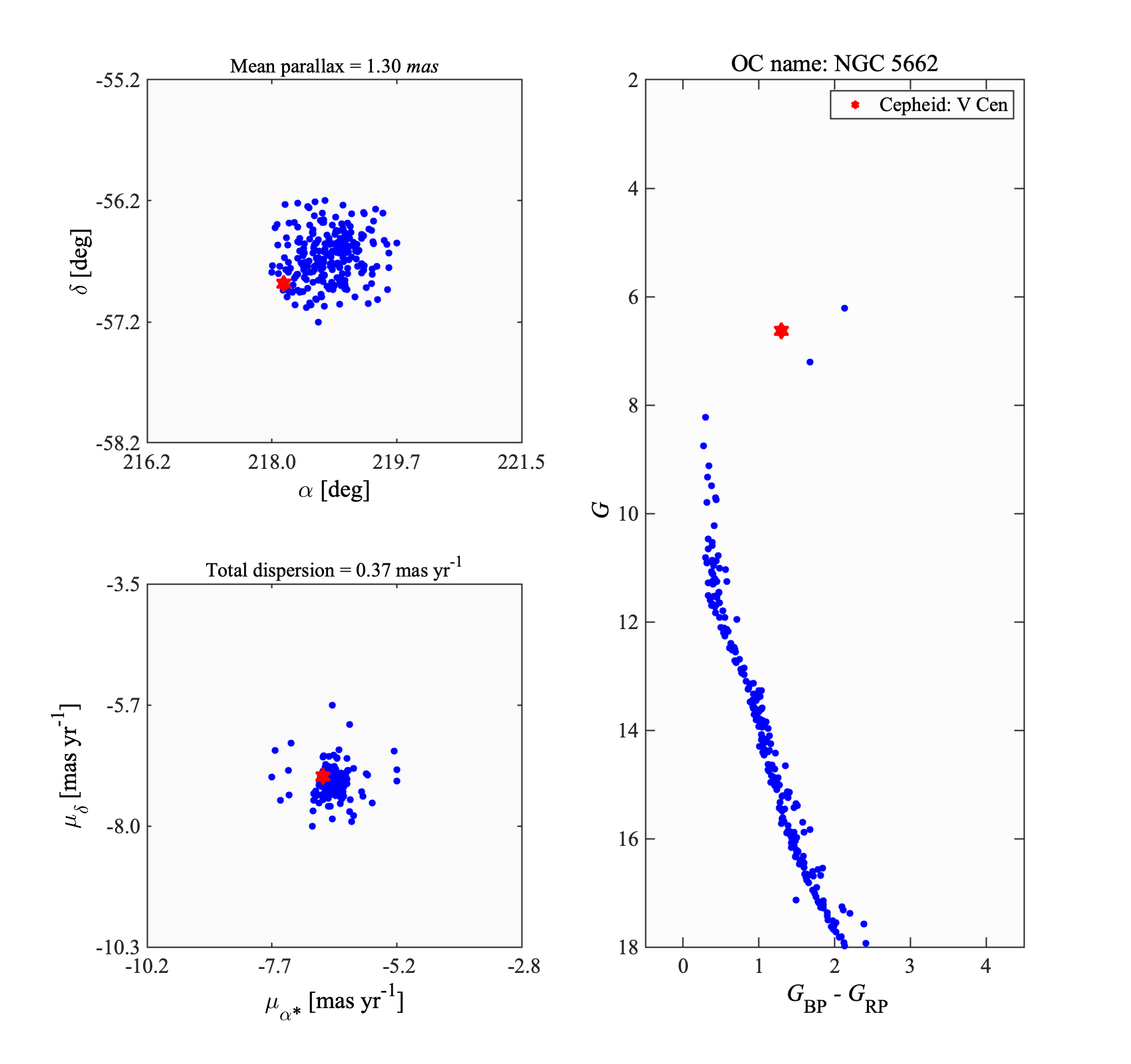}  \hspace{0.0cm}
\includegraphics[width=0.327\linewidth]{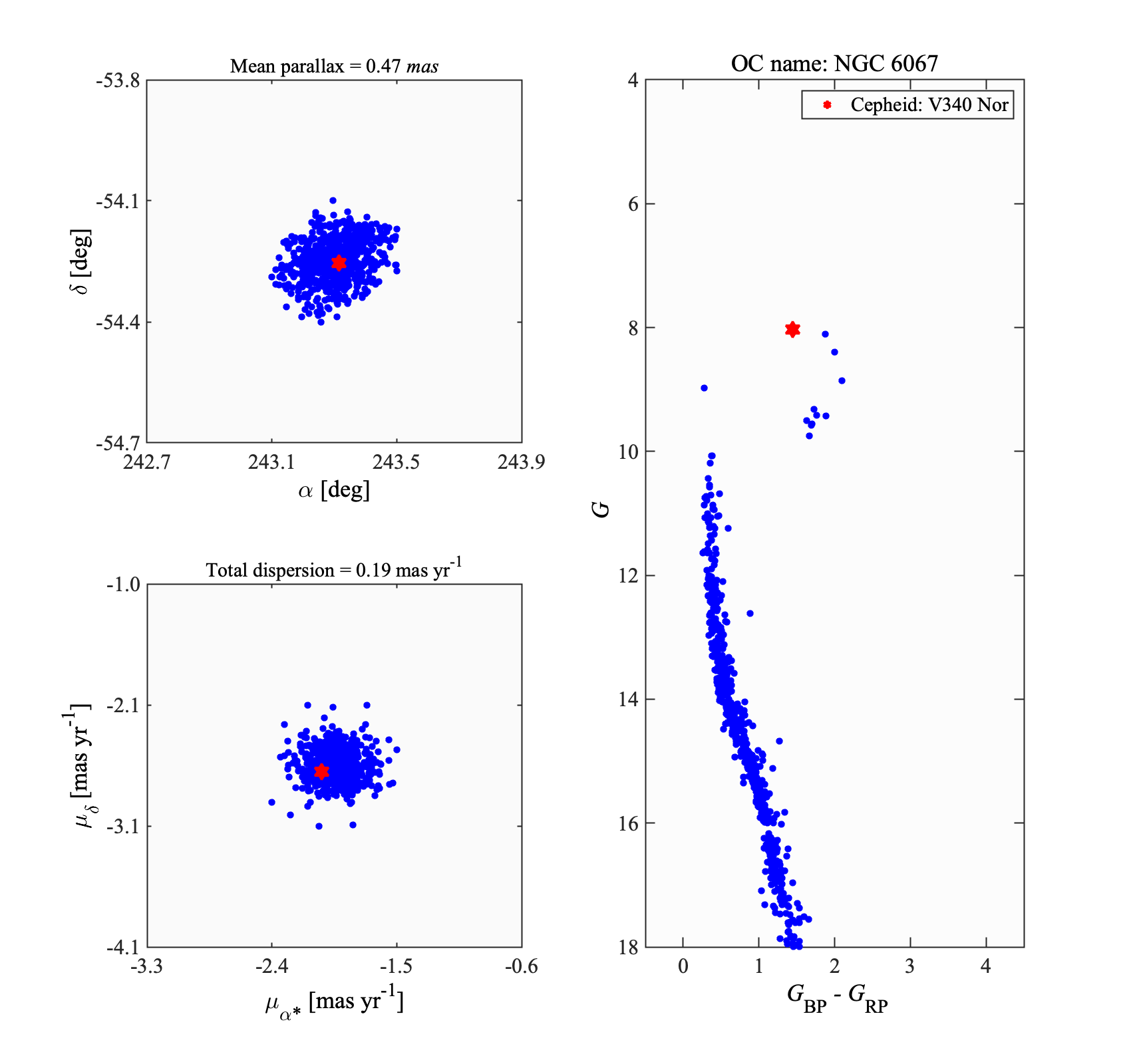} \hspace{0.0cm}
\includegraphics[width=0.327\linewidth]{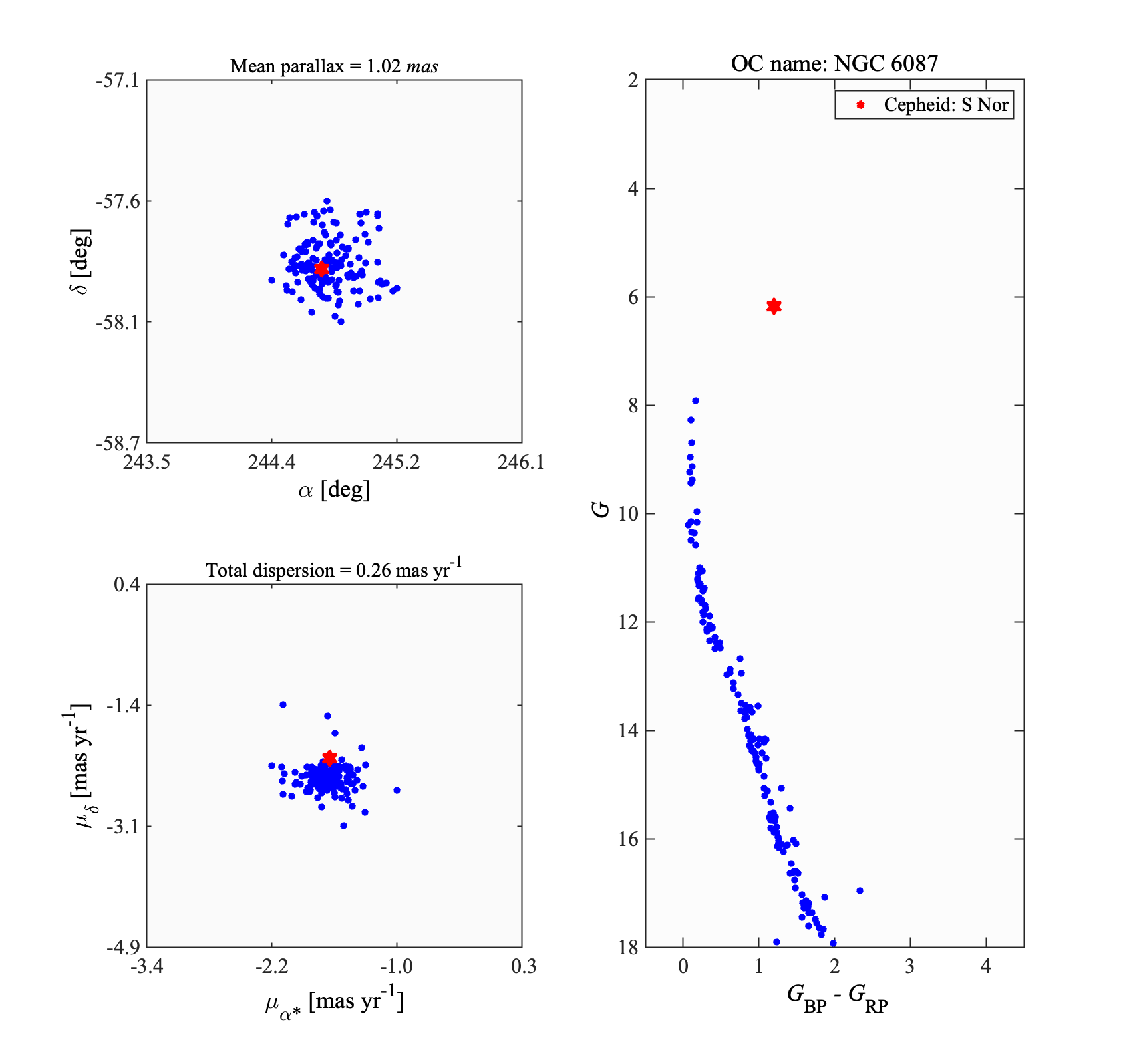}  \hspace{0.0cm}
\includegraphics[width=0.327\linewidth]{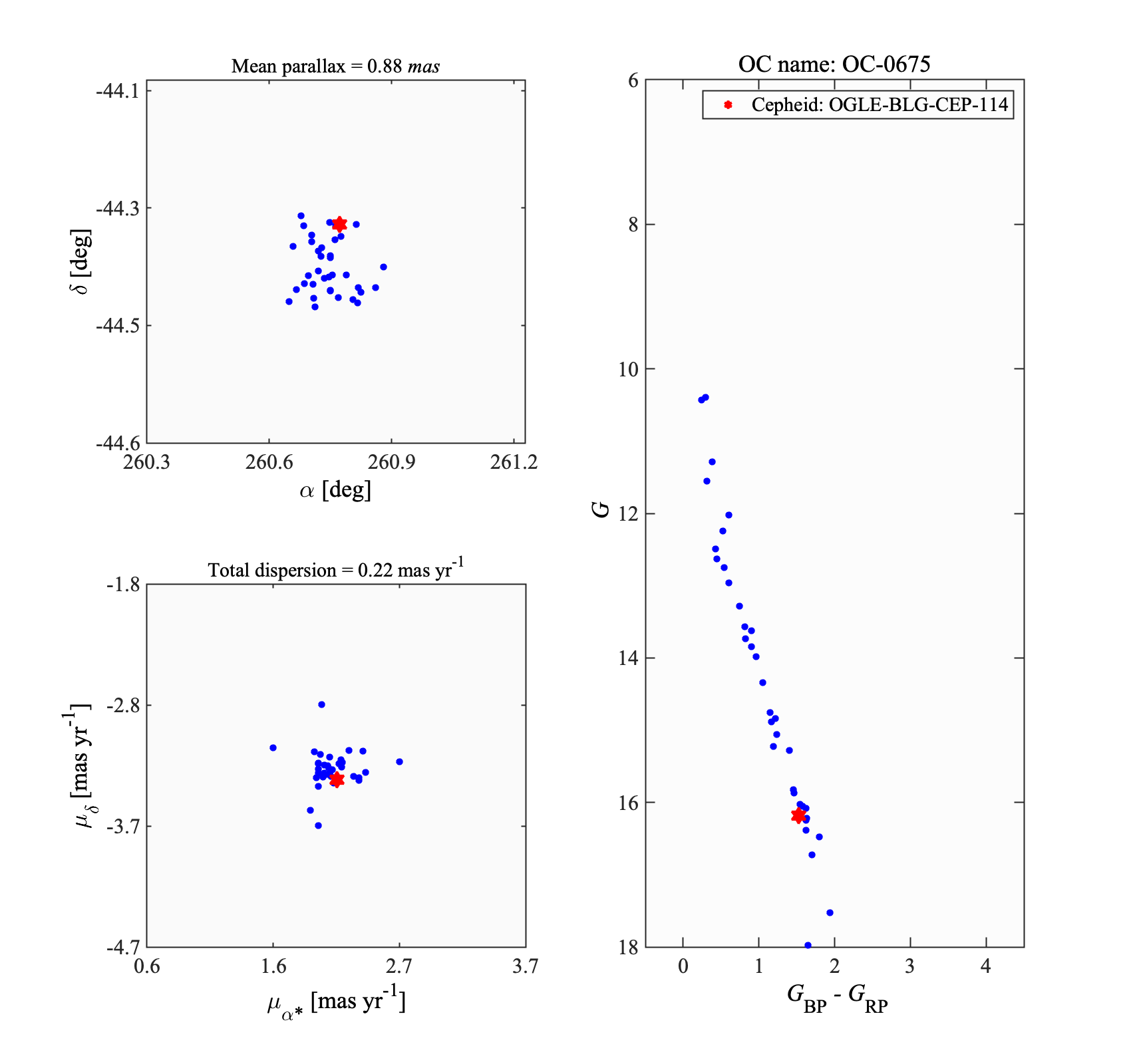} \hspace{0.0cm}
\includegraphics[width=0.327\linewidth]{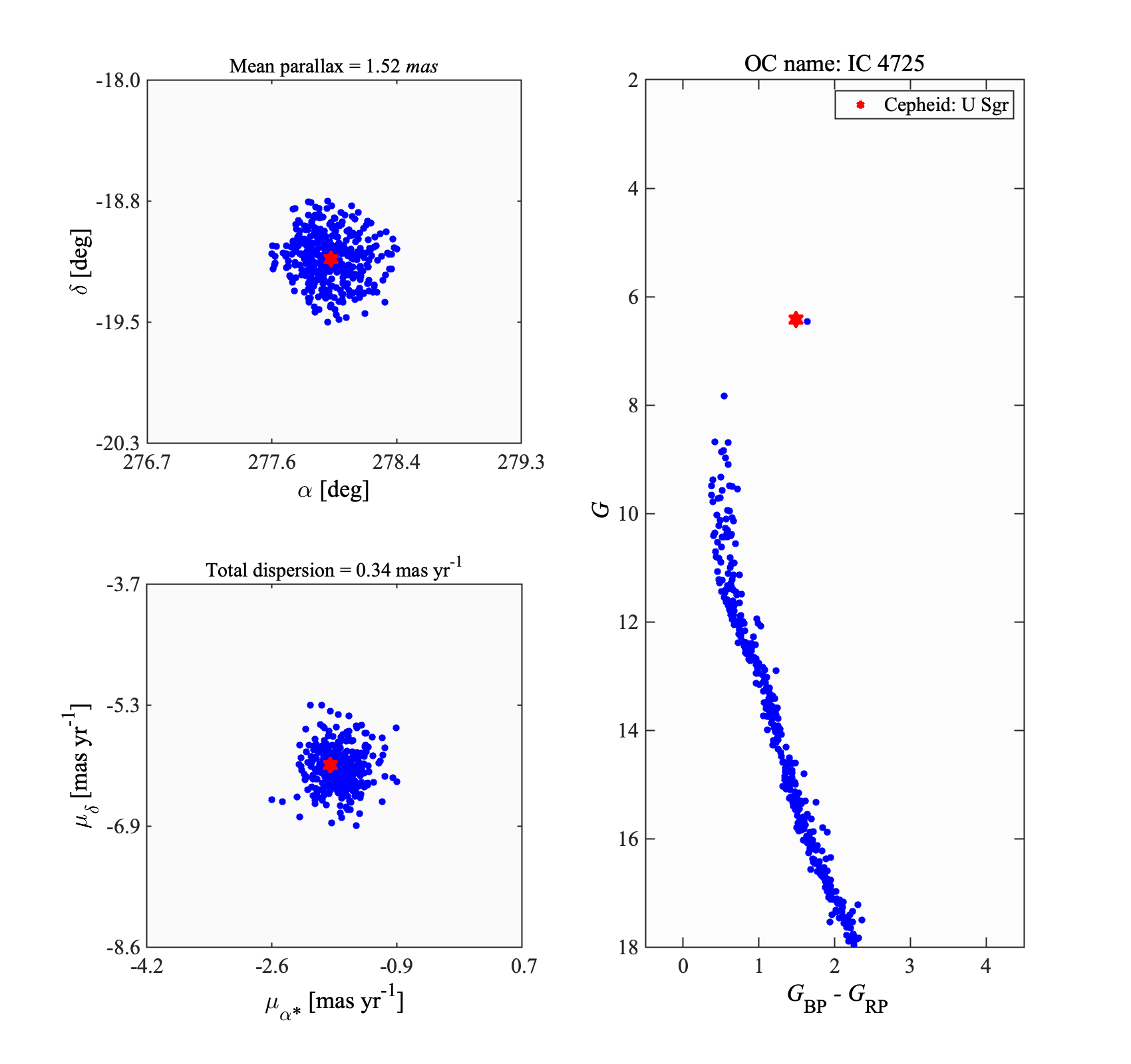}  \hspace{0.0cm}
\caption{Open clusters (blue dots) harboring classical Cepheids (red hexagram). The columns of each panel represent the distributions of the member stars of OCs and classical Cepheids for position in $RA$ ($\alpha$) and $Dec$ ($\delta$), proper motions in $\mu_{\alpha^{*}}$ and $\mu_{\delta}$, and the CMD in $G$ vs. $G_{\rm BP}$ $-$ $G_{\rm RP}$, as well as the mean parallax and total proper-motion dispersion of OCs. Here, the listed OCs are Alessi 95, FSR 0951, vdBergh 1, OC-0717, UBC 229, UBC 1424, Ruprecht 79, 
NGC 5662, NGC 6067, NGC 6087, OC-0675, and IC 4725.}
\label{fig:OCs_Ceps_s1}
\end{figure*}

%%%%%%%%%%%%%%%%%%%%%%%%%%%%%%%%%%%%%%%%%%%%%%% Fig. 1
\addtocounter{figure}{-1}
 \begin{figure*}
 \centering
\includegraphics[width=0.327\linewidth]{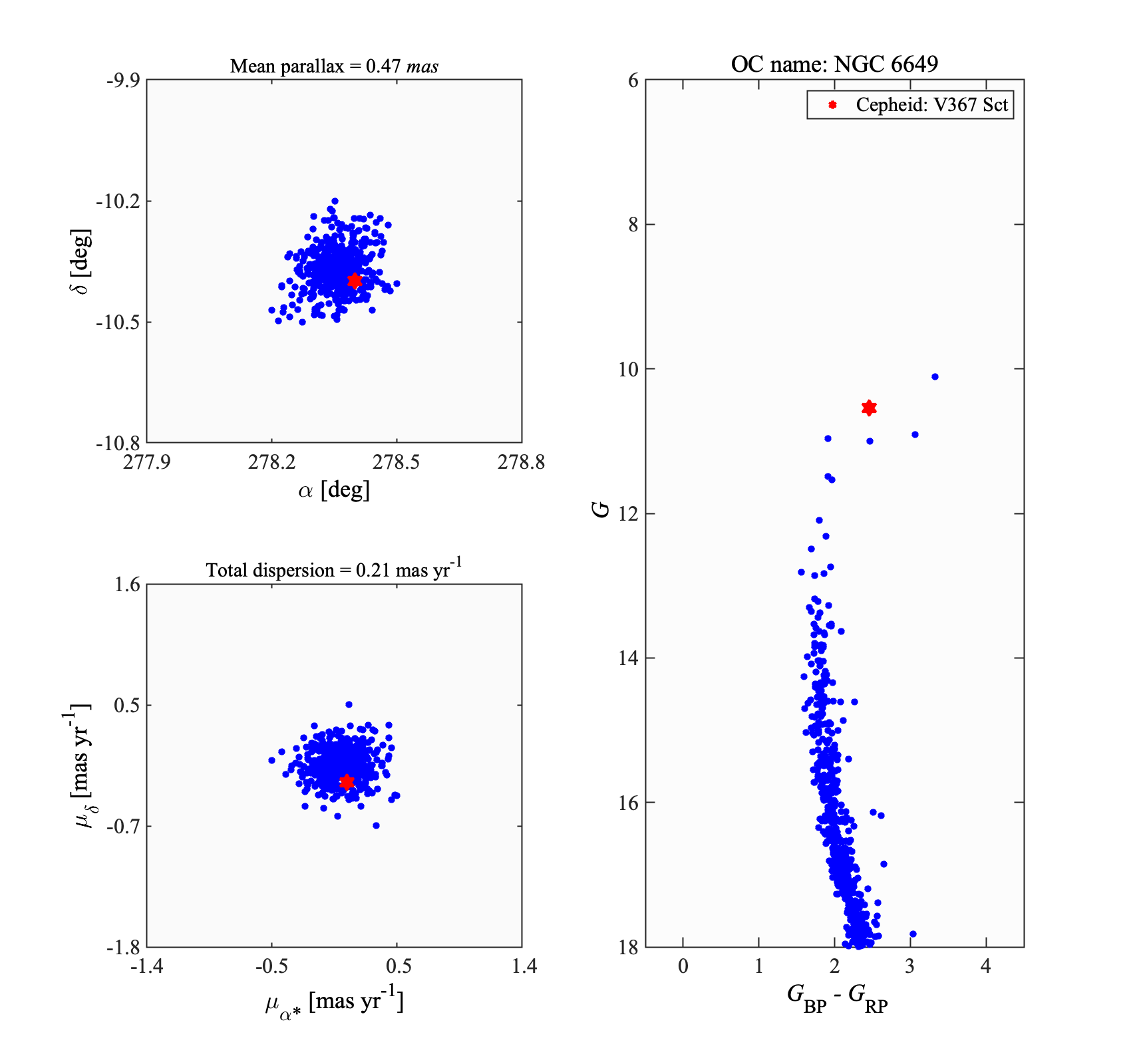} \hspace{0.0cm}
\includegraphics[width=0.327\linewidth]{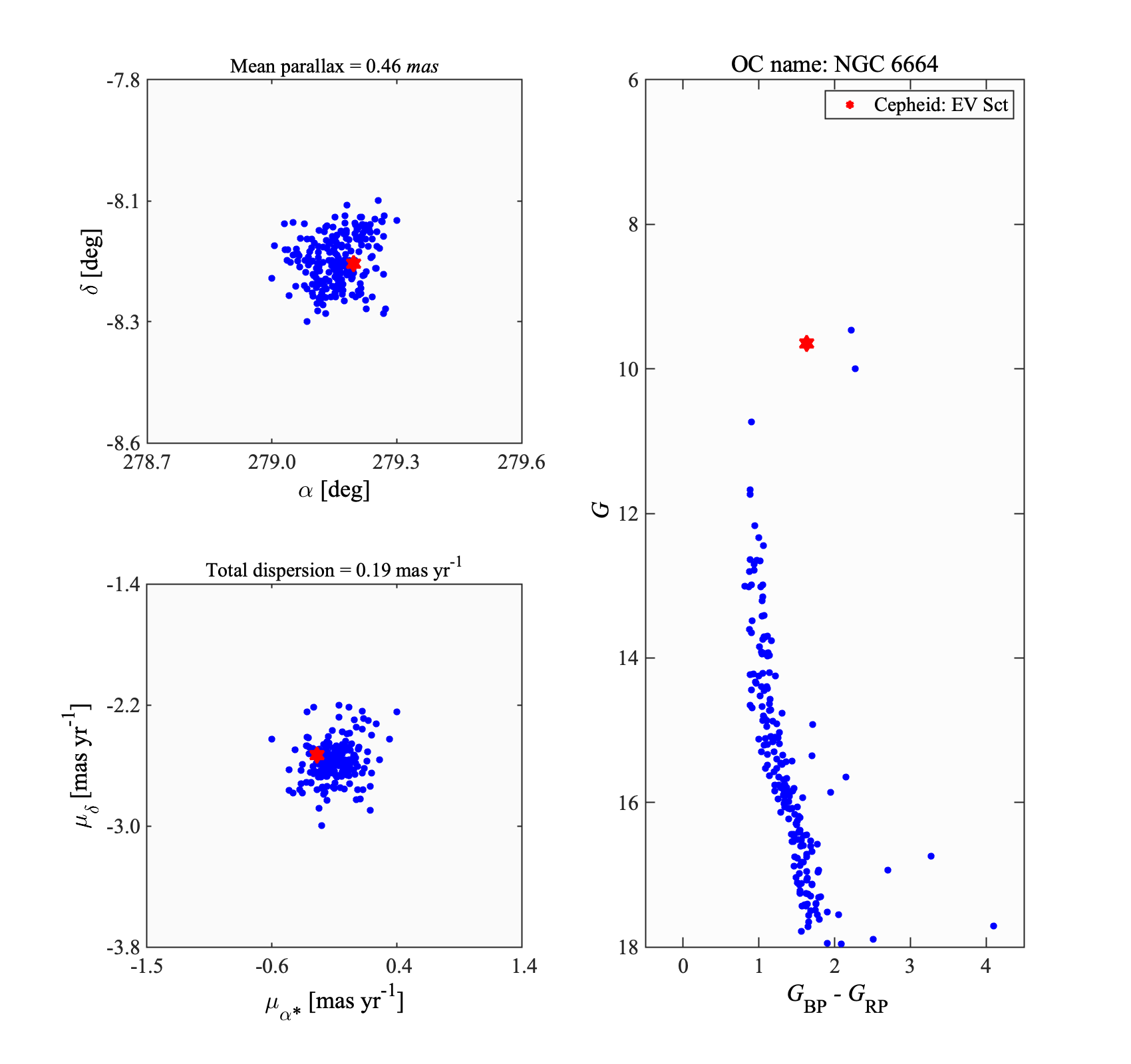} \hspace{0.0cm}
\includegraphics[width=0.327\linewidth]{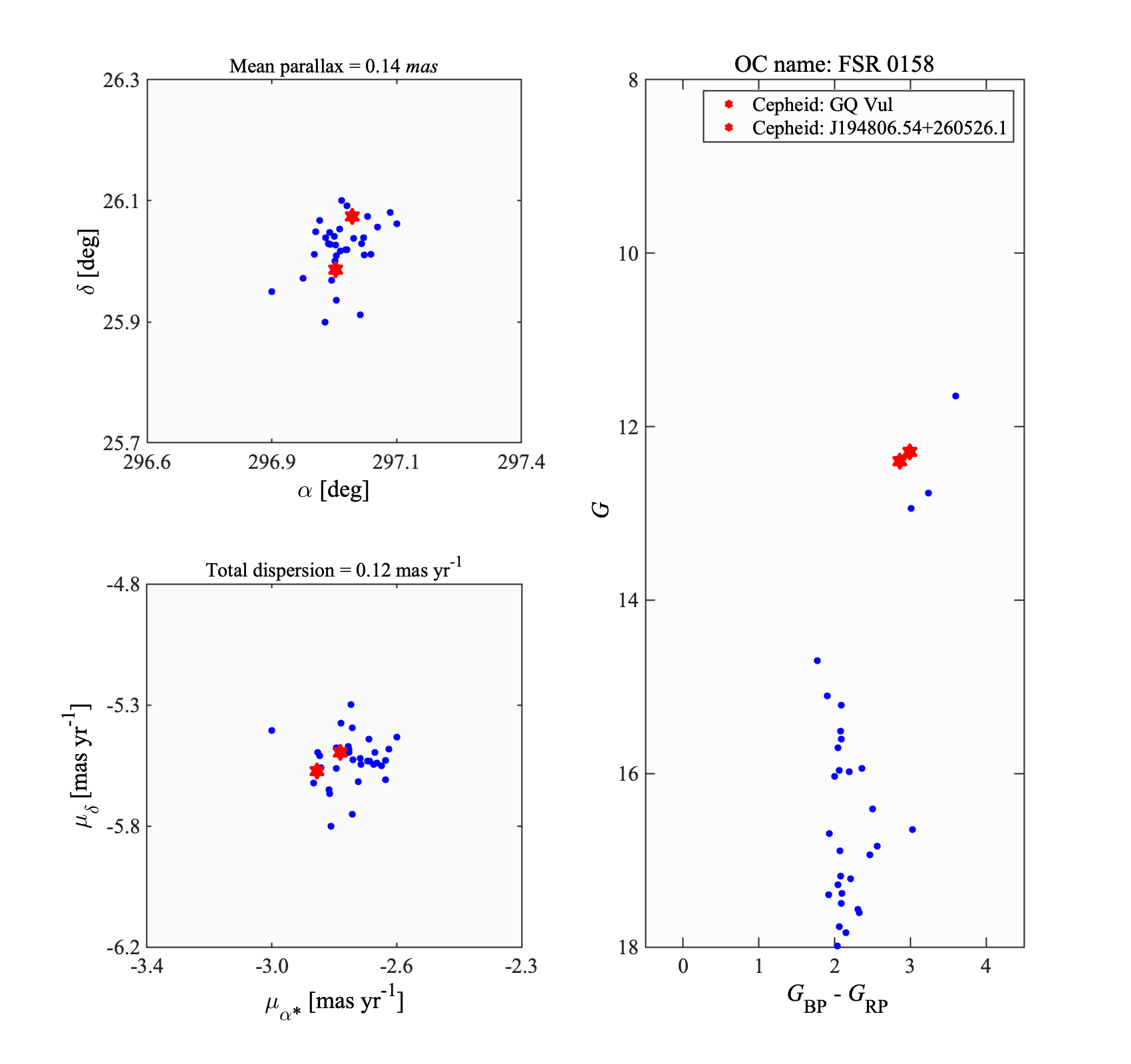} \hspace{0.0cm}
\includegraphics[width=0.327\linewidth]{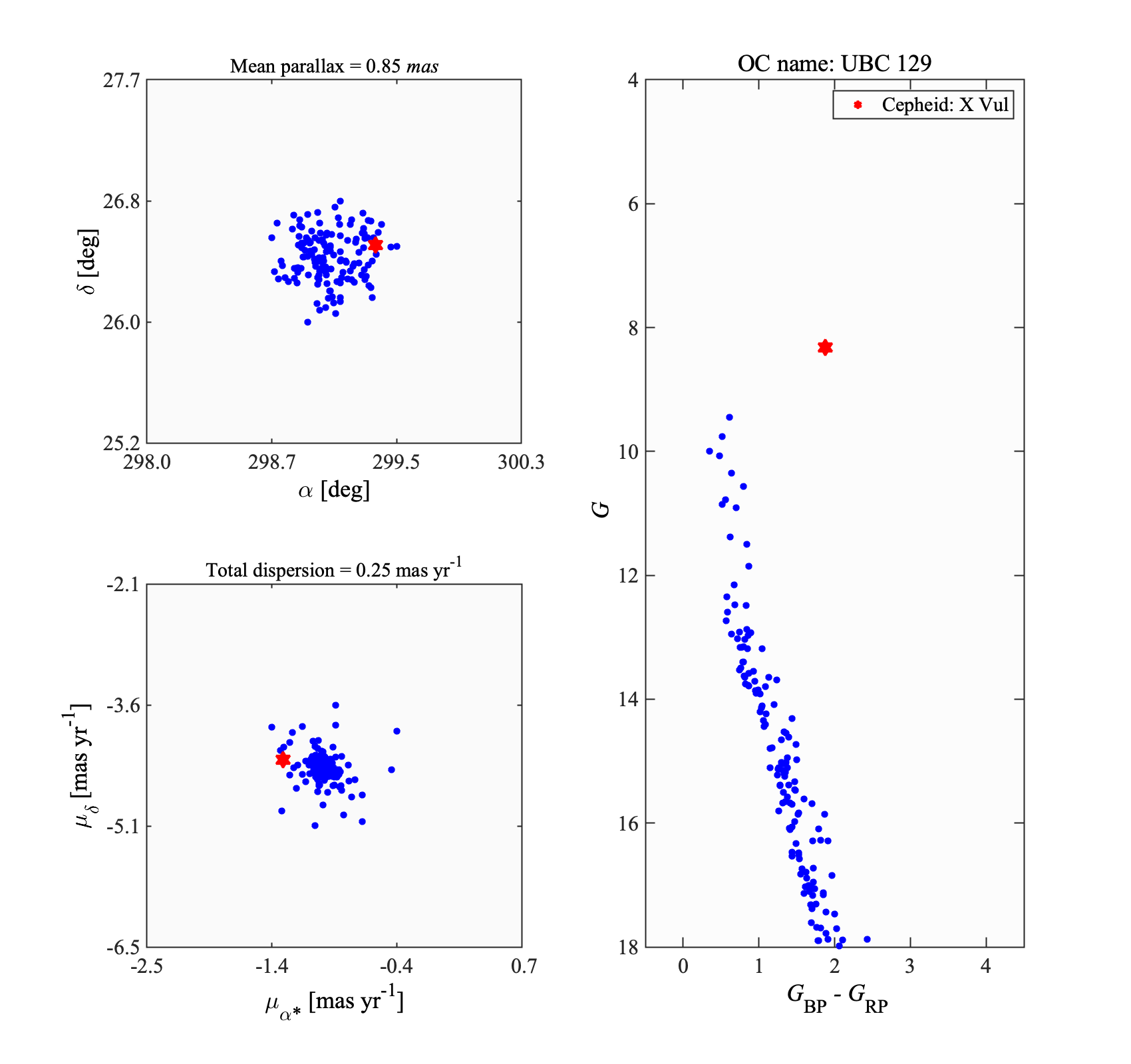} \hspace{0.0cm}
\includegraphics[width=0.327\linewidth]{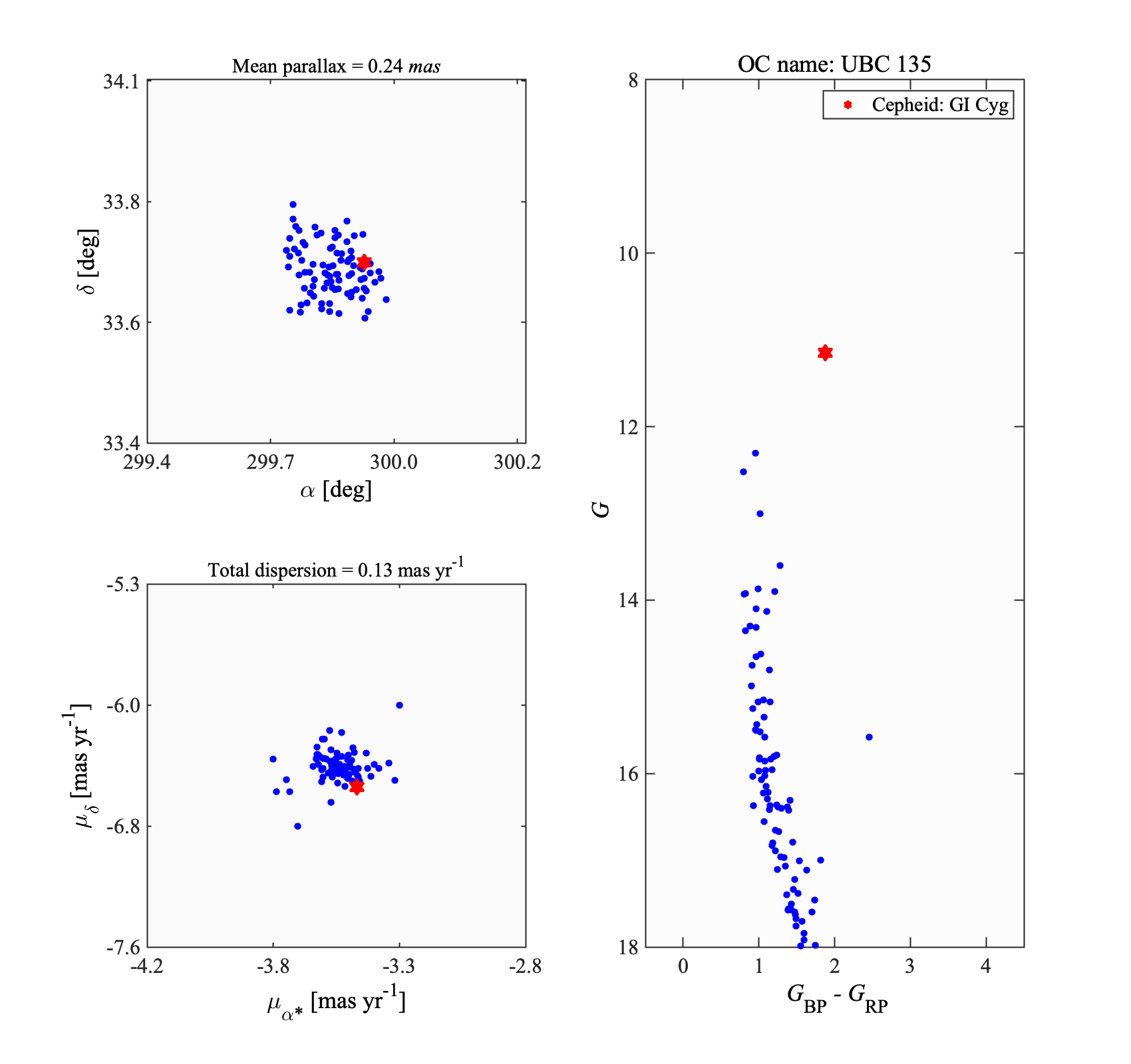} \hspace{0.0cm}
\includegraphics[width=0.327\linewidth]{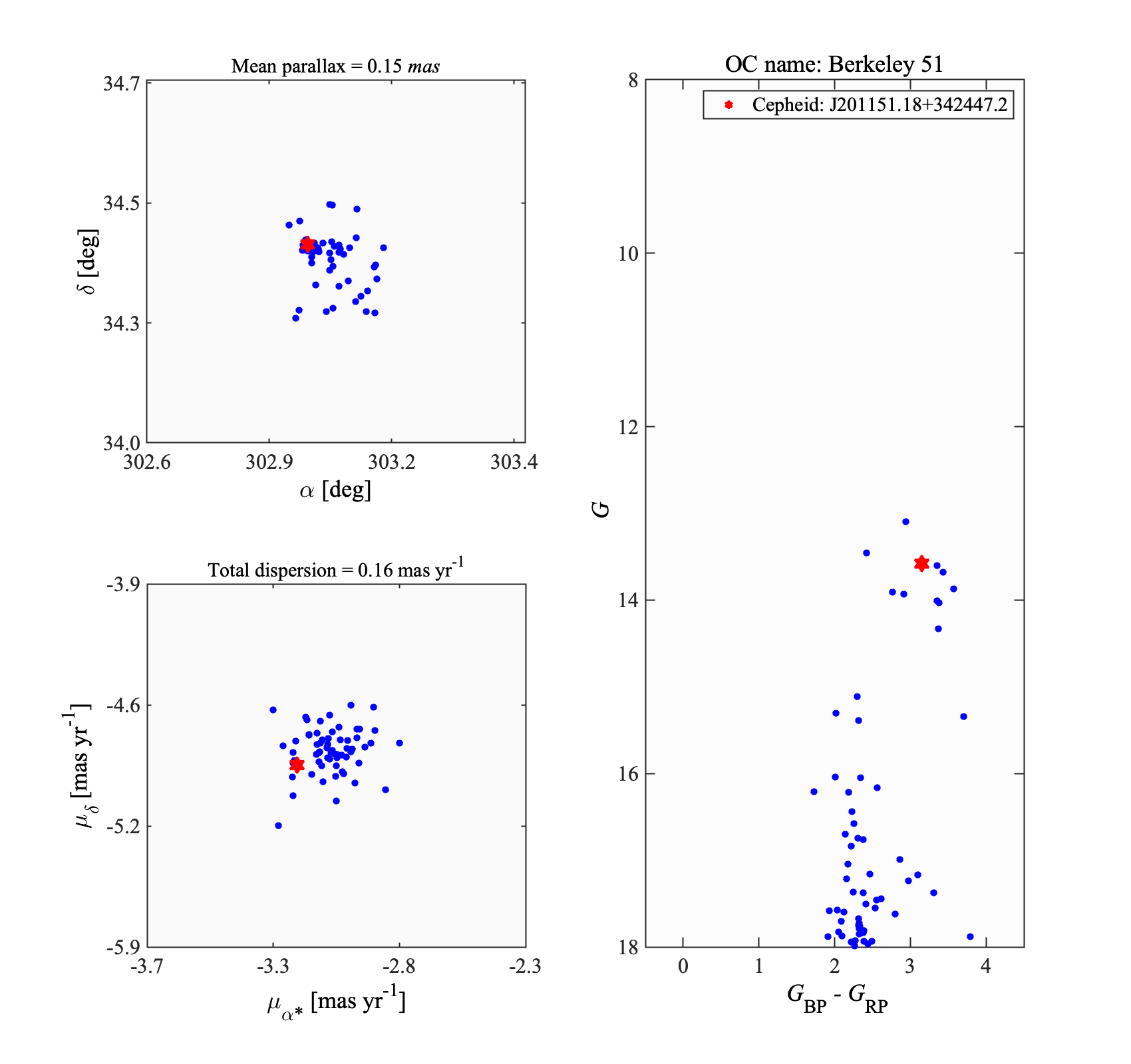}  \hspace{0.0cm}
\includegraphics[width=0.327\linewidth]{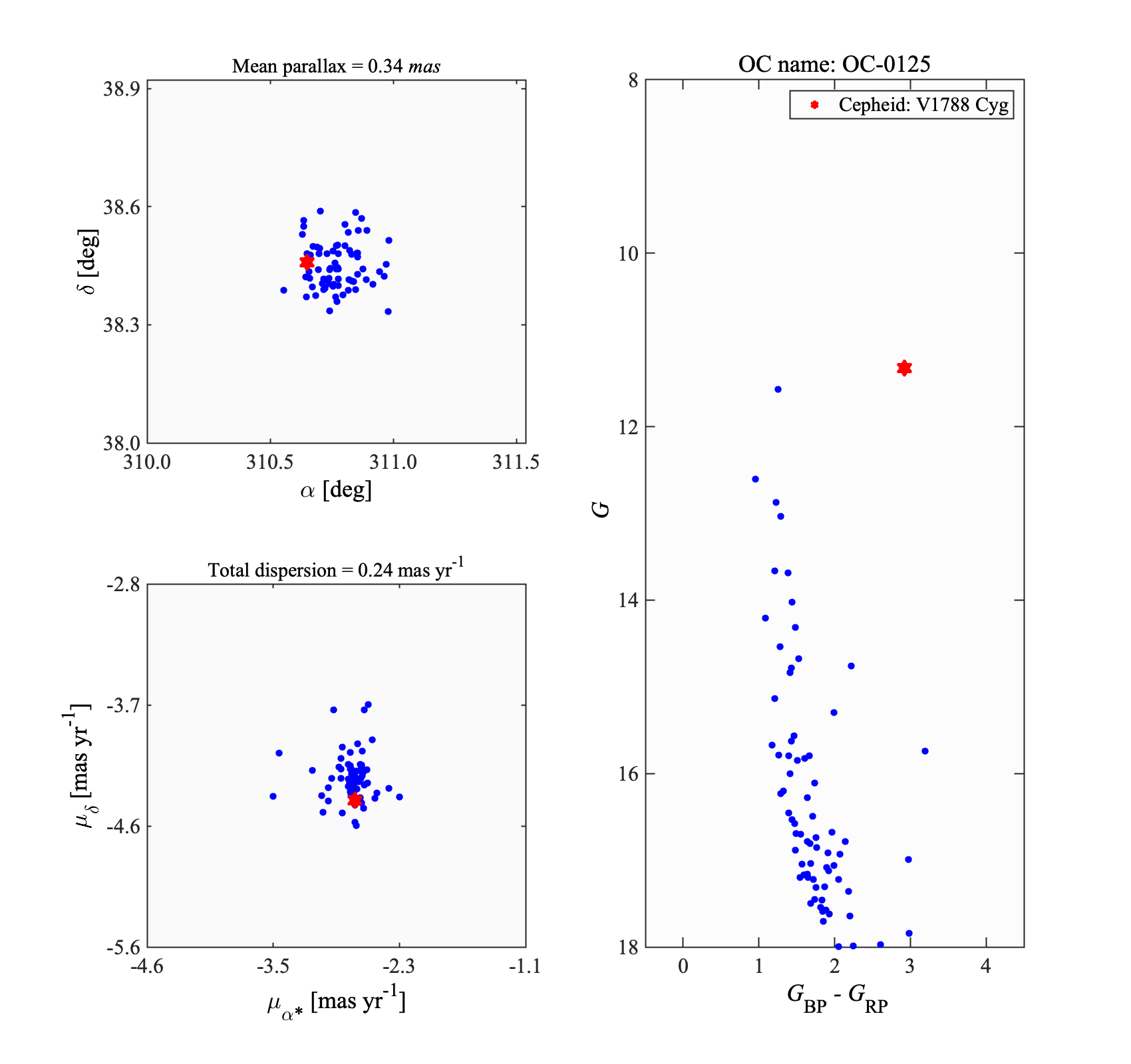} \hspace{0.0cm}
\includegraphics[width=0.327\linewidth]{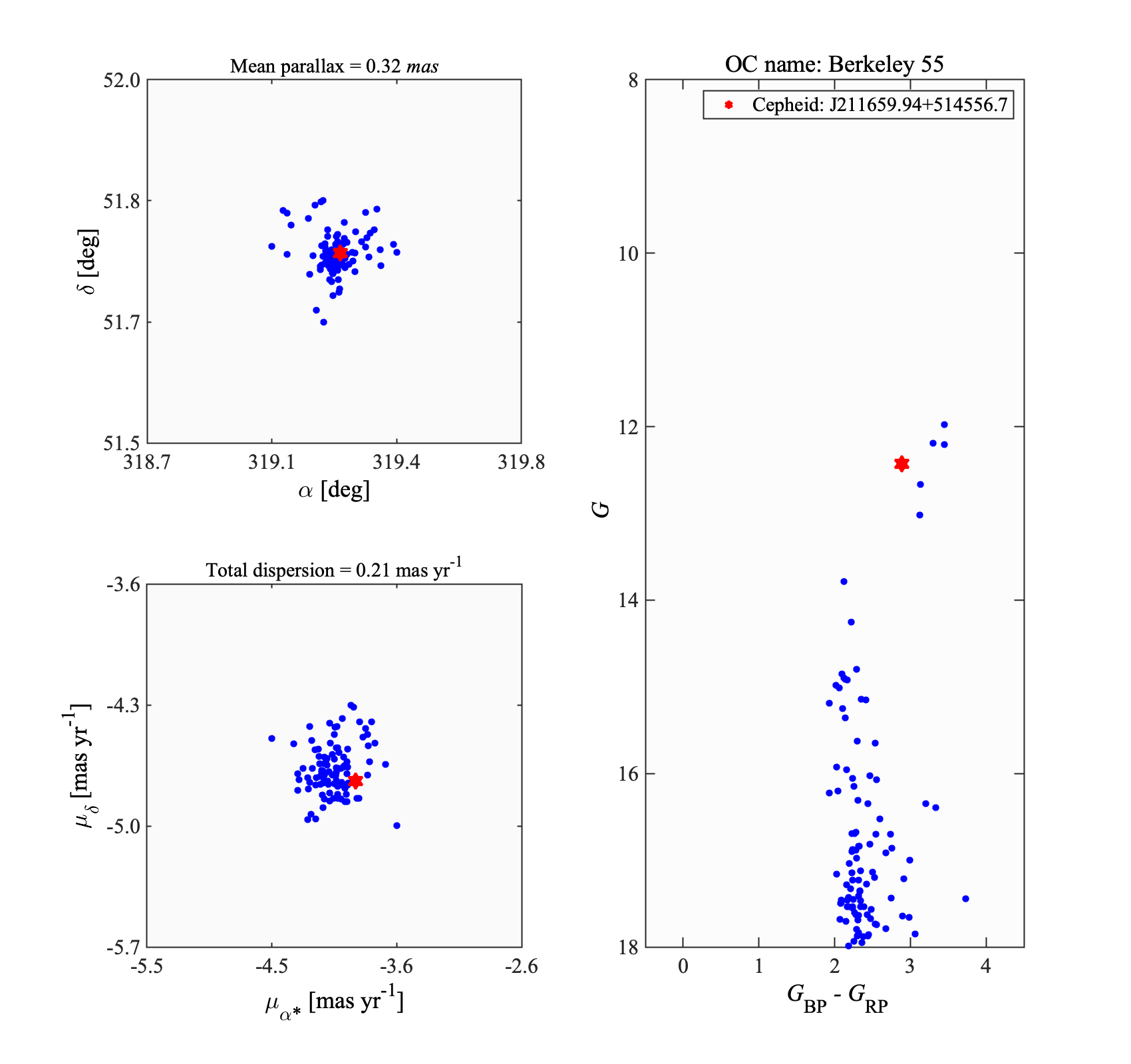} \hspace{0.0cm}
\includegraphics[width=0.327\linewidth]{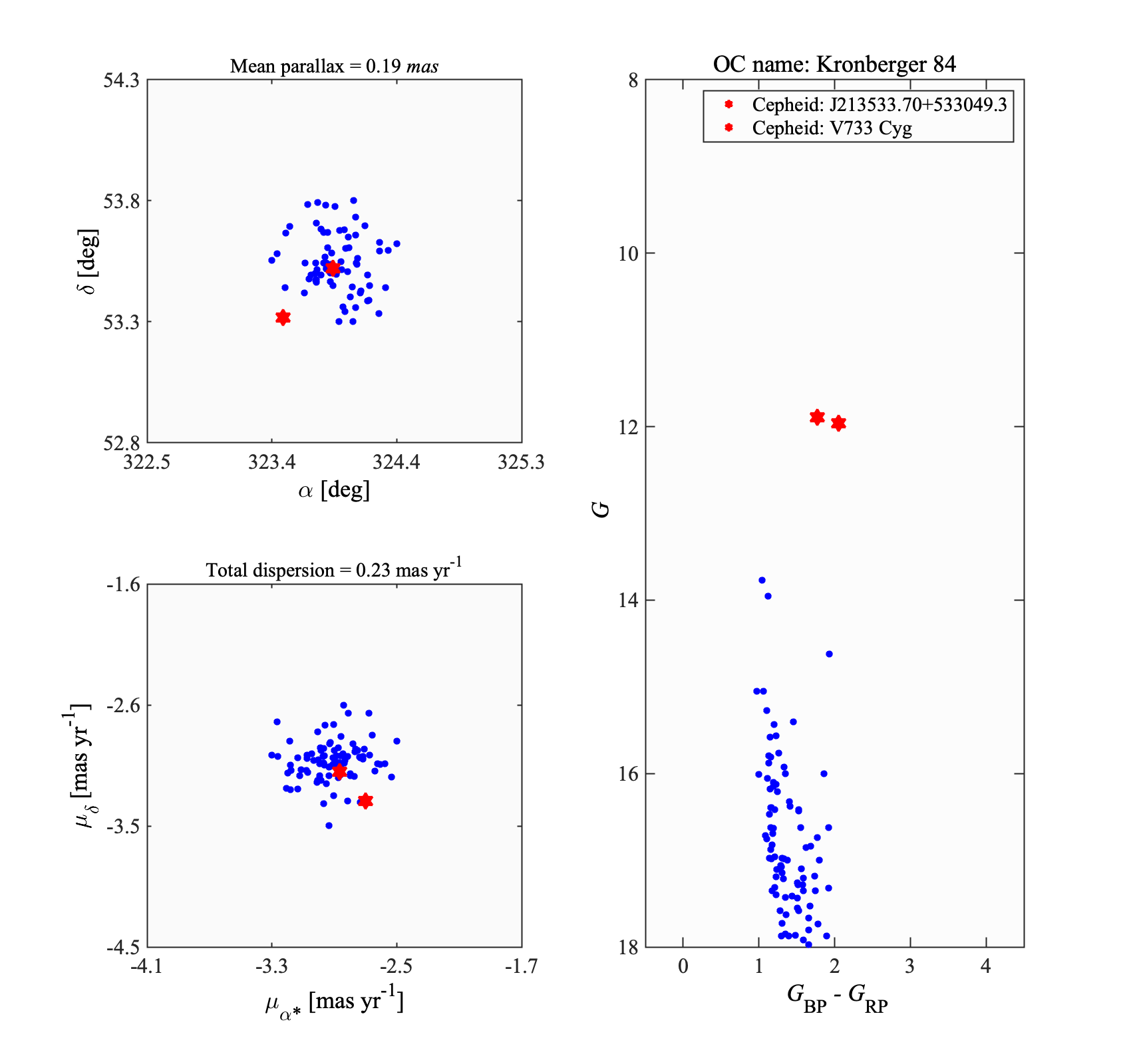}  \hspace{0.0cm}
\includegraphics[width=0.327\linewidth]{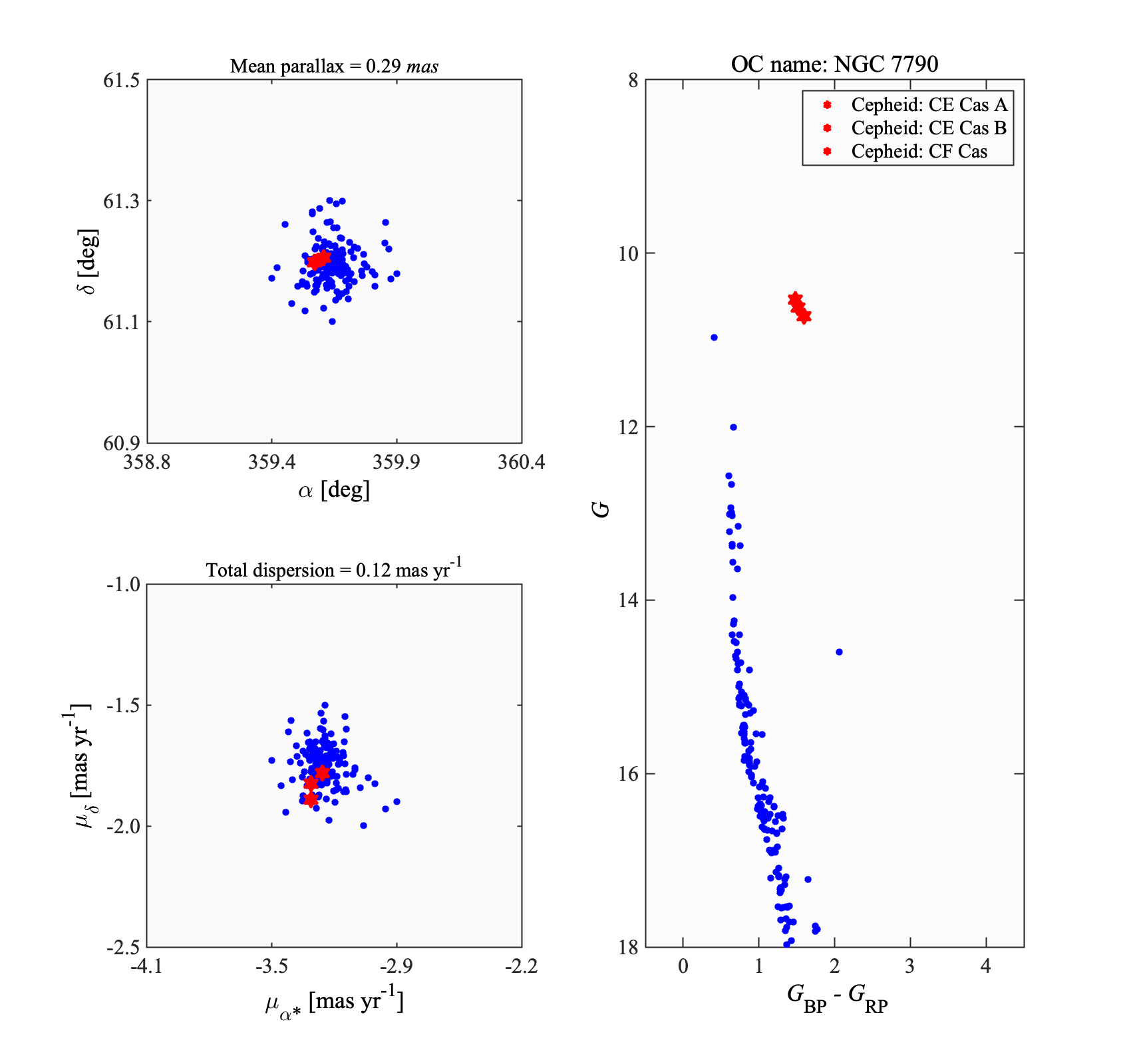} \hspace{0.0cm}
\includegraphics[width=0.327\linewidth]{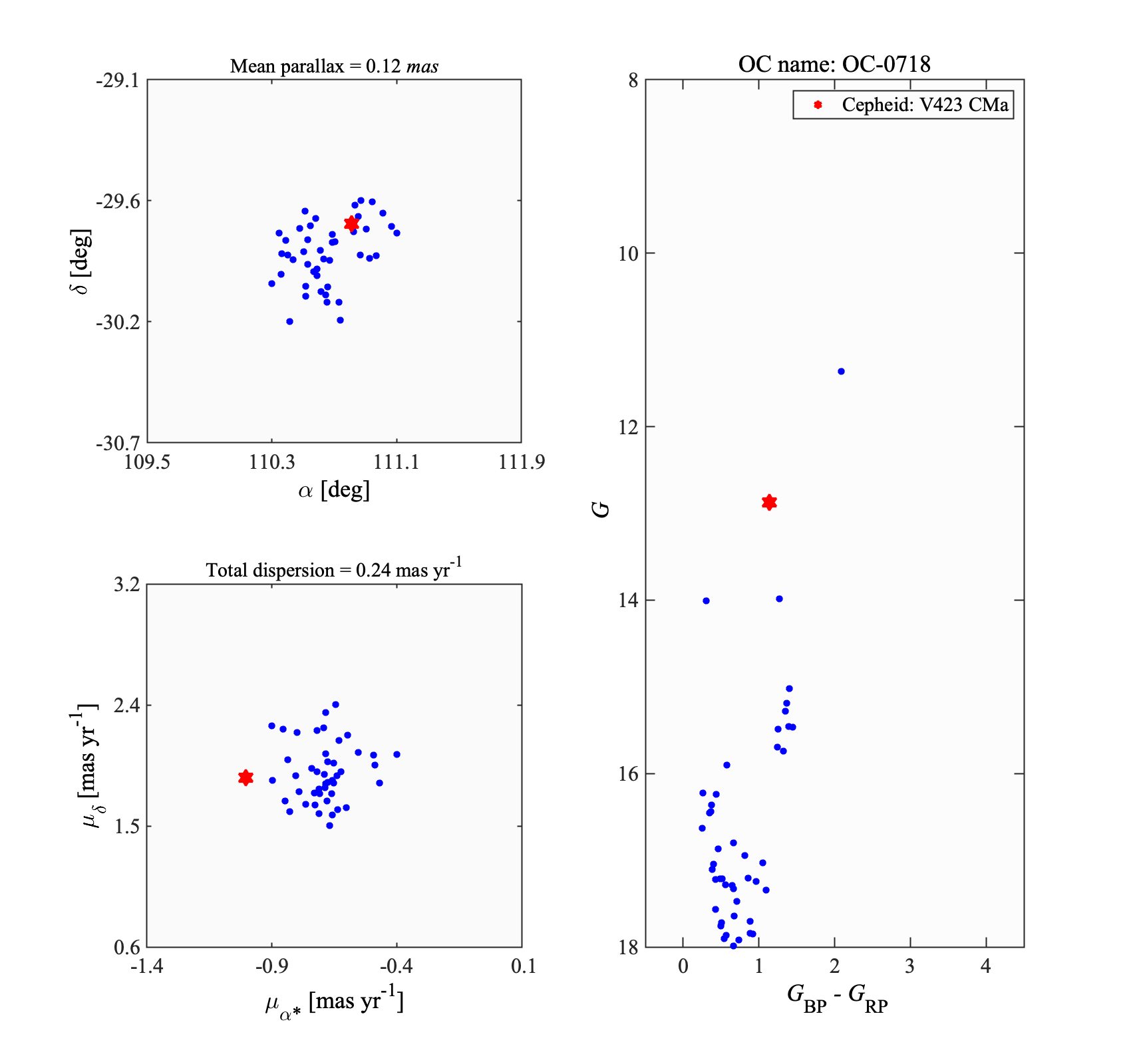} \hspace{0.0cm}
\includegraphics[width=0.327\linewidth]{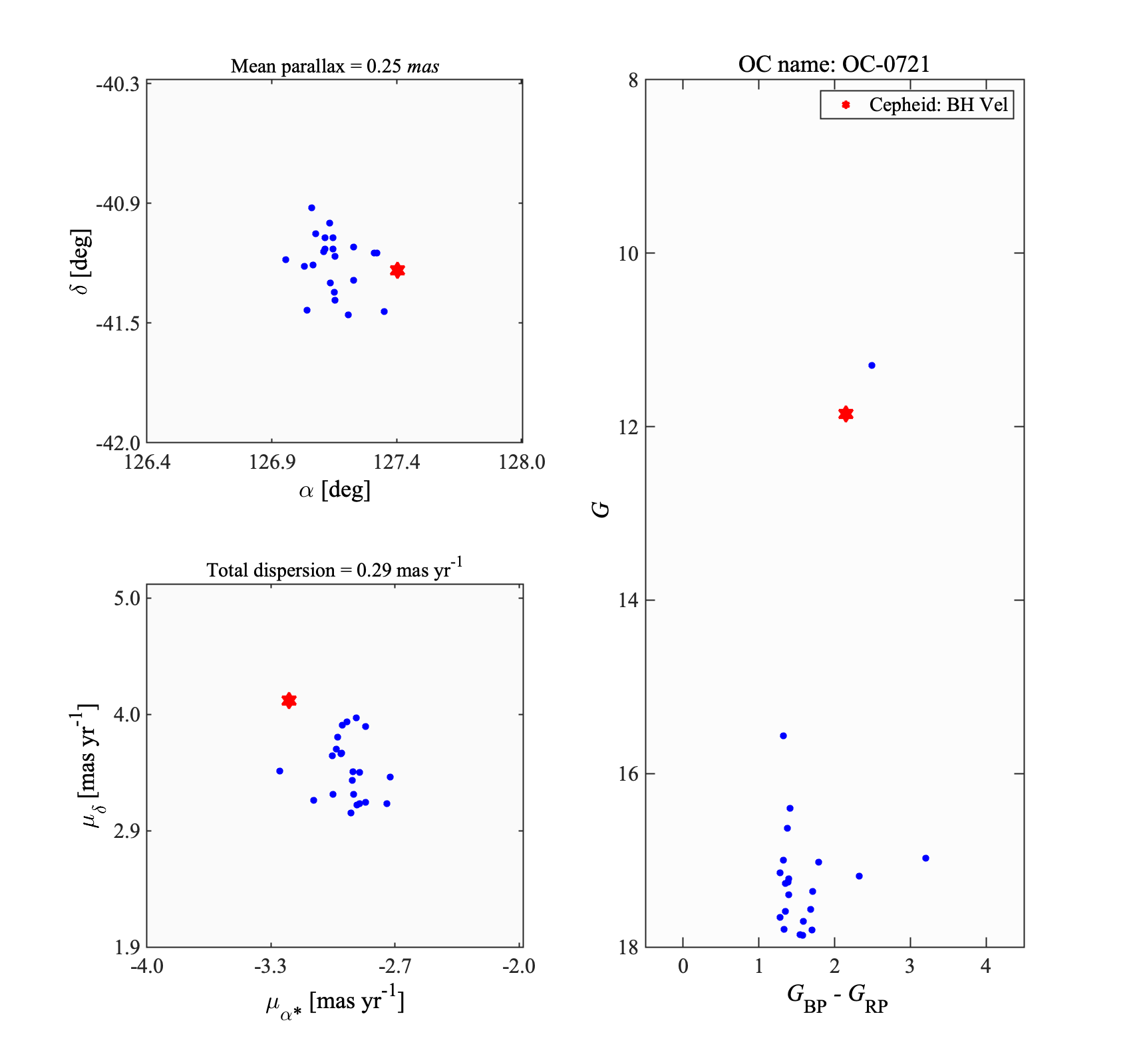} \hspace{0.0cm}
 \caption{continued. Here, the listed OCs are 
 NGC 6649, NGC 6664, FSR 0158, UBC 129, UBC 135, Berkeley 51, OC-0125, Berkeley 55, Kronberger 84, NGC 7790, OC-0718, and OC-0721.}
 %\label{}
 \end{figure*}

%%%%%%%%%%%%%%%%%%%%%%%%%%%%%%%%%%%%%%%%%%%%%%% Fig. 1
\addtocounter{figure}{-1}
 \begin{figure*}
 \centering
\includegraphics[width=0.327\linewidth]{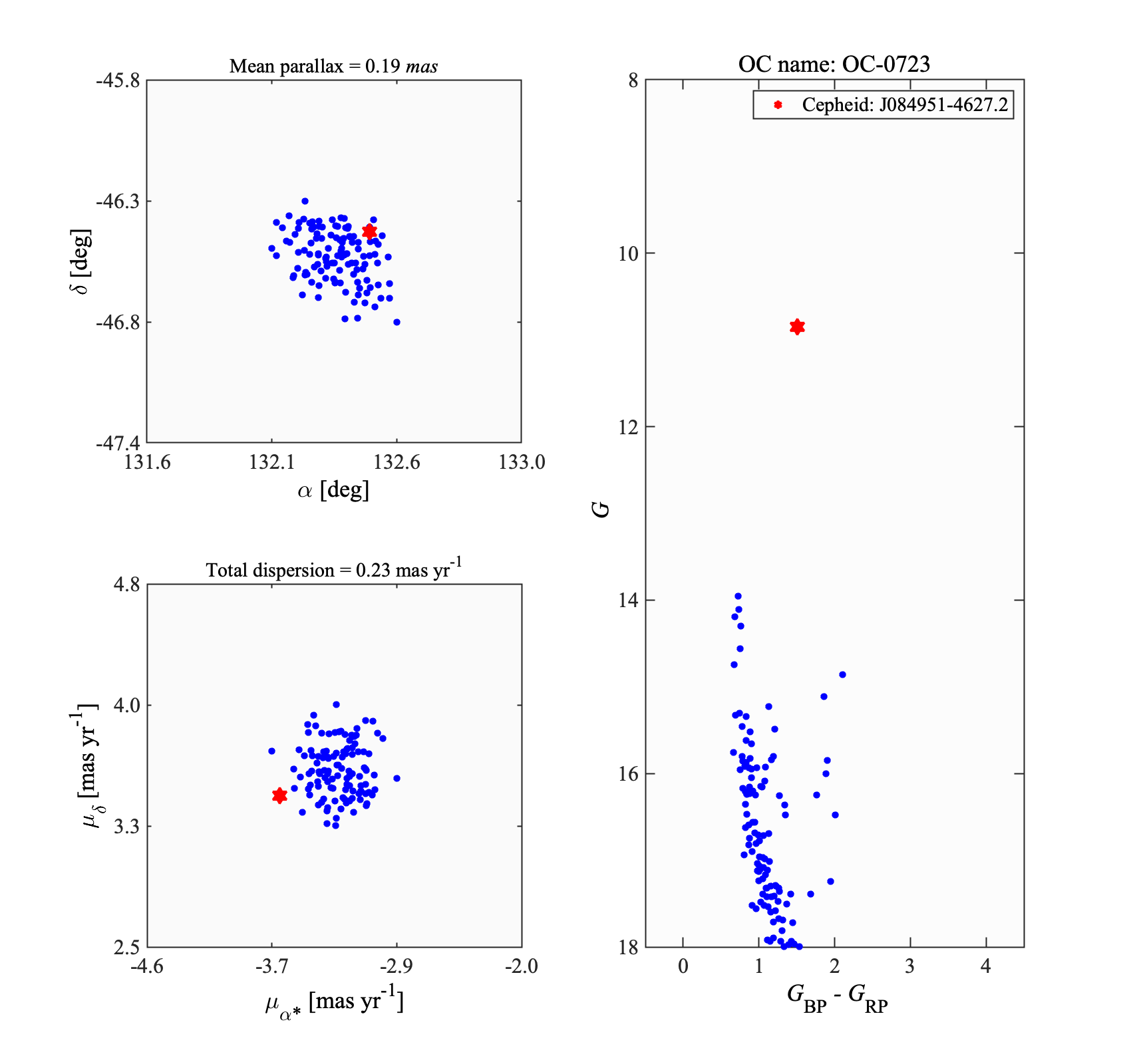} \hspace{0.0cm}
\includegraphics[width=0.327\linewidth]{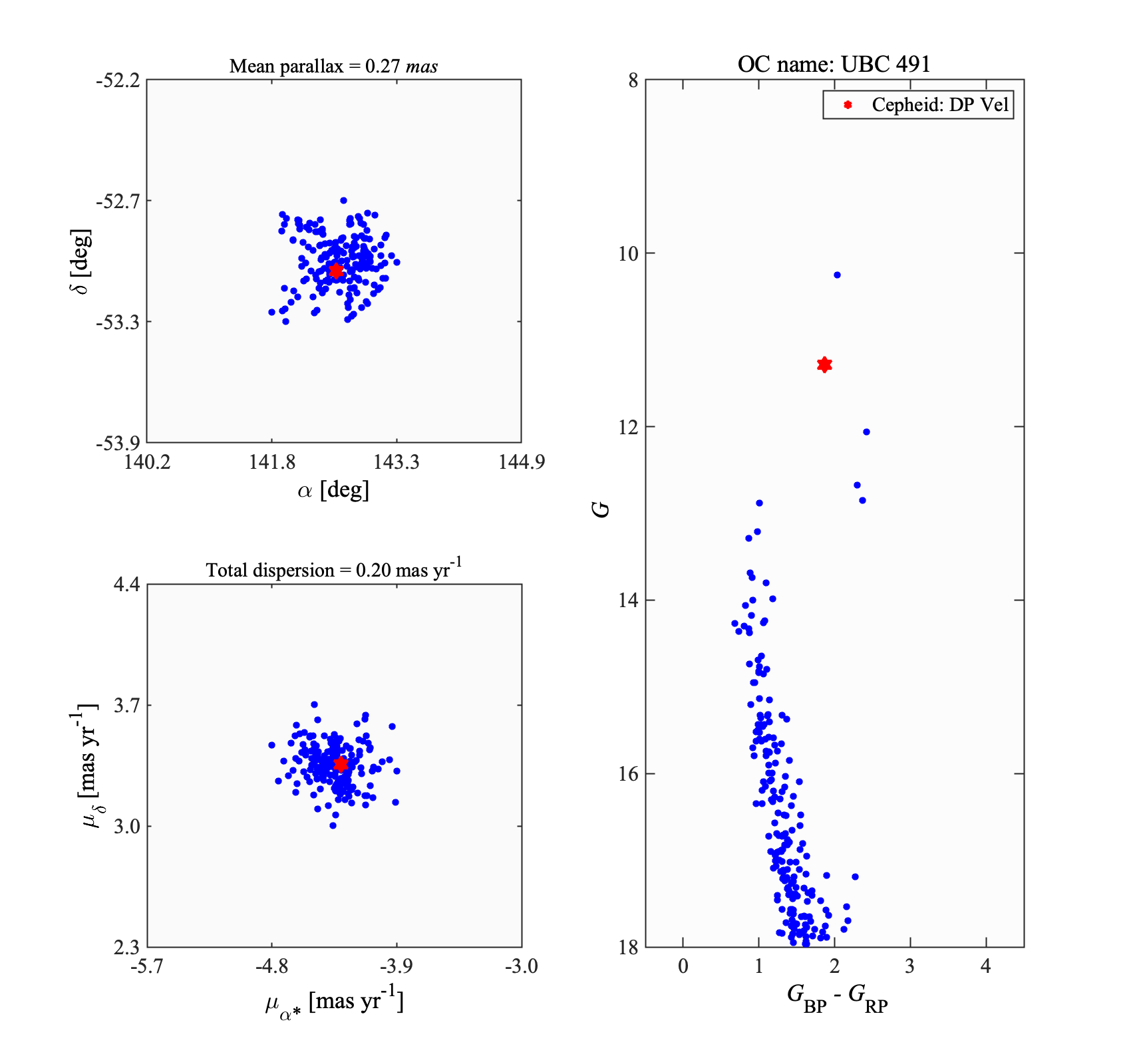} \hspace{0.0cm}
\includegraphics[width=0.327\linewidth]{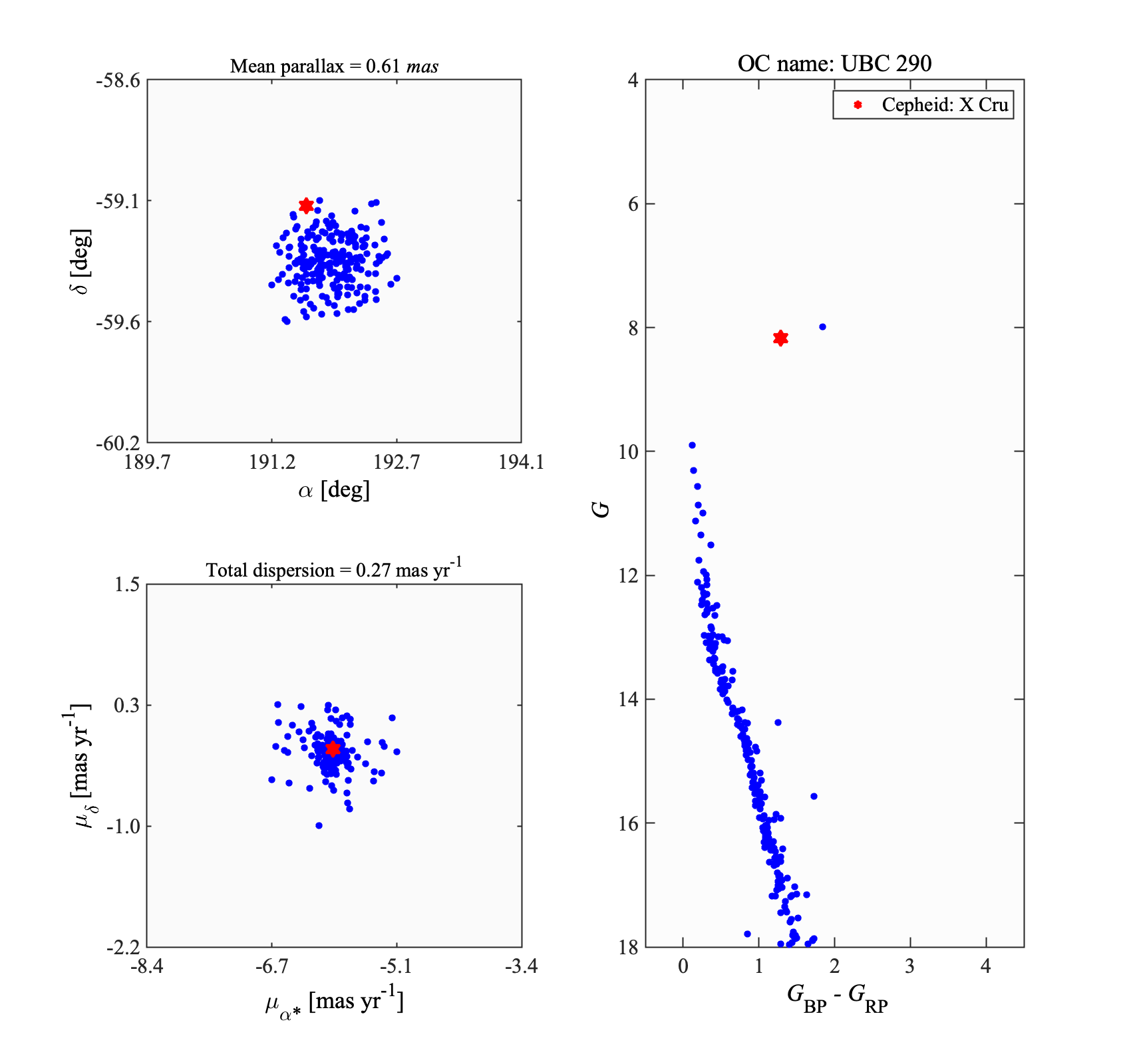}  \hspace{0.0cm}
\includegraphics[width=0.327\linewidth]{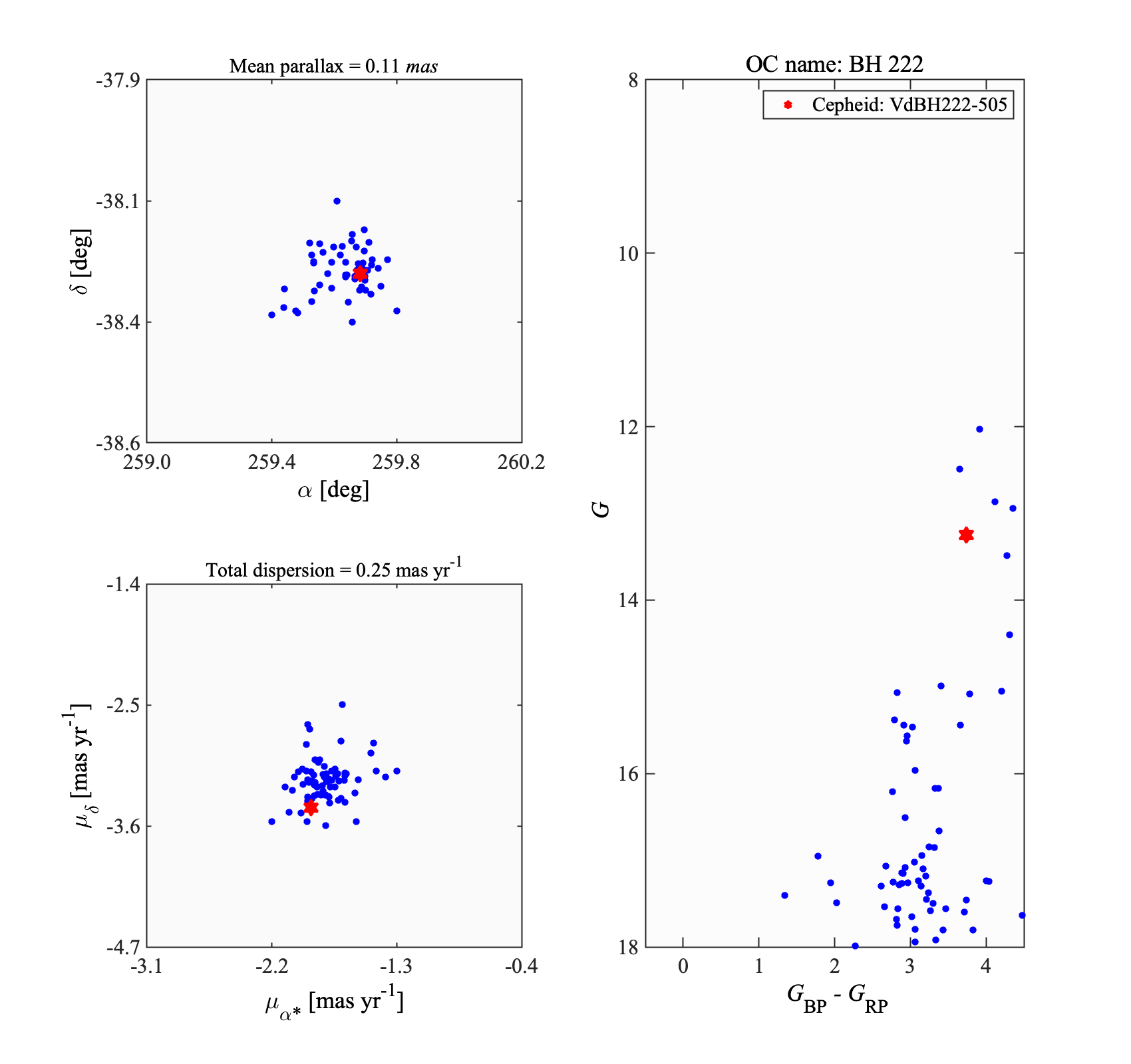}  \hspace{0.0cm}
\includegraphics[width=0.327\linewidth]{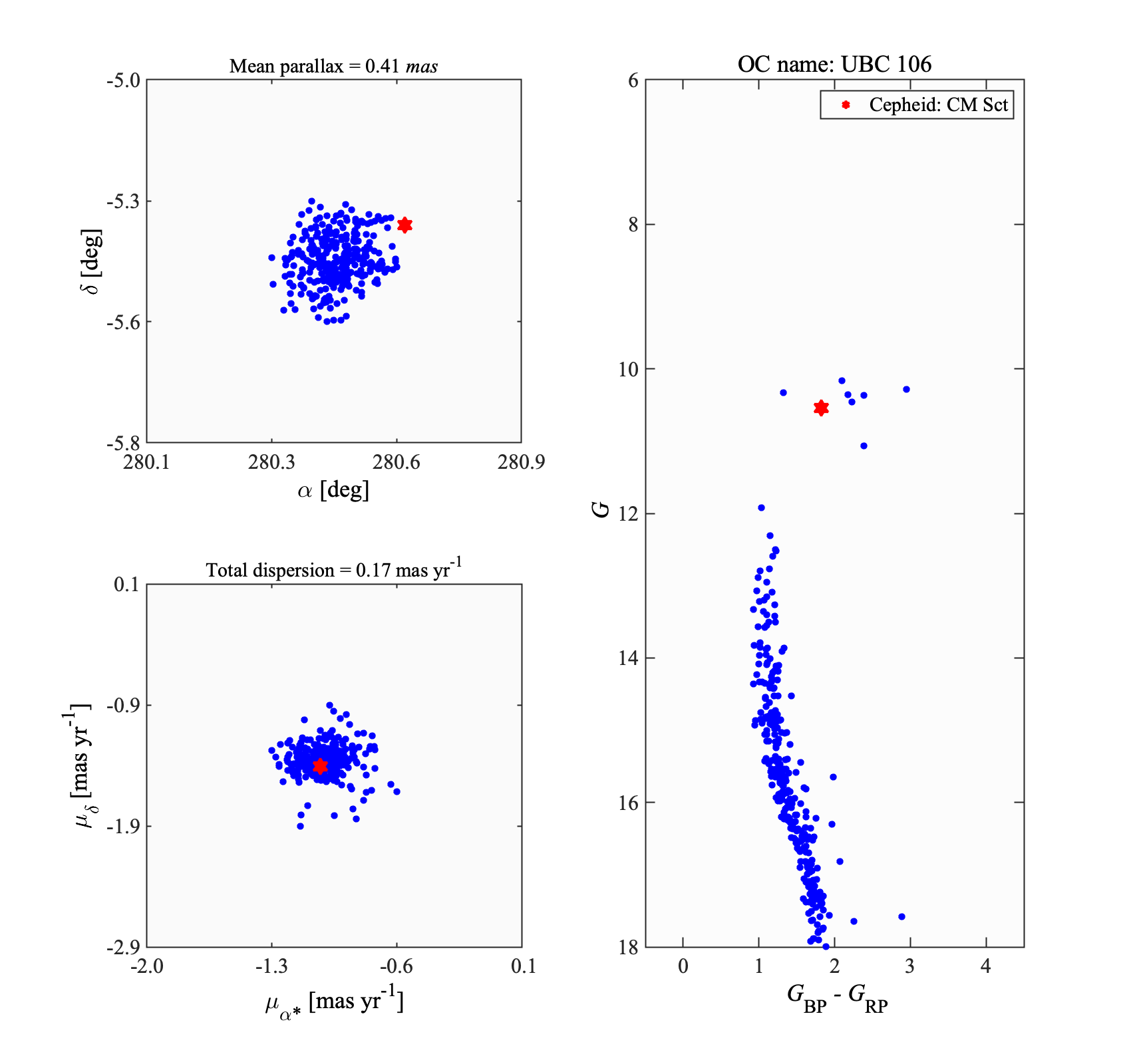} \hspace{0.0cm}
\includegraphics[width=0.327\linewidth]{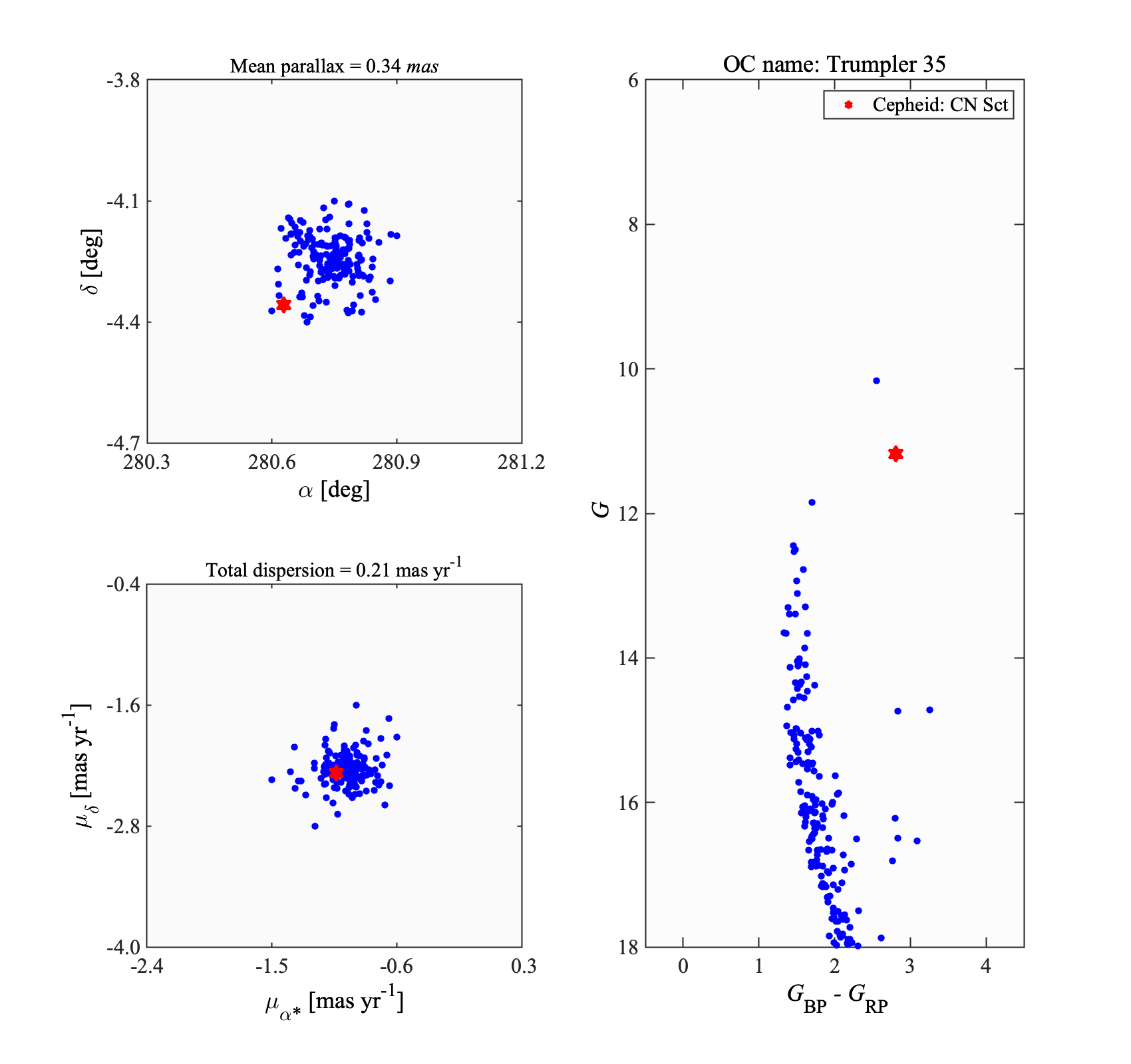}  \hspace{0.0cm}
\includegraphics[width=0.327\linewidth]{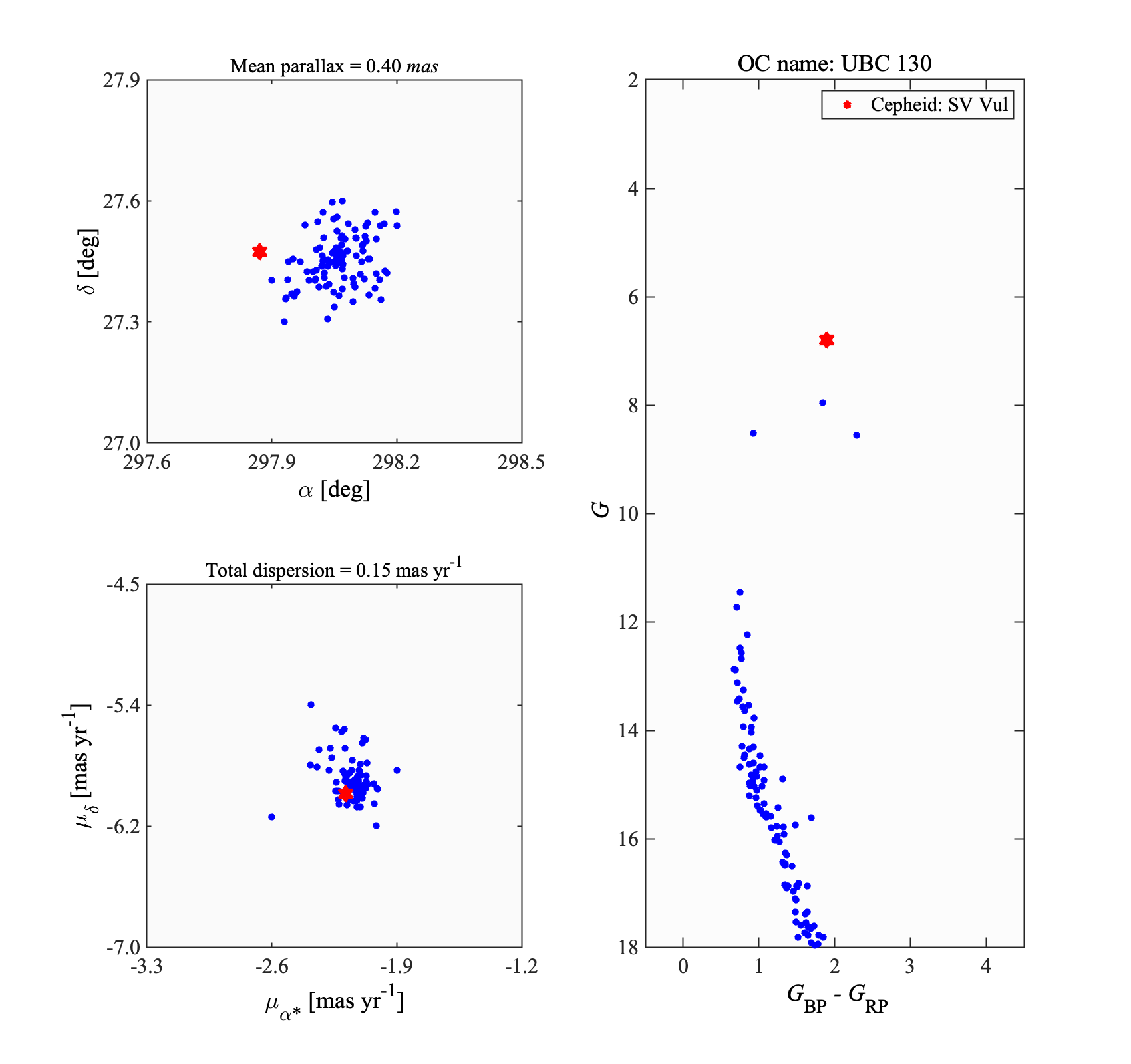} \hspace{0.0cm}
\includegraphics[width=0.327\linewidth]{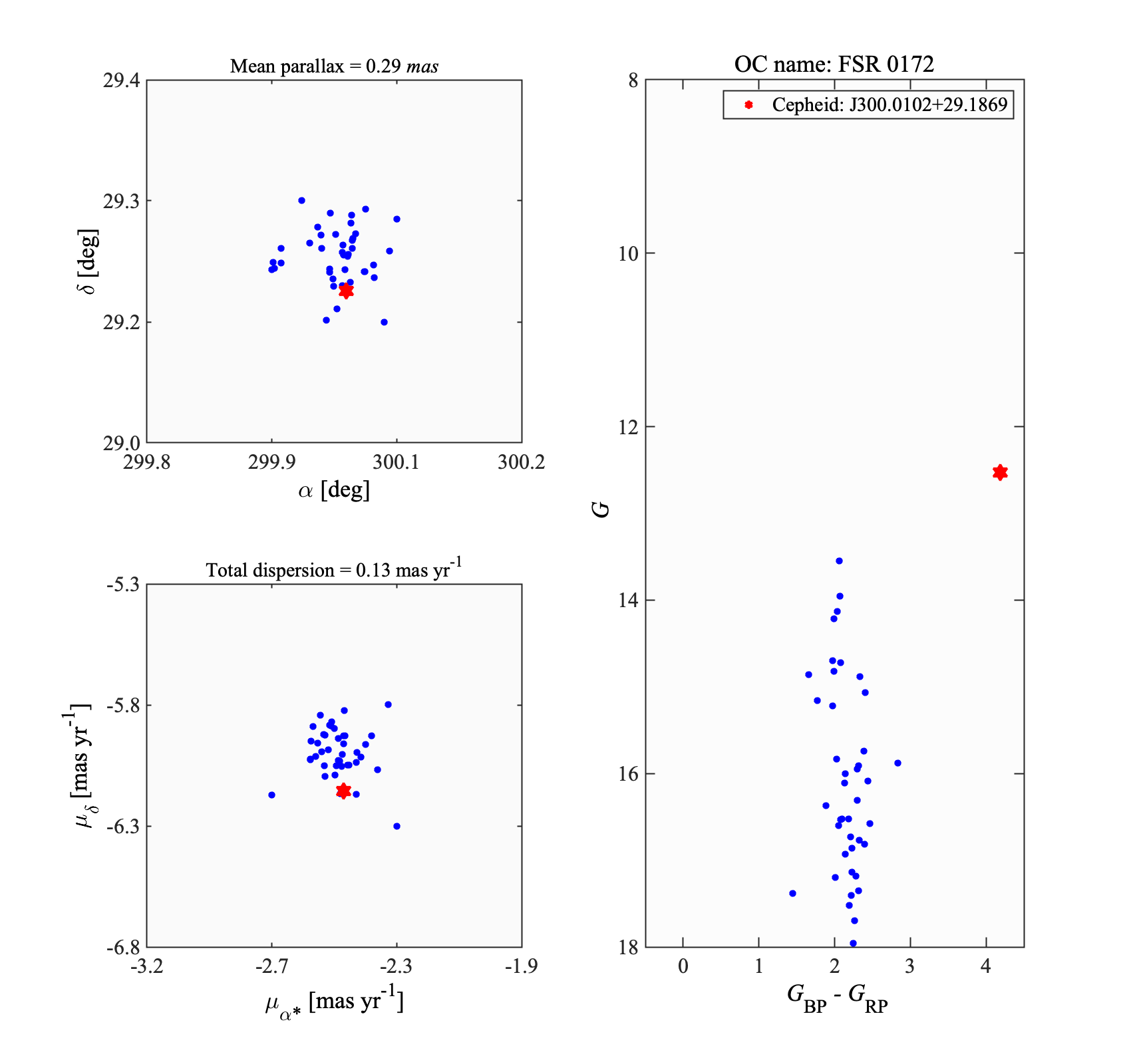} \hspace{0.0cm}
\includegraphics[width=0.327\linewidth]{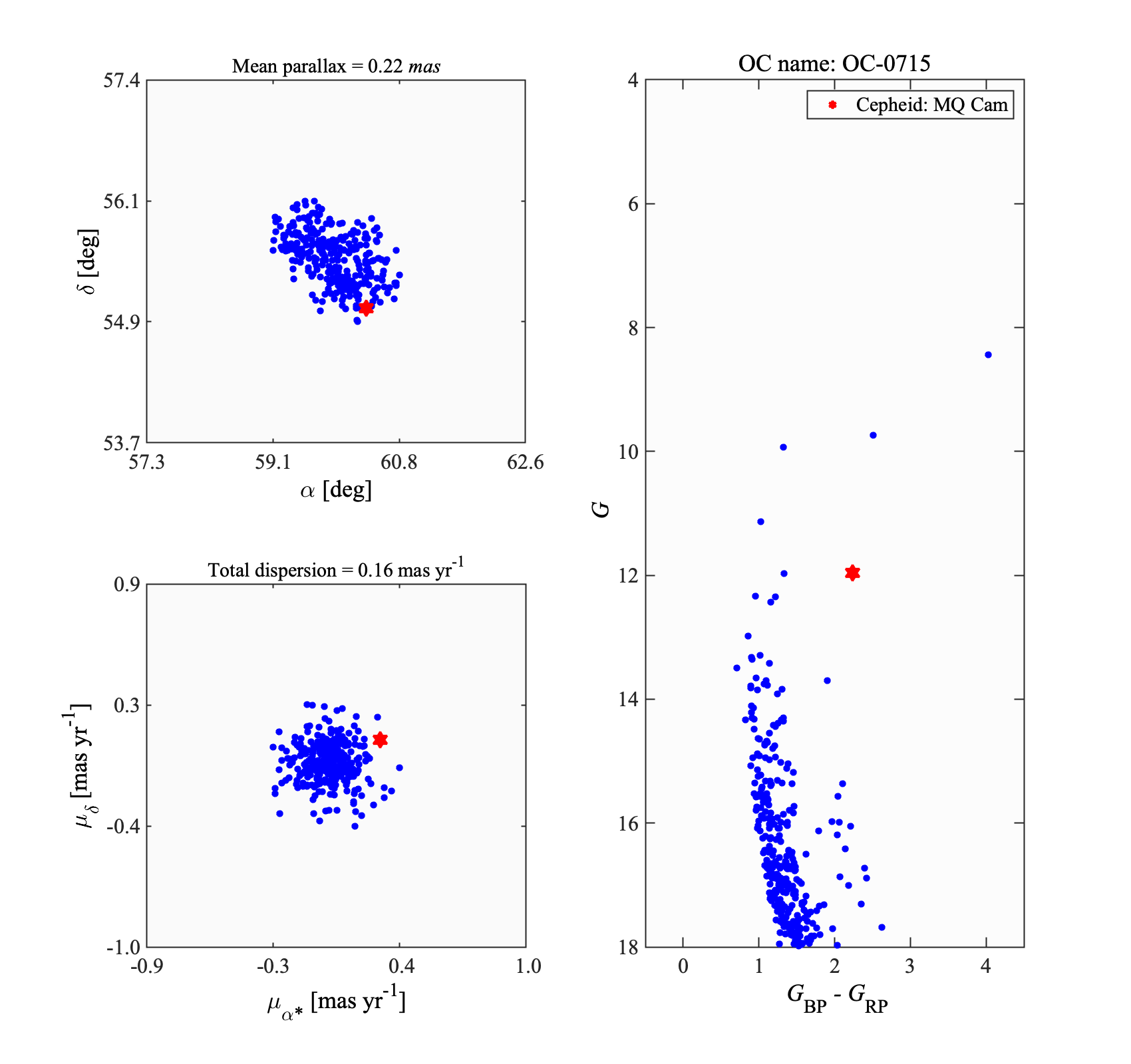}  \hspace{0.0cm}
\includegraphics[width=0.327\linewidth]{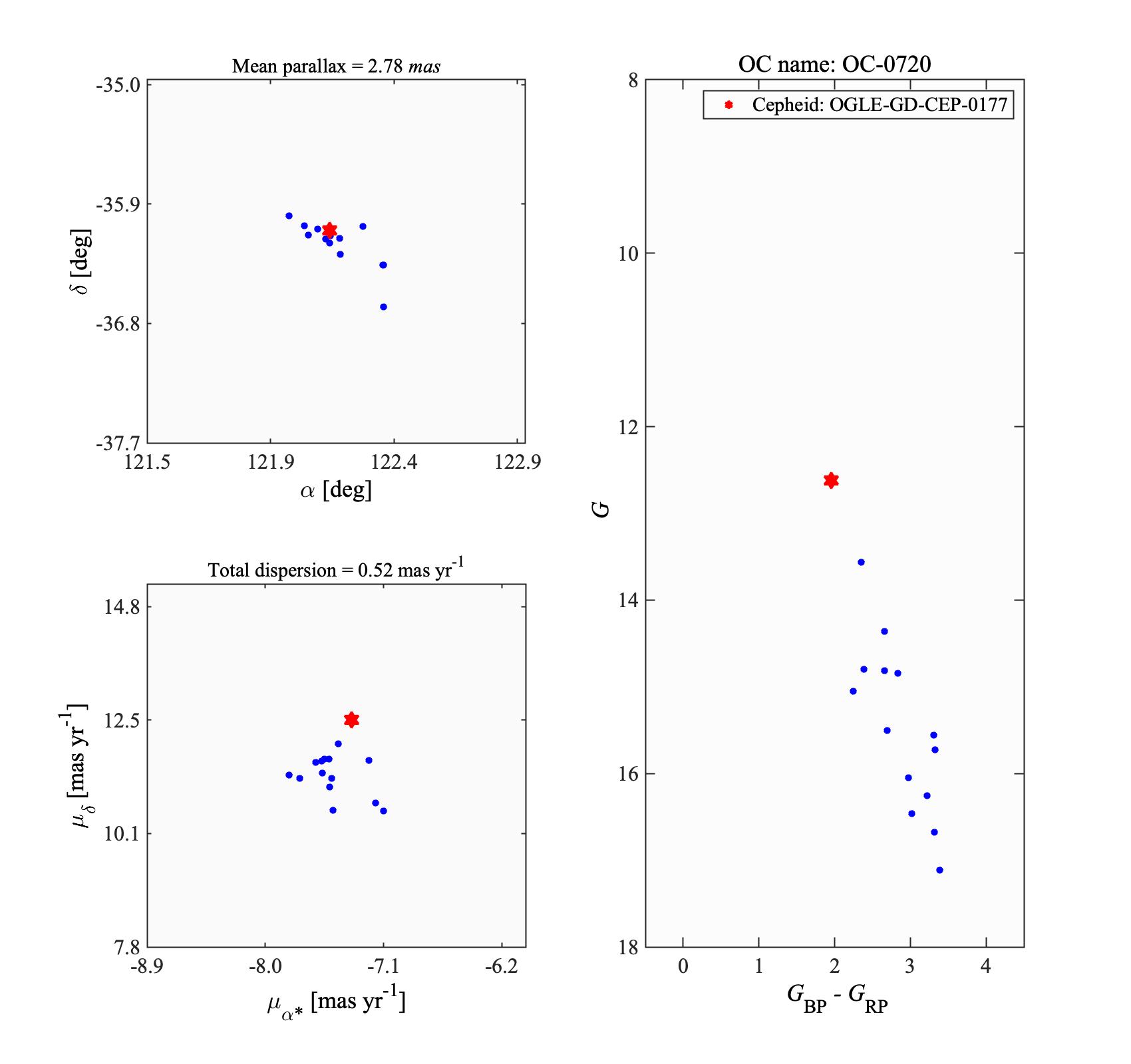} \hspace{0.0cm}
\includegraphics[width=0.327\linewidth]{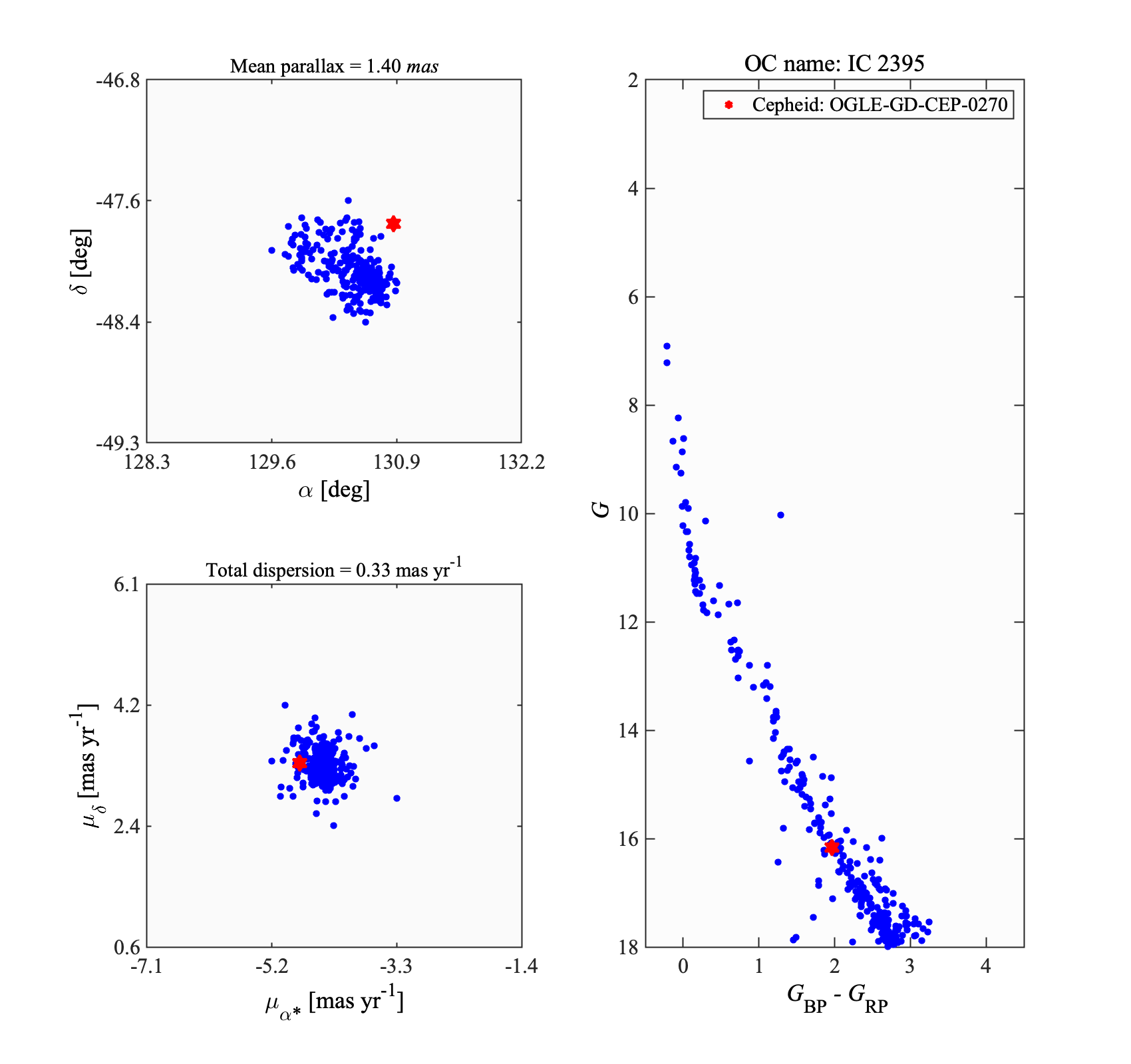} \hspace{0.0cm}
\includegraphics[width=0.327\linewidth]{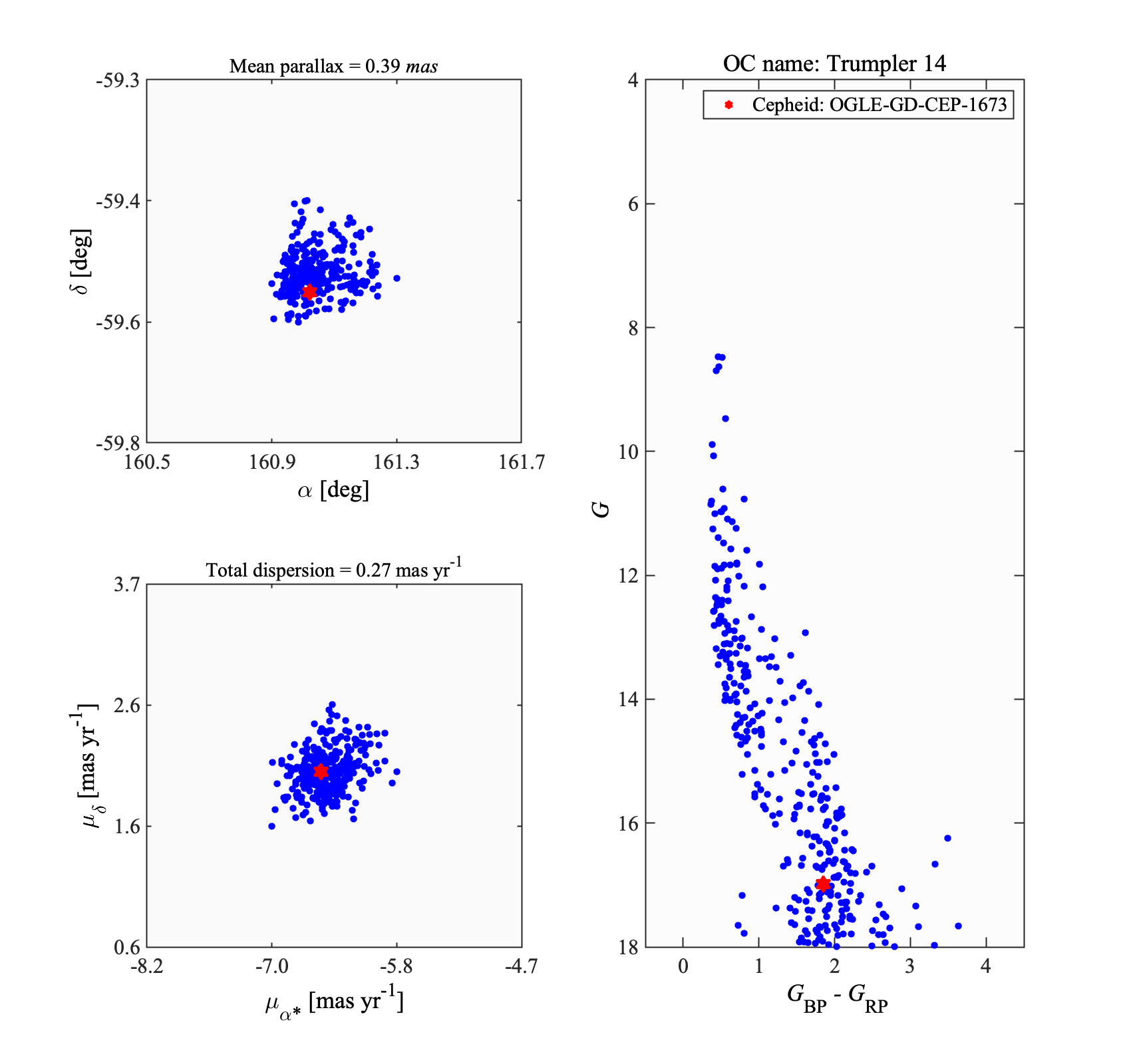} \hspace{0.0cm}
 \caption{continued. Here, the listed OCs are 
OC-0723, UBC 491, UBC 290, BH 222, UBC 106, Trumpler 35, UBC 130, FSR 0172, OC-0715, OC-0720, IC 2395, and Trumpler 14.}
 %\label{}
 \end{figure*}

%%%%%%%%%%%%%%%%%%%%%%%%%%%%%%%%%%%%%%%%%%%%%%% Fig. 1
\addtocounter{figure}{-1}
 \begin{figure*}
 \centering
\includegraphics[width=0.327\linewidth]{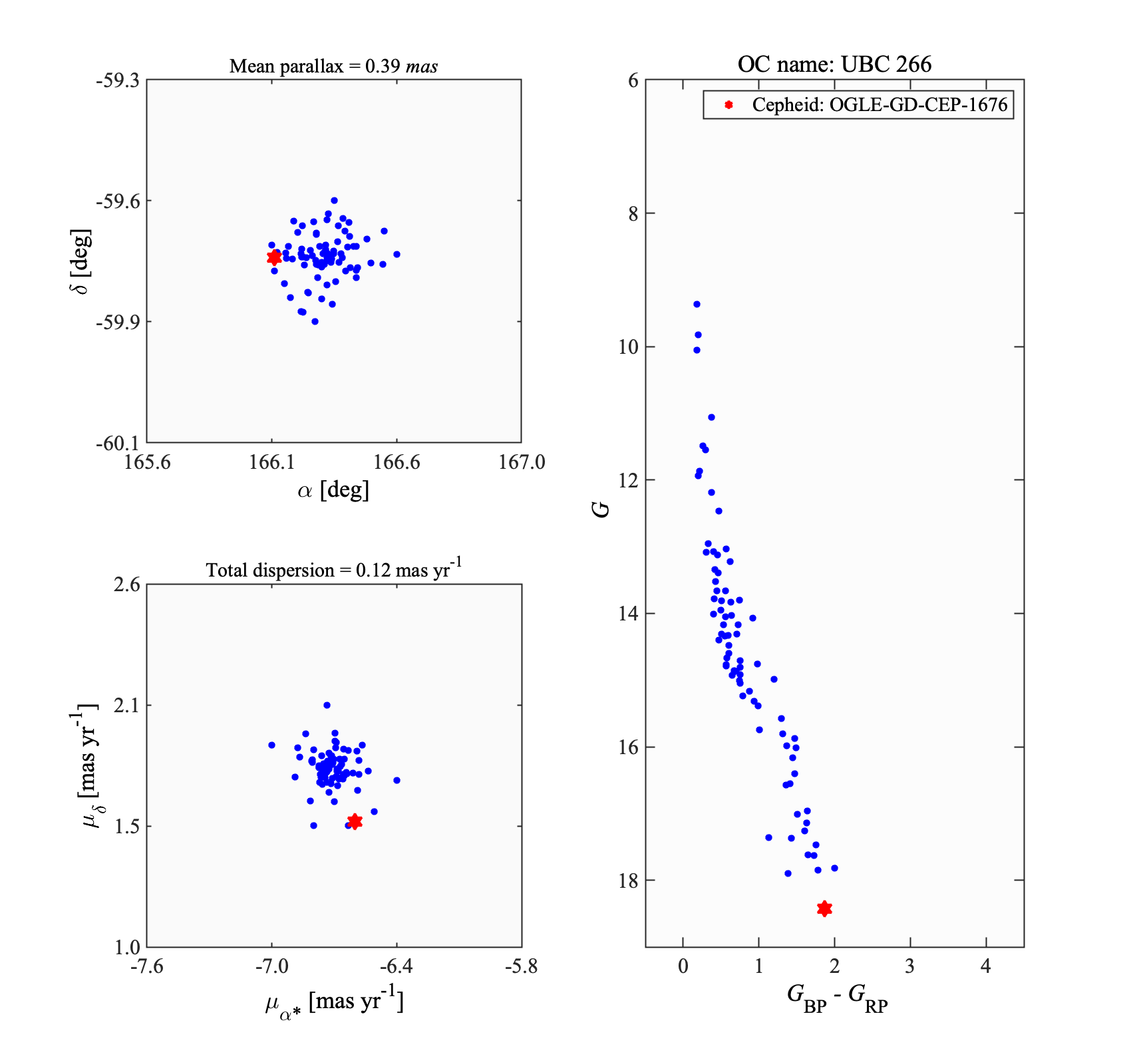} \hspace{0.0cm}
\includegraphics[width=0.327\linewidth]{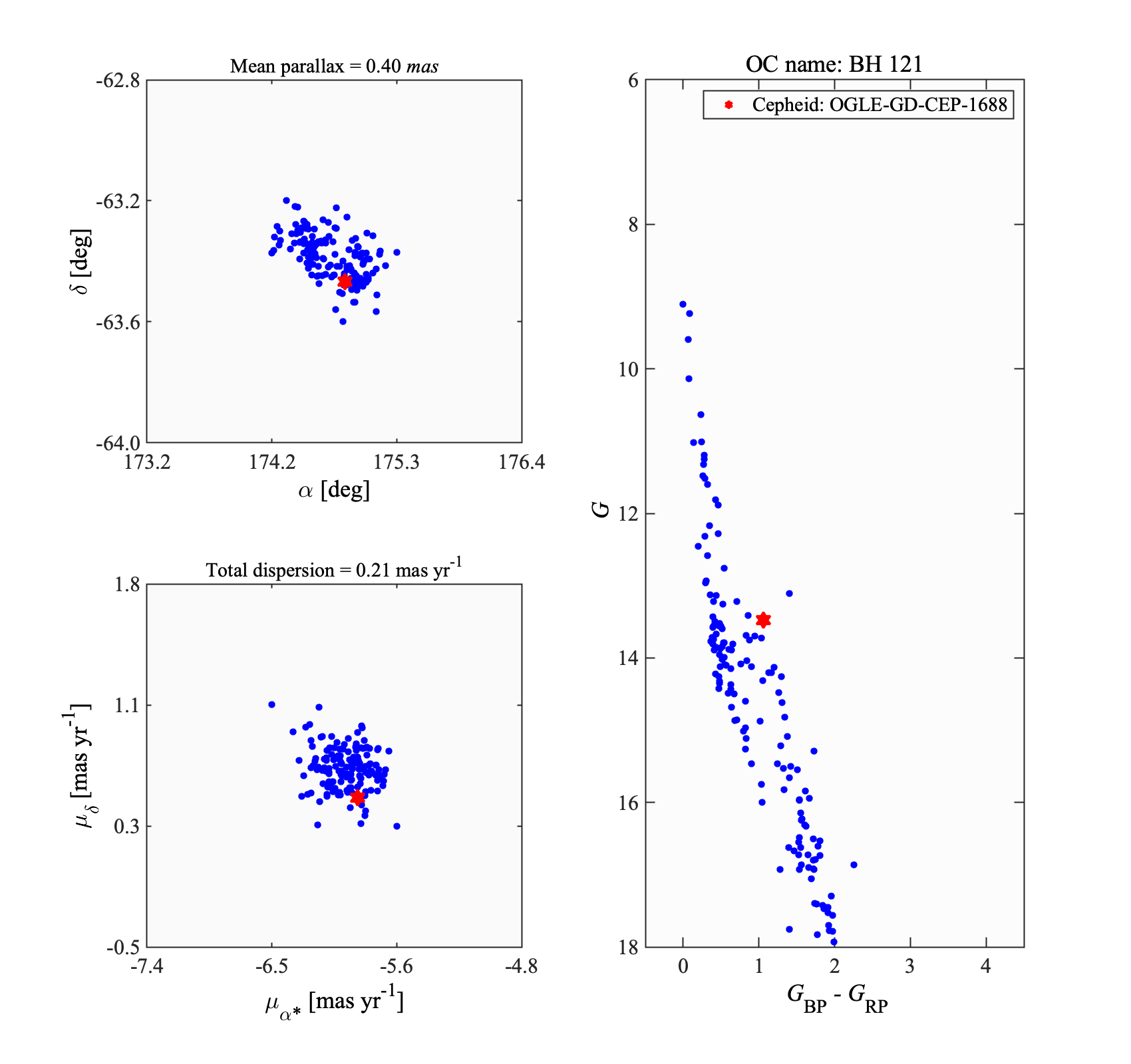} \hspace{0.0cm}
\includegraphics[width=0.327\linewidth]{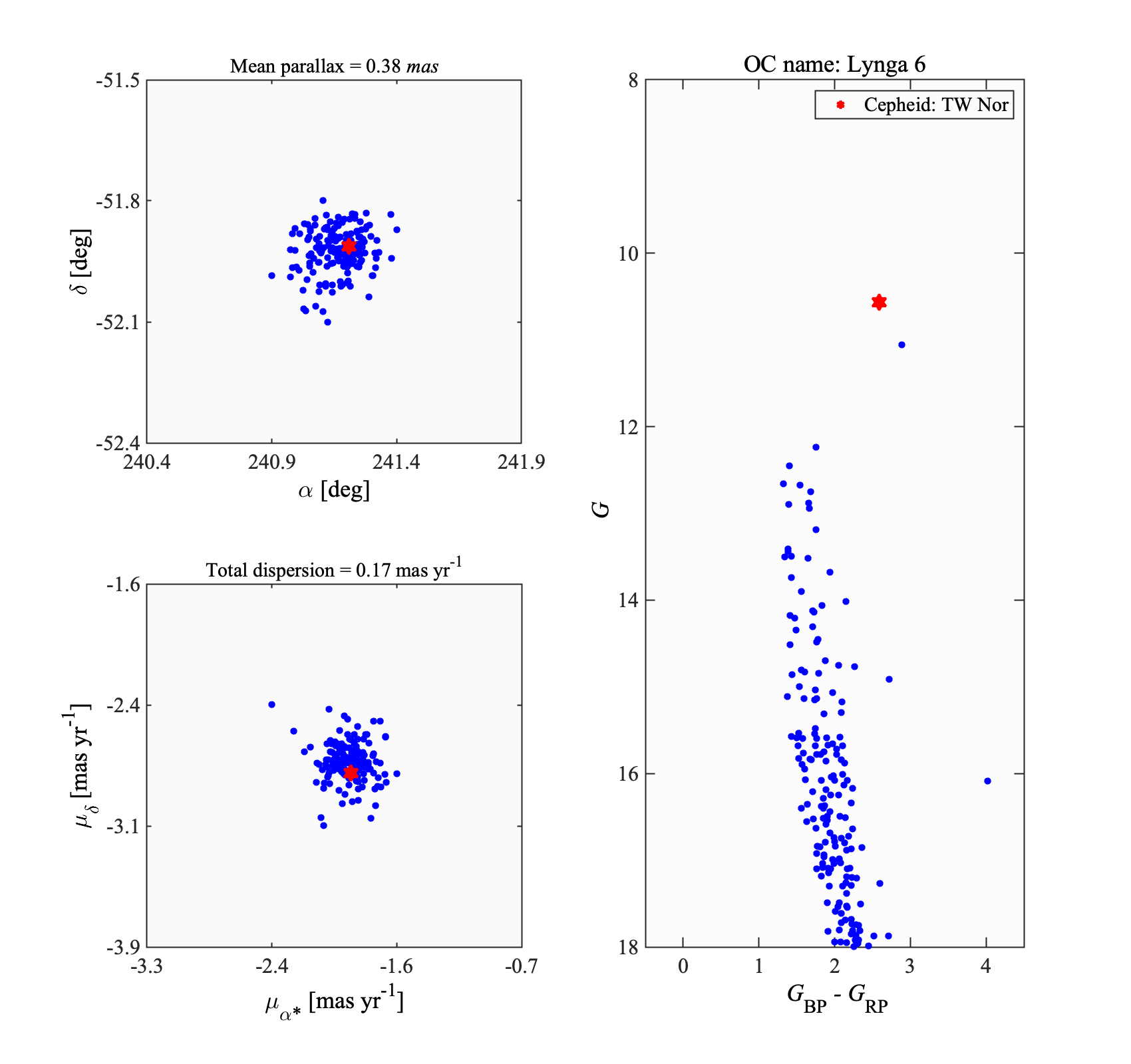} \hspace{0.0cm}
\includegraphics[width=0.327\linewidth]{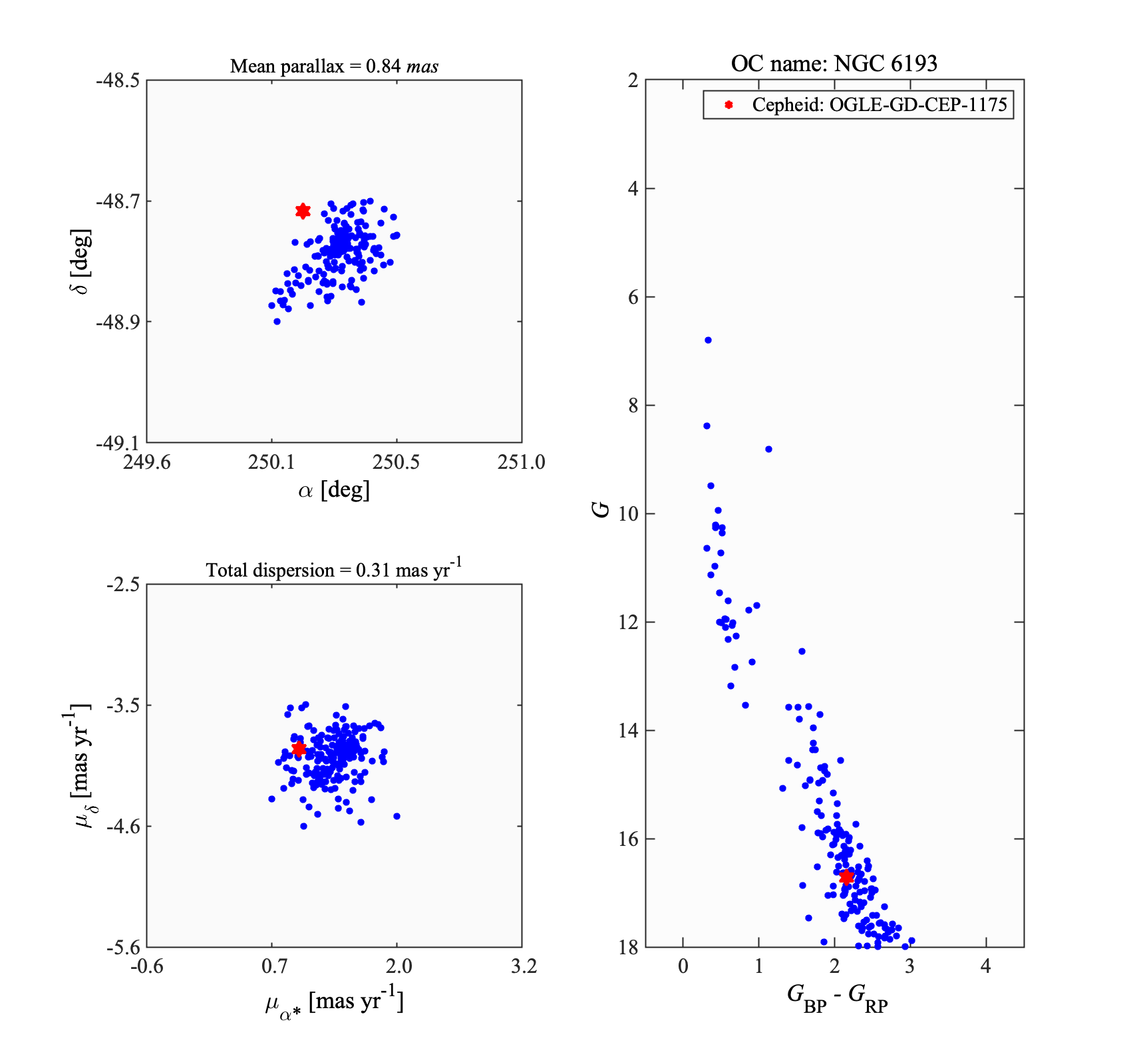} \hspace{0.0cm}
\includegraphics[width=0.327\linewidth]{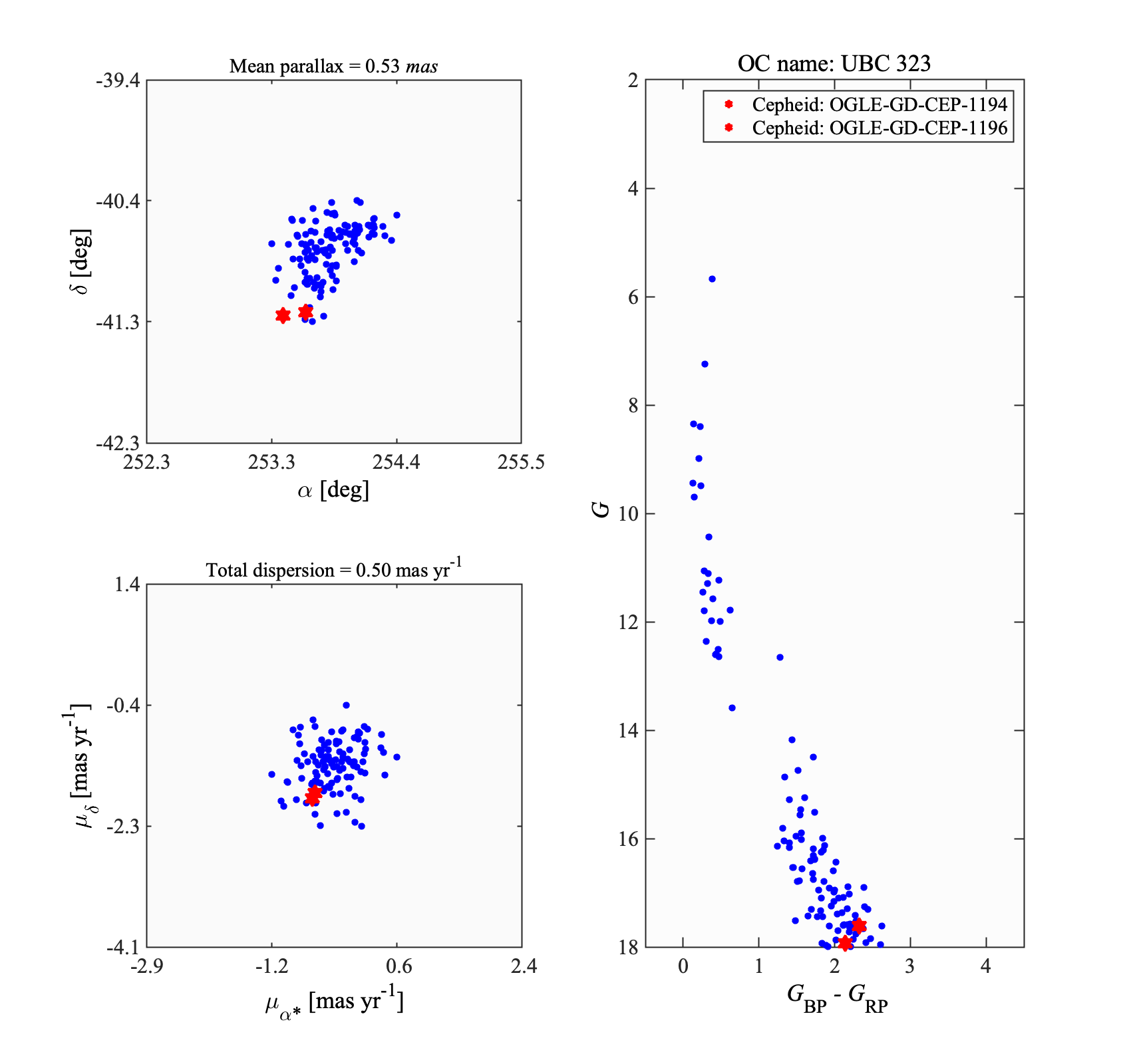} \hspace{0.0cm}
\includegraphics[width=0.327\linewidth]{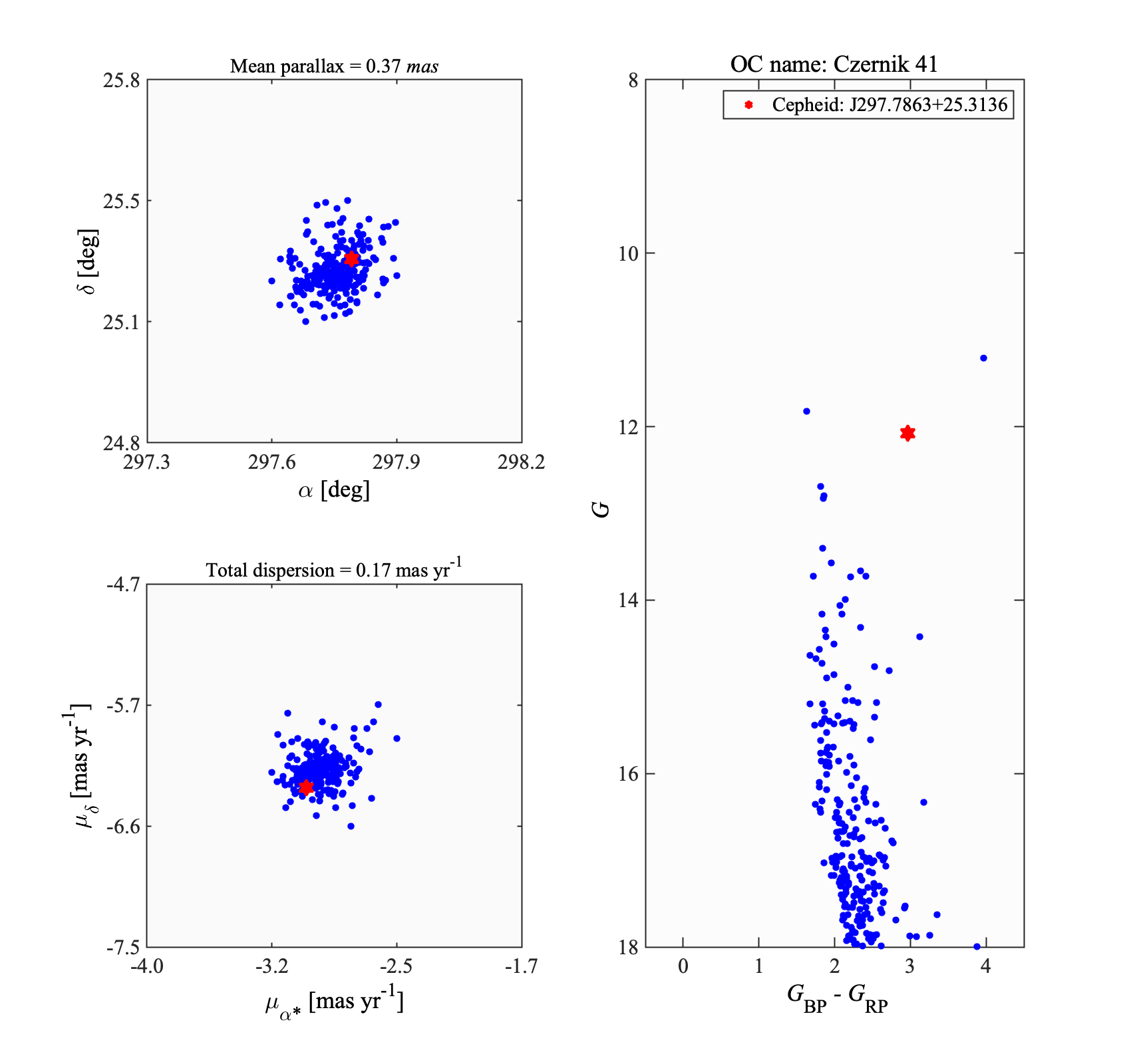} \hspace{0.0cm}
 \caption{ccontinued. Here, the listed OCs are 
UBC 266, BH 121, Lynga 6, NGC 6193, UBC 323, and Czernik 41.}
 %\label{}
 \end{figure*}
%%%%%%%%%%%%%%%%%%%%%%%%%%%%%%%%%%%%%%%%%%%%%%%
%%%%%%%%%%%%%%%%%%%%%%%%%%%%%%%%%%%%%%%%%%%%

\end{appendix}

%%%%%%%%%%%%%%%%%%%%%%%%%%%%%%%%%%%%%%%%%%%%

%%%%%%%%%%%%%%%%%%%%%%%%%%%%%%%%%%%%%%%%%%%%
\end{document}